\RequirePackage{fix-cm}
\documentclass[twocolumn]{svjour3}          
\smartqed  
\usepackage{graphicx}
	\usepackage[utf8]{inputenc}
	\usepackage[english]{babel}              
	\usepackage{amssymb,amsmath,amsfonts}
	\SetSymbolFont{letters}{bold}{OML}{cmm}{b}{it}
	\SetSymbolFont{operators}{bold}{OT1}{cmr}{bx}{n}
	\usepackage{color}
	\usepackage{mathrsfs}
	\usepackage{mathabx}
	\usepackage[normalem]{ulem} 
\usepackage{array,multirow,makecell}
\setcellgapes{1pt}
\makegapedcells

	\newcommand{\be}{\begin{equation}}
	\newcommand{\ee}{\end{equation}}
	\newcommand{\bes}{\begin{equation*}}
	\newcommand{\ees}{\end{equation*}}
	\newcommand{\bpm}{\begin{pmatrix}}
	\newcommand{\epm}{\end{pmatrix}}
	\newcommand{\qtext}[1]{\quad \mbox{ #1 } \quad}
	
	\newcommand{\comment}[1]{}

	\newcommand{\der }[2]{\frac{\rd#1}{\rd#2}}
	\newcommand{\dron}[2]  {\frac{\partial#1   }{\partial#2}}
	
	\newcommand{\norm}[1]{\left\Vert#1 \right\Vert}
	
	\newcommand{\modu}[1]{\left\vert#1 \right\vert}
	
	\newcommand\eps   {{\varepsilon}}
	
	\makeatletter
	\newcommand*\bigcdot{\mathpalette\bigcdot@{.6}}
	\newcommand*\bigcdot@[2]{\mathbin{\vcenter{\hbox{\scalebox{#2}{$\m@th#1\bullet$}}}}}
	\makeatother
	\newcommand{\leqp}{{\,\leq \! \bigcdot \,}}
	\newcommand{\pleq}{{\,\bigcdot \! \leq\,}}
	\newcommand{\geqp}{{\,\geq \! \bigcdot \,}}

	\newcommand{\eqp}{{\, =\!\bigcdot\,}}
	
	\newcommand{\md}[1]{{\text{\d{$#1$}}}}
	\newcommand{\und}[1]{\md{#1}}

	\DeclareTextSymbol{\degre}{OT1}{23}
	
	\newcommand{\Iden}{{\rI\rd}}
		
		\newcommand{\CC}{{\mathbb C}}
		\newcommand\sC{{\mathscr C}}
		
		\newcommand\fD{{\mathfrak D} }

		\newcommand\sD{{\mathscr D}}
		\newcommand\sDb{{\overline{ \sD}}}
		\newcommand\rd {{\mathrm d }}

		\newcommand\fE{{\mathfrak E}}

		\newcommand\bF   {{\bf{F} }}
		\newcommand\bpf   {{\bf{f} }}
		\newcommand\cF  {{\mathcal F }}

		\newcommand\rF {{\mathrm F }}
		
		\newcommand\bG   {{\bf{G} }}
		\newcommand\cG  {{\mathcal G }}
		
		\newcommand\Hb   {\overline{H} }
		
		\newcommand\Hh   {\hat{H} }

		\newcommand\rH  {{\mathrm H }}

		\newcommand\cH  {{\mathcal H }}

		\newcommand\rI {{\mathrm I }}

		\newcommand\rK  {{\mathrm K }}

		\newcommand\lt {{\tilde l }}

		\newcommand\cL  {{\mathcal L }}
		\newcommand\sL  {{\mathscr L}}
		
		\newcommand\rL  {{\mathrm L }}
		\newcommand\brL{\boldsymbol{\rL}}

		\newcommand\mt{{\tilde m }}
		
		\newcommand{\NN}{{\mathbb N}}

		\newcommand\rN  {{\mathrm N }}

		\newcommand\cO{{\mathcal O} }
		
		\newcommand\rP  {{\mathrm P }}

		\newcommand\br{{\bf r} }
		\newcommand\bR{{\bf R} }
		\newcommand\bRt{{\tilde{\bR}} }
		\newcommand\brd{{\dot{\br}}}
		\newcommand\bRd{{\dot{\bR}}}
		\newcommand{\RR}{{\mathbb{R}}}
		
		\newcommand\fR  {{\mathfrak R }}

		\newcommand\cS  {{\mathcal S }}

		\newcommand\rS  {{\mathrm S }}
		
		\newcommand{\TT}{{\mathbb T}}
		\newcommand\cT {{\mathcal T }}
		
		\newcommand\bW  {{\bf {W} }}
		\newcommand\bWt  {{\tilde {\bW} }}
		
		\newcommand\xt   {{\tilde{x} }}
		\newcommand\xb   {{\overline{x} }}

		\newcommand\bXt  {{\tilde{\bX} }}
		\newcommand\bX  {{\bf{X} }}

		\newcommand\yt   { {\tilde{y}{}} }
		\newcommand\yb   { {\overline{y}{}} }

		\newcommand{\ZZ}{{\mathbb Z}}

		\newcommand\Deltat{\tilde{\Delta}}

		\newcommand\lam   {\lambda }
		\newcommand\Lam   {\Lambda }

		\newcommand\Xih	   {\hat{\Xi}}

		\newcommand\Ups{\Upsilon}
		\newcommand\Upsb{\overline{\Upsilon}}

		\newcommand\Upsc{\check{\Upsilon}}
		\newcommand\Upsh{\hat{\Upsilon}}

		\newcommand\om    {\omega }
		\newcommand\Om    {\Omega }

\begin{document}

\title{Revisiting the averaged problem
	in the case of mean-motion resonances in the restricted three-body problem}

\subtitle{Global rigorous treatment and application to the co-orbital motion}

\author{Alexandre Pousse  \and Elisa Maria Alessi} 


\institute{A. Pousse \at
              Istituto di Matematica Applicata e Tecnologie Informatiche ‘‘Enrico Magenes" -- Consiglio Nazionale delle Ricerche (IMATI-CNR), via Alfonso Corti 12, 20133 Milano, Italy \\
              \email{alexandre@mi.imati.cnr.it}           
           \and
           E.~M. Alessi \at
              Istituto di Matematica Applicata e Tecnologie Informatiche ‘‘Enrico Magenes" -- Consiglio Nazionale delle Ricerche (IMATI-CNR), via Alfonso Corti 12, 20133 Milano, Italy 
              \\
              Istituto di Fisica Applicata ‘‘Nello Carrara" -- Consiglio Nazionale delle Ricerche (IFAC-CNR), Via Madonna del Piano 10, 50019 Sesto Fiorentino (FI), Italy \\
               \email{elisamaria.alessi@cnr.it}
}

\date{Received: date / Accepted: date}

\maketitle

\begin{abstract}
	A classical approach to the restricted three-body problem is 
		to analyze the dynamics of the massless body in the synodic reference frame. 
	A different approach is represented by the perturbative treatment: 
		in particular the averaged problem of a mean-motion resonance
		allows to investigate the long-term behavior of the solutions 
			through a suitable approximation
			that focuses on a particular region of the phase space.
	In this paper, we intend to 
		bridge a gap between the two approaches in the specific case of mean-motion resonant dynamics,
		establish the limit of validity of the averaged problem,
		and take advantage of its results
			in order to compute trajectories in the synodic reference frame.
	After the description of each approach,
		we develop a rigorous treatment of the averaging process,
			estimate the size of the transformation
			and prove that the averaged problem is a suitable approximation of the restricted three-body problem
			as long as the solutions are located outside the Hill's sphere of the secondary. 
	In such a case, a rigorous theorem of stability over finite but large timescales can be proven.
	We establish that a solution of the averaged problem provides 
		an accurate approximation of the trajectories on the synodic reference frame
		within a finite time that depend on the minimal distance to the Hill's sphere of the secondary.
	The last part of this work is devoted to the co-orbital motion (i.e., the dynamics in 1:1 mean-motion resonance) 
	in the circular-planar case.
	In this case, an interpretation of the solutions of the averaged problem in the synodic reference frame is detailed
		and a method that allows to compute co-orbital trajectories is displayed.

\end{abstract}

\section{Introduction}

	This work focuses on the restricted three-body problem, that is the study of the motion of a massless body
		affected by the gravitational attraction of two massive bodies.
	More precisely, we will consider the situation for which the mass of the secondary body is treated as a small quantity.
	Since the planetary three-body problem will also be mentioned,  
		we recall that it corresponds to the study of the motion of two massive bodies orbiting
		a more massive one, the three bodies being governed only by their mutual gravitational interactions.
			
	The analysis of the dynamics in the synodic reference frame,
		that is the frame rotating with the mean longitude of the secondary,
	is the classical approach adopted for the restricted three-body problem.
	Usually, periodic orbit families and the dynamics located in their neighborhood are computed 
		by using Poincar\'e maps and continuation methods (see, e.g., \cite{1967Sz,2001GoMo}).
	 
	Perturbative treatments provide another approach.
	They allow to investigate specific regions of the phase space
		through a proper approximation.
	Among them, averaging methods are common techniques in order to study the long-term dynamics of the solutions.
	For instance, the secular problem studies
		the long-term deformation of the ellipse of the massless body
		as well as the evolution of its orientation in the tridimensional space.
	It is obtained by the averaging of the Hamiltonian 
		over the mean longitudes of the secondary and of the massless body.
	More precisely, it corresponds to a symplectic transformation, 
		that is supposed to be close to the identity,
	ant that maps the original Hamiltonian to the secular one. 
	
	Lagrange \cite{1778La} introduced the secular problem in the framework of the stability of the Solar System 
	and 
	the expression of the secular Hamiltonian of the planetary three-body problem was given 
		by Poincar\'e \cite{1892Po}.
	Precise estimates on the size of the transformation of averaging were required 
		in order to prove theorems of stability like KAM theory
		and were provided, especially by Arnol'd \cite{1963Ar}, F\'ejoz \cite{2004Fe}, 
		Chierchia and Pinzari \cite{2011ChPi}.
	
	When the massless body is in mean-motion resonance with the secondary, 
		that is, when their orbital periods are commensurable,
		the transformation leading to the secular Hamiltonian is no more close to the identity
		and the solutions of the secular problem do not provide a good representation of the real motion. 
	In such a case, it is still possible to use averaging techniques: 
		the averaging process is performed over one mean longitude, 
			generally the one of the secondary,
		and after the introduction of a resonant angle, 
			that is a particular linear combination of the two mean longitudes
			which characterizes the mean-motion resonance.
	This defines the averaged problem that will  be considered in this work.

	Many authors investigated mean-motion resonances through an averaged Hamiltonian
		and the literature on this subject has become so rich that it is impossible to cite all the articles here.
	Nevertheless, let us mention the important series of works realized by Schubart \cite{1964Sc,1968Sc,1978Sc},
		which took advantage of the canonical variables and method suggested by Poincar\'e \cite{1902Po} 
		and applied an averaging process 
			in order to get the interesting part of the Hamiltonian for mean-motion resonances.
	Likewise, the second fundamental model of resonance, developed by Henrard and Lemaitre \cite{1983HeLe},
		follows the strategy of Poincar\'e \cite{1902Po},
		and is commonly used in order to study mean-motion resonances.
	Moons \cite{1994Mo} extended the work of Schubart
		and presented an integrator adapted to the solutions of the averaged problem.
	Being the latter not valid for the 1:1 mean-motion resonance,
		Nesvorn\'y et al. \cite{2002NeThFe} adapted the algorithm
		with a different choice of canonical variables.

	The co-orbital motion, or equivalently, the trajectories in 1:1 mean-motion resonance with the secondary,
		has been intensively studied in the framework of the averaged problem 
		(see, e.g., \cite{2001Mo,2006MiInWi,2014SiNeAr,2020Si}).
	In such a case, 
		since the semi-major axis of the massless body is almost the same as the one of the secondary,
		the issue generated by periodical close encounters arises, even for quasi-circular trajectories.
	In particular, 
		Robutel and Pousse \cite{2013RoPo} and Pousse et al. \cite{2017PoRoVi} highlighted,
		with the help of a frequency analysis, 
		that the averaged Hamiltonian reflects poorly the dynamics close to the singularity 
			associated with the collision between the secondary and the massless body.
	Rigorous estimates on the averaging process have been given by Robutel et al. \cite{2016RoNiPo}.
	More precisely, they allowed the authors to prove in the planetary three-body problem, 
		that the averaged problem is valid 
			for two co-orbital bodies on quasi-circular orbits 
			that stand at a mutual distance larger than their respective Hill's radius.	
			
	The limit of validity of the averaged problem,
		is not specific to case of the co-orbital motion
		and can also occur for other resonant trajectories 
		that cross the orbit of the secondary (i.e., trajectories with a non negligible eccentricity).
	This weakness was already outlined in the works of Schubart \cite{1964Sc} and Moons \cite{1994Mo}.
	Therefore, in the present paper, 
		we intend to generalize the result given by Robutel et al. \cite{2016RoNiPo}
		and provide rigorous estimates on the averaging process
		in order to define a domain of validity of the averaged problem,
			in the case of a generic mean-motion resonance,
			and for any value of inclination and eccentricities (massless body as well as secondary).
			
	 According to the Poincar\'e classification (see, e.g., \cite{2012Ch} for more details),
	some of the periodic families described in the synodic reference frame are related to mean-motion resonances
		and thus can also be tackled in the averaged problem as defined here.
	For that reason, we also intend to bridge a gap between 
		the classical approach in the synodic reference frame and the averaged problem 
		with a unified Hamiltonian formalism 
		that allows to represent solutions in both approaches.
	The underlying idea of this work is to 
		understand the limit of validity of the averaged problem
		and take advantage of its solutions (e.g., initial conditions, types of motion, frequencies)
		for the computation of trajectories in the synodic reference frame.
	
	The paper is structured as follows.
	Section \ref{sec:def} introduces the restricted three-body problem through the classical approach, 
		recalls some remarkable solutions in the synodic reference frame
		and presents the reasoning that led to the averaged problem.

	In Sect.~\ref{sec:validity}, the size of the transformation of averaging is estimated.
	This allows to define a domain of validity of the approximation 
		and to prove a rigorous theorem of stability over finite times.

	In Sect.~\ref{sec:coorb}, we focus on the co-orbital motion in the circular-planar case
		and detail the correspondence between a solution of the averaged problem
		and its corresponding trajectory in the synodic reference frame.
	In particular, we will recover the remarkable solutions described in Sect.~\ref{sec:def}.
	Finally, a method that allows to compute co-orbital trajectories in the synodic reference frame will be described.	
	
	Appendix \ref{sec:appendix}  gives the proof of the theorems and lemma used in our reasonings.



\section{Two approaches for the restricted three-body problem}
\label{sec:def}


	\subsection{The restricted three-body problem}


		\subsubsection{Definition in the heliocentric reference frame}

		Let $(\br, \brd)$ be, respectively, the heliocentric position and velocity vector in $\RR^3$ 
			of a massless body (particle, spacecraft or asteroid),  
			that is  affected by the gravitational attraction of 
			a massive primary (the Sun or  a planet)  of mass $1-\eps>1/2$,
			and a secondary (a planet or a moon) of mass $\eps>0$. 
	
		The motion of the two massive bodies, respectively denoted as Sun and planet, 
			follows a solution of the two-body problem.
		Hence, the trajectory of the planet, denoted $\br'(t)$ in the heliocentric reference frame,
			lies on an ellipse that can be defined by the orbital elements 
			${(a', e', I', \Omega', \om', v')}$, 
			i.e., respectively 
				semi-major axis, 
				eccentricity, 
				inclination, 
				longitude of the node,
				argument of the periaster,
				and 				
				true anomaly.
		\emph{Without loss of generality, the scale and orientation of the orbit are arbitrarily chosen such that
					\bes(a',  I', \Omega', \om') = (1, 0, 0, 0).\ees 
				Likewise, the orbital period of the planet is fixed to $2\pi$ (and therefore its mean motion to 1) 
					which imposes the gravitational constant to be equal to 1.}
		The eccentricity $e'$ is a parameter of the problem associated with the shape of the planet's orbit
			while the angle $v'$ stands for its position on the ellipse.
		Instead of using the true anomaly, the mean longitude $\lam'$ will be adopted in order to take advantage 
			of its proportionality to time since 
		\bes
			t = \lam'(t) + 2k\pi \quad\mbox{where $k\in \ZZ$.}
		\ees

		In the heliocentric reference frame, the equations of motion of the particle read (see \cite{2002Mo}):
		\be
				\ddot{\br} =	- (1-\eps) 		\frac{\br}{\norm{\br}^3}  
							- \eps  		\frac{\br - \br'(\lam')}{\norm{\br - \br'(\lam')}^3}
							- \eps		\frac{\br'(\lam')}{\norm{\br'(\lam')}^3}
\label{eq:EMotion}
		\ee
		where ``$\norm{\,\cdot\,}$" is the Euclidean norm associated with the scalar product denoted ``$\,\bigcdot\,$" 
		in what follows.
	 	The two first terms are respectively the gravitational force of the Sun and of the planet. 
		The third term is associated with the acceleration of the heliocentric reference frame 
			generated by the Sun-planet gravitational interactions.


		\subsubsection{Hamiltonian formalism}

		Since the heliocentric vectors $\br$ and $\brd$ are canonical variables, 
			then the Hamiltonian function
		\be
			\cH(\br, \brd, \lam') 	=  \frac{\norm{\brd}^2}{2} -  \frac{1-\eps}{\norm{\br}} 
							- \frac{\eps}{\norm{\br-\br'(\lam')}}
							+ \eps \frac{\br \bigcdot \br'(\lam')}{\norm{\br'(\lam')}^3}
\label{eq:Helio_Ham}
		\ee 
			provides the equations of motion \eqref{eq:EMotion}.		
		As $\cH$ depends on the periodicity of the planet, it is non-autonomous.
		Moreover, the system just written, that describes the dynamics of the particle, has 3 degrees of freedom
		associated with the position and velocity vectors in the tridimensional space.
		A classical technique that allows to overcome the non-autonomous character of the system 
			consists in extending the phase space with the addition of a generic variable $\Xih$ 
			conjugated with $\lam'$.
		In this extended phase space,
			the Hamiltonian reads $\cH + \Xih$, 
			it has 4 degrees of freedom
			and describes the coupled motion of the particle and the planet, namely	
		\bes
			\der{}{t}(\br,\brd, \lam', \Xih)  = \left(\dron{\cH}{\brd}, -\dron{\cH}{\br},  1, -\dron{\cH}{\lam'}\right).
		\ees	
		As a consequence, investigating the restricted three-body problem consists in studying an autonomous ODE 
			whose solutions belong to a 8-dimensional phase space 
		(position and velocity in the tridimensional space, 
			the mean longitude of the planet 
			and its conjugated action).
		A classical approach in order to simplify the analysis is to consider the behavior of the particle 
			in the synodic reference frame,
			that is the frame that rotates with $\lam'$  in the orbital plane of the planet.
		

	\subsection{A classical approach: the synodic reference frame}

			
		\subsubsection{The synodic reference frame}
\label{sec:SF}

		Let us denote $\fR_k(\alpha)$, the rotation matrix  of an angle $\alpha$ about the $k$-axis ($k\in\{1,2,3\}$),
			and 
		\bes
			\brL(\br, \brd)  = \br \times \brd,
		\ees
			the angular momentum of the particle in the heliocentric reference frame.
		We recall that due to the influence of the planet, $\brL(\br, \brd)$ 
			is not a conserved quantity of the restricted three-body problem. 	

		With the help of the Hamiltonian formalism, 
			the symplectic transformation associated with the synodic reference frame reads
			$$\Ups_{\rS\rF}: (\bR, \bRt, \lam', \Xi_{\rS\rF}) \mapsto (\br, \brd, \lam', \Xih)$$ with
		\begin{eqnarray*}
					\bR    		&=& \fR_3(-\lam') \br	=(X,Y,Z),\\
					\bRt   		&=& \fR_3(-\lam') \brd	=(p_X, p_Y, p_Z),\\
					\Xih  			&=& \Xi_{\rS\rF} - \brL_3(\bR,\bRt)
								= \Xi_{\rS\rF} + Yp_X - Xp_Y.
		\end{eqnarray*}
		It provides the Hamiltonian  \bes{(\cH + \Xih) \circ \Ups_{\rS\rF} = \cH_{\rS\rF}+ \Xi_{\rS\rF}}\ees 
		such that
	\begin{eqnarray*}
			\cH_{\rS\rF}(\bR, \bRt, \lam') 
				 	&=& \frac{1}{2}\norm{\bRt}^2
							- \frac{1-\eps}{\norm{\bR}}
							- \frac{\eps}{\norm{\bR- \bR'(\lam')}}\\
					&\phantom{=}& + \eps \frac{\bR \bigcdot \bR'(\lam')}{\norm{\bR'(\lam')}^3}
							- \brL_3(\bR,  \bRt)
	 \end{eqnarray*}
		with $ \bR'(\lam') = \fR_3(-\lam')\br'(\lam')$ 
			that corresponds to the position of the planet.
		Moreover, the velocity of the particle in the synodic reference frame is deduced as follows:
		\bes
			\bRd =  \dron{}{\bRt}\cH_{\rS\rF}(\bR, \bRt, \lam') = \bRt + (Y, -X, 0).
		\ees

		This framework gives rise to an important reduction in the context of the circular case ($e' = 0$):
		since the planet is a fixed point located in $\bR'  = (1, 0, 0)$,
			the Hamiltonian does not depend on $\lam'$
			and $\Xi_{\rS\rF}$ is an integral of motion that can be dropped.
		Hence, the dimension of the phase space to explore is reduced by two units.
		Moreover, $\cH_{\rS\rF}$ is related to the Jacobi constant 	that defines the energy of the particle.
		In the synodic reference frame centered on the Sun, 
		the Jacobi constant can be written as follows:
		\be
		\begin{aligned}
			\sC(\bR, \bRd)	
								&=  - 2\cH_{\rS\rF}(\bR, \bRt) + \eps\\
								&= 	(X-\eps)^2 + Y^2 
									+ \eps(1-\eps)\\
								&\phantom{=}
									+ 2\left( \frac{\eps}{\norm{\bR - (1,0,0)}} +\frac{1-\eps}{\norm{\bR}}\right)
									 -\norm{\bRd}^2.				
		\end{aligned}
\label{eq:Jacobi}	
		\ee
		Thus, for a given value of $\sC$, 
			the corresponding isoenergetic hypersurface is a manifold of dimension 5.

\begin{sloppypar}		
		$\sC$ being the only conserved quantity (see \cite{1967Sz}), 
			it is not possible to reduce the problem through another global transformation.		
		A further way to simplify the study is to consider the particle's motion restricted to the orbital plane of the planet.
		Hence, the phase space to explore can be reduced by two units.
		Consequently, investigating the restricted three-body problem in the circular-planar case is equivalent to
			explore a one-parameter family of 3-dimensional manifolds parametrized by the energy.
		Without too much details, the following part is dedicated to  
			some remarkable solutions of the circular-planar case that are relevant for the scope of this work.
\end{sloppypar}

		\subsubsection{Some remarkable solutions in the circular-planar case ($e'=0$)}	
\label{sec:rem_sol}

\begin{figure*}[!h]
	\begin{center}
		\tiny
		\def\svgwidth{0.326\textwidth}
\begingroup%
  \makeatletter%
  \providecommand\color[2][]{%
    \errmessage{(Inkscape) Color is used for the text in Inkscape, but the package 'color.sty' is not loaded}%
    \renewcommand\color[2][]{}%
  }%
  \providecommand\transparent[1]{%
    \errmessage{(Inkscape) Transparency is used (non-zero) for the text in Inkscape, but the package 'transparent.sty' is not loaded}%
    \renewcommand\transparent[1]{}%
  }%
  \providecommand\rotatebox[2]{#2}%
  \newcommand*\fsize{\dimexpr\f@size pt\relax}%
  \newcommand*\lineheight[1]{\fontsize{\fsize}{#1\fsize}\selectfont}%
  \ifx\svgwidth\undefined%
    \setlength{\unitlength}{359.95033264bp}%
    \ifx\svgscale\undefined%
      \relax%
    \else%
      \setlength{\unitlength}{\unitlength * \real{\svgscale}}%
    \fi%
  \else%
    \setlength{\unitlength}{\svgwidth}%
  \fi%
  \global\let\svgwidth\undefined%
  \global\let\svgscale\undefined%
  \makeatother%
  \begin{picture}(1,1.00221353)%
    \lineheight{1}%
    \setlength\tabcolsep{0pt}%
    \put(0,0){\includegraphics[width=\unitlength]{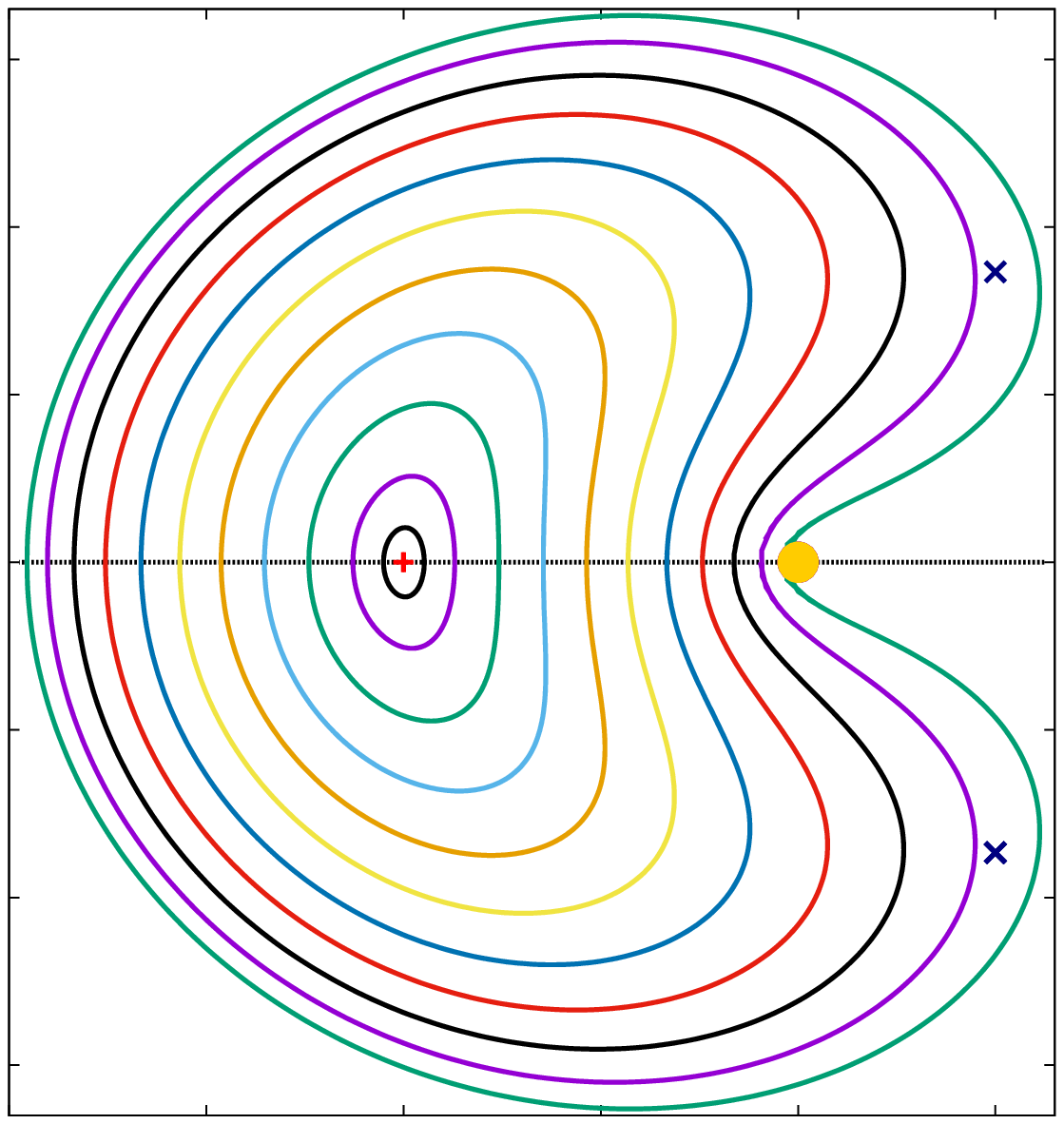}}%
    \put(0.06940403,0.07967286){\color[rgb]{0,0,0}\makebox(0,0)[rt]{\lineheight{1.25}\smash{\begin{tabular}[t]{r}-1.5\end{tabular}}}}%
    \put(0.06590537,0.22066453){\color[rgb]{0,0,0}\makebox(0,0)[rt]{\lineheight{1.25}\smash{\begin{tabular}[t]{r}-1\end{tabular}}}}%
    \put(0.06940403,0.36179511){\color[rgb]{0,0,0}\makebox(0,0)[rt]{\lineheight{1.25}\smash{\begin{tabular}[t]{r}-0.5\end{tabular}}}}%
    \put(0.0696393,0.50278678){\color[rgb]{0,0,0}\makebox(0,0)[rt]{\lineheight{1.25}\smash{\begin{tabular}[t]{r}0\end{tabular}}}}%
    \put(0.06961152,0.64377845){\color[rgb]{0,0,0}\makebox(0,0)[rt]{\lineheight{1.25}\smash{\begin{tabular}[t]{r}0.5\end{tabular}}}}%
    \put(0.06465815,0.78490904){\color[rgb]{0,0,0}\makebox(0,0)[rt]{\lineheight{1.25}\smash{\begin{tabular}[t]{r}1\end{tabular}}}}%
    \put(0.06817903,0.92590071){\color[rgb]{0,0,0}\makebox(0,0)[rt]{\lineheight{1.25}\smash{\begin{tabular}[t]{r}1.5\end{tabular}}}}%
    \put(0.08274263,0.01521952){\color[rgb]{0,0,0}\makebox(0,0)[t]{\lineheight{1.25}\smash{\begin{tabular}[t]{c}-2\end{tabular}}}}%
    \put(0.24870189,0.01521952){\color[rgb]{0,0,0}\makebox(0,0)[t]{\lineheight{1.25}\smash{\begin{tabular}[t]{c}-1.5\end{tabular}}}}%
    \put(0.41294768,0.01521952){\color[rgb]{0,0,0}\makebox(0,0)[t]{\lineheight{1.25}\smash{\begin{tabular}[t]{c}-1\end{tabular}}}}%
    \put(0.58069213,0.01521952){\color[rgb]{0,0,0}\makebox(0,0)[t]{\lineheight{1.25}\smash{\begin{tabular}[t]{c}-0.5\end{tabular}}}}%
    \put(0.74985901,0.01521952){\color[rgb]{0,0,0}\makebox(0,0)[t]{\lineheight{1.25}\smash{\begin{tabular}[t]{c}0\end{tabular}}}}%
    \put(0.91584024,0.01521952){\color[rgb]{0,0,0}\makebox(0,0)[t]{\lineheight{1.25}\smash{\begin{tabular}[t]{c}0.5\end{tabular}}}}%
    \put(0.89589226,0.75731665){\makebox(0,0)[rt]{\lineheight{1.25}\smash{\begin{tabular}[t]{r}$L_4$\end{tabular}}}}%
    \put(0.89589226,0.26558215){\makebox(0,0)[rt]{\lineheight{1.25}\smash{\begin{tabular}[t]{r}$L_5$\end{tabular}}}}%
    \put(0.96046034,0.52502689){\makebox(0,0)[rt]{\lineheight{1.25}\smash{\begin{tabular}[t]{r}$\{Y=0\}$\end{tabular}}}}%
    \put(0.40398206,0.52490209){\makebox(0,0)[rt]{\lineheight{1.25}\smash{\begin{tabular}[t]{r}$L_3$\end{tabular}}}}%
    \put(0.12078489,0.93482142){\makebox(0,0)[lt]{\lineheight{1.25}\smash{\begin{tabular}[t]{l}a.\end{tabular}}}}%
  \end{picture}%
\endgroup%
		\def\svgwidth{0.326\textwidth}
\begingroup%
  \makeatletter%
  \providecommand\color[2][]{%
    \errmessage{(Inkscape) Color is used for the text in Inkscape, but the package 'color.sty' is not loaded}%
    \renewcommand\color[2][]{}%
  }%
  \providecommand\transparent[1]{%
    \errmessage{(Inkscape) Transparency is used (non-zero) for the text in Inkscape, but the package 'transparent.sty' is not loaded}%
    \renewcommand\transparent[1]{}%
  }%
  \providecommand\rotatebox[2]{#2}%
  \newcommand*\fsize{\dimexpr\f@size pt\relax}%
  \newcommand*\lineheight[1]{\fontsize{\fsize}{#1\fsize}\selectfont}%
  \ifx\svgwidth\undefined%
    \setlength{\unitlength}{360bp}%
    \ifx\svgscale\undefined%
      \relax%
    \else%
      \setlength{\unitlength}{\unitlength * \real{\svgscale}}%
    \fi%
  \else%
    \setlength{\unitlength}{\svgwidth}%
  \fi%
  \global\let\svgwidth\undefined%
  \global\let\svgscale\undefined%
  \makeatother%
  \begin{picture}(1,1)%
    \lineheight{1}%
    \setlength\tabcolsep{0pt}%
    \put(0,0){\includegraphics[width=\unitlength]{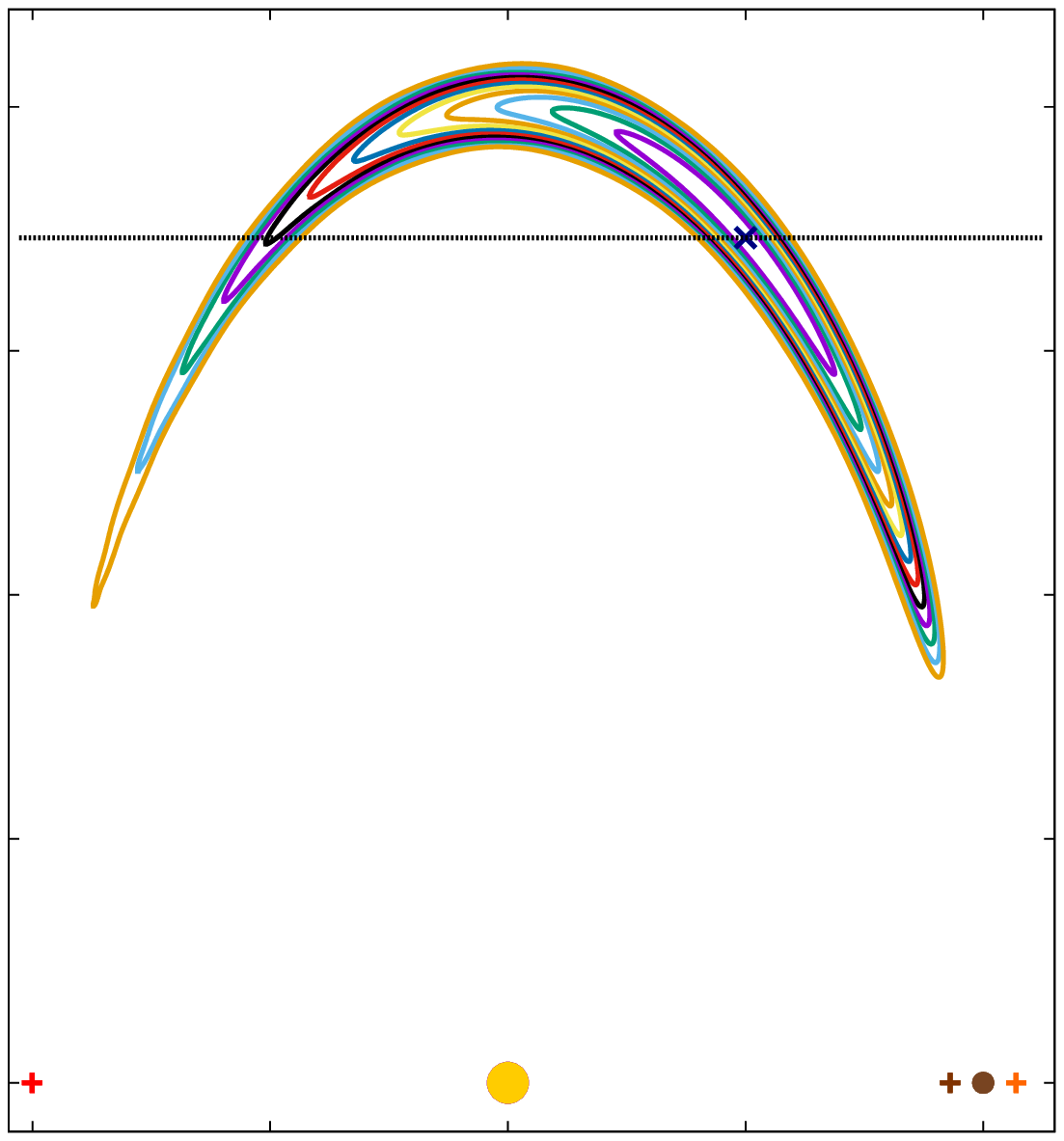}}%
    \put(0.0819908,0.07748994){\color[rgb]{0,0,0}\makebox(0,0)[rt]{\lineheight{1.25}\smash{\begin{tabular}[t]{r}0\end{tabular}}}}%
    \put(0.08195747,0.27985104){\color[rgb]{0,0,0}\makebox(0,0)[rt]{\lineheight{1.25}\smash{\begin{tabular}[t]{r}0.25\end{tabular}}}}%
    \put(0.08196303,0.48221215){\color[rgb]{0,0,0}\makebox(0,0)[rt]{\lineheight{1.25}\smash{\begin{tabular}[t]{r}0.5\end{tabular}}}}%
    \put(0.08195747,0.68457326){\color[rgb]{0,0,0}\makebox(0,0)[rt]{\lineheight{1.25}\smash{\begin{tabular}[t]{r}0.75\end{tabular}}}}%
    \put(0.07701034,0.88693437){\color[rgb]{0,0,0}\makebox(0,0)[rt]{\lineheight{1.25}\smash{\begin{tabular}[t]{r}1\end{tabular}}}}%
    \put(0.11302961,0.01498994){\color[rgb]{0,0,0}\makebox(0,0)[t]{\lineheight{1.25}\smash{\begin{tabular}[t]{c}-1\end{tabular}}}}%
    \put(0.31186202,0.01498994){\color[rgb]{0,0,0}\makebox(0,0)[t]{\lineheight{1.25}\smash{\begin{tabular}[t]{c}-0.5\end{tabular}}}}%
    \put(0.51225556,0.01498994){\color[rgb]{0,0,0}\makebox(0,0)[t]{\lineheight{1.25}\smash{\begin{tabular}[t]{c}0\end{tabular}}}}%
    \put(0.709325,0.01498994){\color[rgb]{0,0,0}\makebox(0,0)[t]{\lineheight{1.25}\smash{\begin{tabular}[t]{c}0.5\end{tabular}}}}%
    \put(0.90393198,0.01498994){\color[rgb]{0,0,0}\makebox(0,0)[t]{\lineheight{1.25}\smash{\begin{tabular}[t]{c}1\end{tabular}}}}%
    \put(0.72254326,0.79867817){\makebox(0,0)[lt]{\lineheight{1.25}\smash{\begin{tabular}[t]{l}$L_4$\end{tabular}}}}%
    \put(0.10485665,0.10215079){\makebox(0,0)[lt]{\lineheight{1.25}\smash{\begin{tabular}[t]{l}$L_3$\end{tabular}}}}%
    \put(0.93411049,0.10631752){\makebox(0,0)[t]{\lineheight{1.25}\smash{\begin{tabular}[t]{c}$L_2$\end{tabular}}}}%
    \put(0.86798325,0.08131746){\makebox(0,0)[rt]{\lineheight{1.25}\smash{\begin{tabular}[t]{r}$L_1$\end{tabular}}}}%
    \put(0.38703316,0.7980873){\makebox(0,0)[lt]{\lineheight{1.25}\smash{\begin{tabular}[t]{l}$\{Y=\sqrt{3}/2\}$\end{tabular}}}}%
    \put(0.12072753,0.93265788){\makebox(0,0)[lt]{\lineheight{1.25}\smash{\begin{tabular}[t]{l}c.\end{tabular}}}}%
  \end{picture}%
\endgroup%
\\
		\def\svgwidth{0.326\textwidth}
\begingroup%
  \makeatletter%
  \providecommand\color[2][]{%
    \errmessage{(Inkscape) Color is used for the text in Inkscape, but the package 'color.sty' is not loaded}%
    \renewcommand\color[2][]{}%
  }%
  \providecommand\transparent[1]{%
    \errmessage{(Inkscape) Transparency is used (non-zero) for the text in Inkscape, but the package 'transparent.sty' is not loaded}%
    \renewcommand\transparent[1]{}%
  }%
  \providecommand\rotatebox[2]{#2}%
  \newcommand*\fsize{\dimexpr\f@size pt\relax}%
  \newcommand*\lineheight[1]{\fontsize{\fsize}{#1\fsize}\selectfont}%
  \ifx\svgwidth\undefined%
    \setlength{\unitlength}{360bp}%
    \ifx\svgscale\undefined%
      \relax%
    \else%
      \setlength{\unitlength}{\unitlength * \real{\svgscale}}%
    \fi%
  \else%
    \setlength{\unitlength}{\svgwidth}%
  \fi%
  \global\let\svgwidth\undefined%
  \global\let\svgscale\undefined%
  \makeatother%
  \begin{picture}(1,1)%
    \lineheight{1}%
    \setlength\tabcolsep{0pt}%
    \put(0,0){\includegraphics[width=\unitlength]{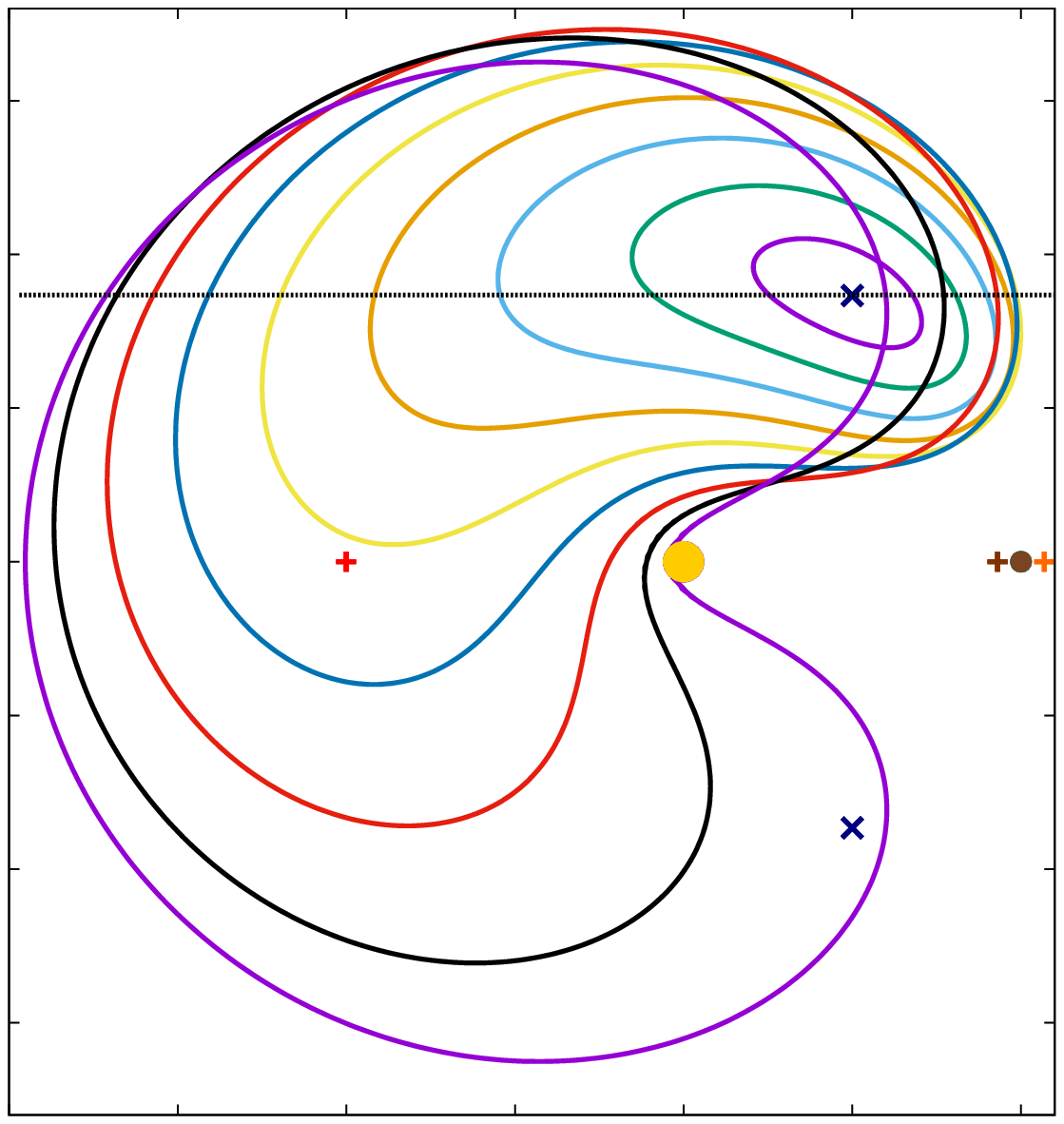}}%
    \put(0.06939446,0.11286438){\color[rgb]{0,0,0}\makebox(0,0)[rt]{\lineheight{1.25}\smash{\begin{tabular}[t]{r}-1.5\end{tabular}}}}%
    \put(0.06589629,0.24203104){\color[rgb]{0,0,0}\makebox(0,0)[rt]{\lineheight{1.25}\smash{\begin{tabular}[t]{r}-1\end{tabular}}}}%
    \put(0.06939446,0.37133659){\color[rgb]{0,0,0}\makebox(0,0)[rt]{\lineheight{1.25}\smash{\begin{tabular}[t]{r}-0.5\end{tabular}}}}%
    \put(0.0696297,0.50064214){\color[rgb]{0,0,0}\makebox(0,0)[rt]{\lineheight{1.25}\smash{\begin{tabular}[t]{r}0\end{tabular}}}}%
    \put(0.06960193,0.6299477){\color[rgb]{0,0,0}\makebox(0,0)[rt]{\lineheight{1.25}\smash{\begin{tabular}[t]{r}0.5\end{tabular}}}}%
    \put(0.06464924,0.75925326){\color[rgb]{0,0,0}\makebox(0,0)[rt]{\lineheight{1.25}\smash{\begin{tabular}[t]{r}1\end{tabular}}}}%
    \put(0.06816963,0.88841992){\color[rgb]{0,0,0}\makebox(0,0)[rt]{\lineheight{1.25}\smash{\begin{tabular}[t]{r}1.5\end{tabular}}}}%
    \put(0.08273122,0.01314216){\color[rgb]{0,0,0}\makebox(0,0)[t]{\lineheight{1.25}\smash{\begin{tabular}[t]{c}-2\end{tabular}}}}%
    \put(0.22463981,0.01314216){\color[rgb]{0,0,0}\makebox(0,0)[t]{\lineheight{1.25}\smash{\begin{tabular}[t]{c}-1.5\end{tabular}}}}%
    \put(0.36469628,0.01314216){\color[rgb]{0,0,0}\makebox(0,0)[t]{\lineheight{1.25}\smash{\begin{tabular}[t]{c}-1\end{tabular}}}}%
    \put(0.50838981,0.01314216){\color[rgb]{0,0,0}\makebox(0,0)[t]{\lineheight{1.25}\smash{\begin{tabular}[t]{c}-0.5\end{tabular}}}}%
    \put(0.65336668,0.01314216){\color[rgb]{0,0,0}\makebox(0,0)[t]{\lineheight{1.25}\smash{\begin{tabular}[t]{c}0\end{tabular}}}}%
    \put(0.79529723,0.01314216){\color[rgb]{0,0,0}\makebox(0,0)[t]{\lineheight{1.25}\smash{\begin{tabular}[t]{c}0.5\end{tabular}}}}%
    \put(0.93476533,0.01314216){\color[rgb]{0,0,0}\makebox(0,0)[t]{\lineheight{1.25}\smash{\begin{tabular}[t]{c}1\end{tabular}}}}%
    \put(0.90536129,0.50900108){\makebox(0,0)[rt]{\lineheight{1.25}\smash{\begin{tabular}[t]{r}$L_1$\end{tabular}}}}%
    \put(0.77780459,0.75066762){\makebox(0,0)[rt]{\lineheight{1.25}\smash{\begin{tabular}[t]{r}$L_4$\end{tabular}}}}%
    \put(0.77780459,0.28816761){\makebox(0,0)[rt]{\lineheight{1.25}\smash{\begin{tabular}[t]{r}$L_5$\end{tabular}}}}%
    \put(0.35280458,0.50900094){\makebox(0,0)[rt]{\lineheight{1.25}\smash{\begin{tabular}[t]{r}$L_3$\end{tabular}}}}%
    \put(0.26954043,0.74890064){\makebox(0,0)[lt]{\lineheight{1.25}\smash{\begin{tabular}[t]{l}$\{Y=\sqrt{3}/2\}$\end{tabular}}}}%
    \put(0.11747233,0.91585286){\makebox(0,0)[lt]{\lineheight{1.25}\smash{\begin{tabular}[t]{l}b.\end{tabular}}}}%
  \end{picture}%
\endgroup%
		\def\svgwidth{0.326\textwidth}
\begingroup%
  \makeatletter%
  \providecommand\color[2][]{%
    \errmessage{(Inkscape) Color is used for the text in Inkscape, but the package 'color.sty' is not loaded}%
    \renewcommand\color[2][]{}%
  }%
  \providecommand\transparent[1]{%
    \errmessage{(Inkscape) Transparency is used (non-zero) for the text in Inkscape, but the package 'transparent.sty' is not loaded}%
    \renewcommand\transparent[1]{}%
  }%
  \providecommand\rotatebox[2]{#2}%
  \newcommand*\fsize{\dimexpr\f@size pt\relax}%
  \newcommand*\lineheight[1]{\fontsize{\fsize}{#1\fsize}\selectfont}%
  \ifx\svgwidth\undefined%
    \setlength{\unitlength}{358.40483093bp}%
    \ifx\svgscale\undefined%
      \relax%
    \else%
      \setlength{\unitlength}{\unitlength * \real{\svgscale}}%
    \fi%
  \else%
    \setlength{\unitlength}{\svgwidth}%
  \fi%
  \global\let\svgwidth\undefined%
  \global\let\svgscale\undefined%
  \makeatother%
  \begin{picture}(1,1.00334448)%
    \lineheight{1}%
    \setlength\tabcolsep{0pt}%
    \put(0,0){\includegraphics[width=\unitlength]{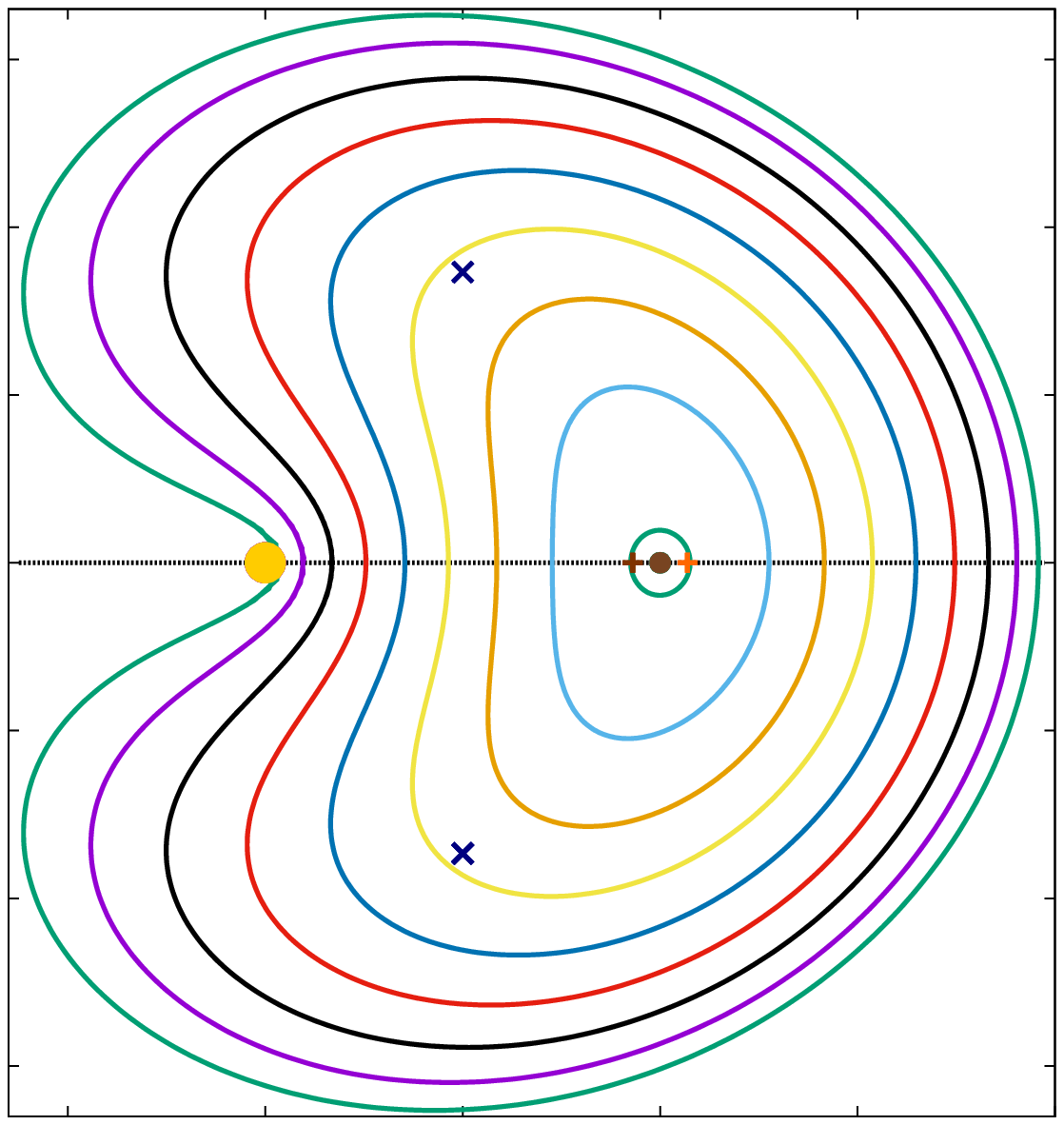}}%
    \put(0.06438157,0.07880292){\color[rgb]{0,0,0}\makebox(0,0)[rt]{\lineheight{1.25}\smash{\begin{tabular}[t]{r}-1.5\end{tabular}}}}%
    \put(0.06086782,0.22040257){\color[rgb]{0,0,0}\makebox(0,0)[rt]{\lineheight{1.25}\smash{\begin{tabular}[t]{r}-1\end{tabular}}}}%
    \put(0.06438157,0.36214173){\color[rgb]{0,0,0}\makebox(0,0)[rt]{\lineheight{1.25}\smash{\begin{tabular}[t]{r}-0.5\end{tabular}}}}%
    \put(0.06461786,0.50374138){\color[rgb]{0,0,0}\makebox(0,0)[rt]{\lineheight{1.25}\smash{\begin{tabular}[t]{r}0\end{tabular}}}}%
    \put(0.06458996,0.64534103){\color[rgb]{0,0,0}\makebox(0,0)[rt]{\lineheight{1.25}\smash{\begin{tabular}[t]{r}0.5\end{tabular}}}}%
    \put(0.05961522,0.7870802){\color[rgb]{0,0,0}\makebox(0,0)[rt]{\lineheight{1.25}\smash{\begin{tabular}[t]{r}1\end{tabular}}}}%
    \put(0.06315129,0.92867985){\color[rgb]{0,0,0}\makebox(0,0)[rt]{\lineheight{1.25}\smash{\begin{tabular}[t]{r}1.5\end{tabular}}}}%
    \put(0.12782471,0.01407165){\color[rgb]{0,0,0}\makebox(0,0)[t]{\lineheight{1.25}\smash{\begin{tabular}[t]{c}-0.5\end{tabular}}}}%
    \put(0.29772106,0.01407165){\color[rgb]{0,0,0}\makebox(0,0)[t]{\lineheight{1.25}\smash{\begin{tabular}[t]{c}0\end{tabular}}}}%
    \put(0.46441803,0.01407165){\color[rgb]{0,0,0}\makebox(0,0)[t]{\lineheight{1.25}\smash{\begin{tabular}[t]{c}0.5\end{tabular}}}}%
    \put(0.62864158,0.01407165){\color[rgb]{0,0,0}\makebox(0,0)[t]{\lineheight{1.25}\smash{\begin{tabular}[t]{c}1\end{tabular}}}}%
    \put(0.79712053,0.01407165){\color[rgb]{0,0,0}\makebox(0,0)[t]{\lineheight{1.25}\smash{\begin{tabular}[t]{c}1.5\end{tabular}}}}%
    \put(0.96457563,0.01407165){\color[rgb]{0,0,0}\makebox(0,0)[t]{\lineheight{1.25}\smash{\begin{tabular}[t]{c}2\end{tabular}}}}%
    \put(0.47343702,0.75836582){\makebox(0,0)[lt]{\lineheight{1.25}\smash{\begin{tabular}[t]{l}$L_4$\end{tabular}}}}%
    \put(0.47343702,0.26451083){\makebox(0,0)[lt]{\lineheight{1.25}\smash{\begin{tabular}[t]{l}$L_5$\end{tabular}}}}%
    \put(0.65340111,0.52817915){\makebox(0,0)[lt]{\lineheight{1.25}\smash{\begin{tabular}[t]{l}$L_2$\end{tabular}}}}%
    \put(0.59962278,0.52817915){\makebox(0,0)[rt]{\lineheight{1.25}\smash{\begin{tabular}[t]{r}$L_1$\end{tabular}}}}%
    \put(0.08461313,0.52507438){\makebox(0,0)[lt]{\lineheight{1.25}\smash{\begin{tabular}[t]{l}$\{Y=0\}$\end{tabular}}}}%
    \put(0.12114226,0.91882283){\makebox(0,0)[lt]{\lineheight{1.25}\smash{\begin{tabular}[t]{l}d.\end{tabular}}}}%
  \end{picture}%
\endgroup%
		\caption{Periodic orbits in the synodic reference frame 
				for a Sun-Jupiter like system (${\eps = 1/1000}$) in the circular-planar case.
			More precisely, the periodic orbits belong to
						(a.) the Lyapunov family $\sL_3$,
					 	(b.) the short-periodic family $\sL_4^s$,
						(c.) the long-periodic family $\sL_4^l$,
						and (d.) the family $f$.
			Their initial conditions are computed with the help of the Poincar\'e maps 
			defined by the following sections:
				(a., d.) ${\Sigma_0 = \{Y = 0, \,\dot{Y}<0 \}}$,
				(b.) ${\Sigma = \{ Y = \sqrt{3}/2, \,\dot{Y}>0\}}$,
				and
				(c.) ${\Sigma\cap\{X>0\}}$.
			}
\label{fig:1}
	\end{center}
\end{figure*}

\begin{figure*}
	\begin{center}
		\tiny
		\def\svgwidth{0.326\textwidth}
\begingroup%
  \makeatletter%
  \providecommand\color[2][]{%
    \errmessage{(Inkscape) Color is used for the text in Inkscape, but the package 'color.sty' is not loaded}%
    \renewcommand\color[2][]{}%
  }%
  \providecommand\transparent[1]{%
    \errmessage{(Inkscape) Transparency is used (non-zero) for the text in Inkscape, but the package 'transparent.sty' is not loaded}%
    \renewcommand\transparent[1]{}%
  }%
  \providecommand\rotatebox[2]{#2}%
  \newcommand*\fsize{\dimexpr\f@size pt\relax}%
  \newcommand*\lineheight[1]{\fontsize{\fsize}{#1\fsize}\selectfont}%
  \ifx\svgwidth\undefined%
    \setlength{\unitlength}{360bp}%
    \ifx\svgscale\undefined%
      \relax%
    \else%
      \setlength{\unitlength}{\unitlength * \real{\svgscale}}%
    \fi%
  \else%
    \setlength{\unitlength}{\svgwidth}%
  \fi%
  \global\let\svgwidth\undefined%
  \global\let\svgscale\undefined%
  \makeatother%
  \begin{picture}(1,1)%
    \lineheight{1}%
    \setlength\tabcolsep{0pt}%
    \put(0,0){\includegraphics[width=\unitlength]{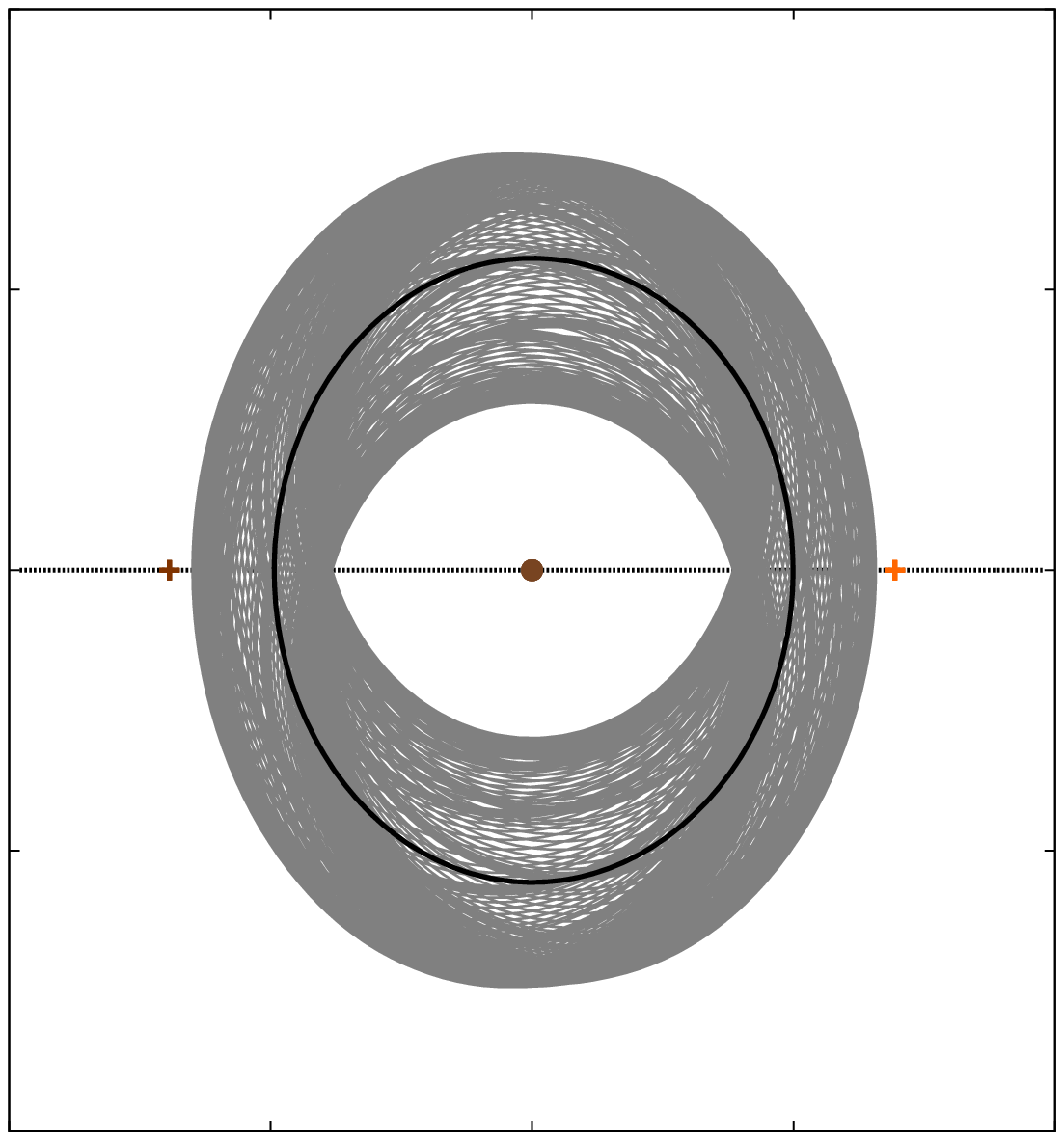}}%
    \put(0.07623889,0.037194){\color[rgb]{0,0,0}\makebox(0,0)[rt]{\lineheight{1.25}\smash{\begin{tabular}[t]{r}-0.1\end{tabular}}}}%
    \put(0.07978151,0.26997177){\color[rgb]{0,0,0}\makebox(0,0)[rt]{\lineheight{1.25}\smash{\begin{tabular}[t]{r}-0.05\end{tabular}}}}%
    \put(0.08002231,0.50261065){\color[rgb]{0,0,0}\makebox(0,0)[rt]{\lineheight{1.25}\smash{\begin{tabular}[t]{r}0\end{tabular}}}}%
    \put(0.07998897,0.73538844){\color[rgb]{0,0,0}\makebox(0,0)[rt]{\lineheight{1.25}\smash{\begin{tabular}[t]{r}0.05\end{tabular}}}}%
    \put(0.07644635,0.96802732){\color[rgb]{0,0,0}\makebox(0,0)[rt]{\lineheight{1.25}\smash{\begin{tabular}[t]{r}0.1\end{tabular}}}}%
    \put(0.09618572,0.01511067){\color[rgb]{0,0,0}\makebox(0,0)[t]{\lineheight{1.25}\smash{\begin{tabular}[t]{c}0.9\end{tabular}}}}%
    \put(0.31304818,0.01511067){\color[rgb]{0,0,0}\makebox(0,0)[t]{\lineheight{1.25}\smash{\begin{tabular}[t]{c}0.95\end{tabular}}}}%
    \put(0.52738016,0.01511067){\color[rgb]{0,0,0}\makebox(0,0)[t]{\lineheight{1.25}\smash{\begin{tabular}[t]{c}1\end{tabular}}}}%
    \put(0.74608202,0.01511067){\color[rgb]{0,0,0}\makebox(0,0)[t]{\lineheight{1.25}\smash{\begin{tabular}[t]{c}1.05\end{tabular}}}}%
    \put(0.96111626,0.01511067){\color[rgb]{0,0,0}\makebox(0,0)[t]{\lineheight{1.25}\smash{\begin{tabular}[t]{c}1.1\end{tabular}}}}%
    \put(0.81755427,0.52763609){\makebox(0,0)[lt]{\lineheight{1.25}\smash{\begin{tabular}[t]{l}$L_2$\end{tabular}}}}%
    \put(0.25269794,0.52763609){\makebox(0,0)[rt]{\lineheight{1.25}\smash{\begin{tabular}[t]{r}$L_1$\end{tabular}}}}%
    \put(0.12076823,0.93261719){\makebox(0,0)[lt]{\lineheight{1.25}\smash{\begin{tabular}[t]{l}a.\end{tabular}}}}%
  \end{picture}%
\endgroup%
		\def\svgwidth{0.326\textwidth}
\begingroup%
  \makeatletter%
  \providecommand\color[2][]{%
    \errmessage{(Inkscape) Color is used for the text in Inkscape, but the package 'color.sty' is not loaded}%
    \renewcommand\color[2][]{}%
  }%
  \providecommand\transparent[1]{%
    \errmessage{(Inkscape) Transparency is used (non-zero) for the text in Inkscape, but the package 'transparent.sty' is not loaded}%
    \renewcommand\transparent[1]{}%
  }%
  \providecommand\rotatebox[2]{#2}%
  \newcommand*\fsize{\dimexpr\f@size pt\relax}%
  \newcommand*\lineheight[1]{\fontsize{\fsize}{#1\fsize}\selectfont}%
  \ifx\svgwidth\undefined%
    \setlength{\unitlength}{360bp}%
    \ifx\svgscale\undefined%
      \relax%
    \else%
      \setlength{\unitlength}{\unitlength * \real{\svgscale}}%
    \fi%
  \else%
    \setlength{\unitlength}{\svgwidth}%
  \fi%
  \global\let\svgwidth\undefined%
  \global\let\svgscale\undefined%
  \makeatother%
  \begin{picture}(1,1)%
    \lineheight{1}%
    \setlength\tabcolsep{0pt}%
    \put(0,0){\includegraphics[width=\unitlength]{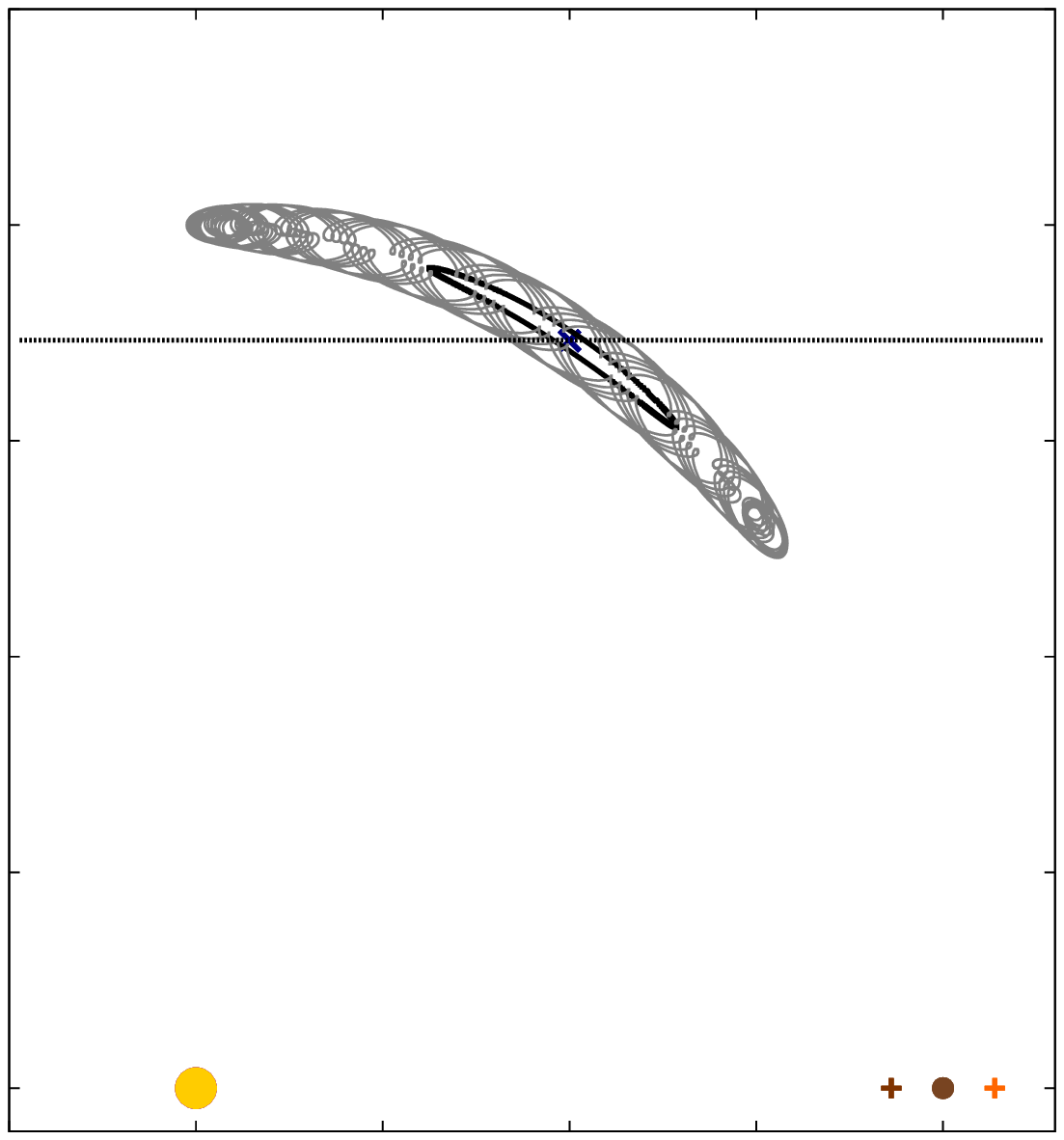}}%
    \put(0.08002231,0.07302733){\color[rgb]{0,0,0}\makebox(0,0)[rt]{\lineheight{1.25}\smash{\begin{tabular}[t]{r}0\end{tabular}}}}%
    \put(0.07998897,0.2520551){\color[rgb]{0,0,0}\makebox(0,0)[rt]{\lineheight{1.25}\smash{\begin{tabular}[t]{r}0.25\end{tabular}}}}%
    \put(0.07999453,0.43094399){\color[rgb]{0,0,0}\makebox(0,0)[rt]{\lineheight{1.25}\smash{\begin{tabular}[t]{r}0.5\end{tabular}}}}%
    \put(0.07998897,0.60997176){\color[rgb]{0,0,0}\makebox(0,0)[rt]{\lineheight{1.25}\smash{\begin{tabular}[t]{r}0.75\end{tabular}}}}%
    \put(0.07504184,0.78899955){\color[rgb]{0,0,0}\makebox(0,0)[rt]{\lineheight{1.25}\smash{\begin{tabular}[t]{r}1\end{tabular}}}}%
    \put(0.07855668,0.96802732){\color[rgb]{0,0,0}\makebox(0,0)[rt]{\lineheight{1.25}\smash{\begin{tabular}[t]{r}1.25\end{tabular}}}}%
    \put(0.0930852,0.01511067){\color[rgb]{0,0,0}\makebox(0,0)[t]{\lineheight{1.25}\smash{\begin{tabular}[t]{c}-0.25\end{tabular}}}}%
    \put(0.25112039,0.01511067){\color[rgb]{0,0,0}\makebox(0,0)[t]{\lineheight{1.25}\smash{\begin{tabular}[t]{c}0\end{tabular}}}}%
    \put(0.40596484,0.01511067){\color[rgb]{0,0,0}\makebox(0,0)[t]{\lineheight{1.25}\smash{\begin{tabular}[t]{c}0.25\end{tabular}}}}%
    \put(0.56096762,0.01511067){\color[rgb]{0,0,0}\makebox(0,0)[t]{\lineheight{1.25}\smash{\begin{tabular}[t]{c}0.5\end{tabular}}}}%
    \put(0.71582595,0.01511067){\color[rgb]{0,0,0}\makebox(0,0)[t]{\lineheight{1.25}\smash{\begin{tabular}[t]{c}0.75\end{tabular}}}}%
    \put(0.86821349,0.01511067){\color[rgb]{0,0,0}\makebox(0,0)[t]{\lineheight{1.25}\smash{\begin{tabular}[t]{c}1\end{tabular}}}}%
    \put(0.83763697,0.10377545){\makebox(0,0)[rt]{\lineheight{1.25}\smash{\begin{tabular}[t]{r}$L_1$\end{tabular}}}}%
    \put(0.90973332,0.10377562){\makebox(0,0)[t]{\lineheight{1.25}\smash{\begin{tabular}[t]{c}$L_2$\end{tabular}}}}%
    \put(0.54117702,0.72044212){\makebox(0,0)[lt]{\lineheight{1.25}\smash{\begin{tabular}[t]{l}$L_4$\end{tabular}}}}%
    \put(0.12072754,0.93265788){\makebox(0,0)[lt]{\lineheight{1.25}\smash{\begin{tabular}[t]{l}c.\end{tabular}}}}%
  \end{picture}%
\endgroup%
		\def\svgwidth{0.326\textwidth}
\begingroup%
  \makeatletter%
  \providecommand\color[2][]{%
    \errmessage{(Inkscape) Color is used for the text in Inkscape, but the package 'color.sty' is not loaded}%
    \renewcommand\color[2][]{}%
  }%
  \providecommand\transparent[1]{%
    \errmessage{(Inkscape) Transparency is used (non-zero) for the text in Inkscape, but the package 'transparent.sty' is not loaded}%
    \renewcommand\transparent[1]{}%
  }%
  \providecommand\rotatebox[2]{#2}%
  \newcommand*\fsize{\dimexpr\f@size pt\relax}%
  \newcommand*\lineheight[1]{\fontsize{\fsize}{#1\fsize}\selectfont}%
  \ifx\svgwidth\undefined%
    \setlength{\unitlength}{360bp}%
    \ifx\svgscale\undefined%
      \relax%
    \else%
      \setlength{\unitlength}{\unitlength * \real{\svgscale}}%
    \fi%
  \else%
    \setlength{\unitlength}{\svgwidth}%
  \fi%
  \global\let\svgwidth\undefined%
  \global\let\svgscale\undefined%
  \makeatother%
  \begin{picture}(1,1)%
    \lineheight{1}%
    \setlength\tabcolsep{0pt}%
    \put(0,0){\includegraphics[width=\unitlength]{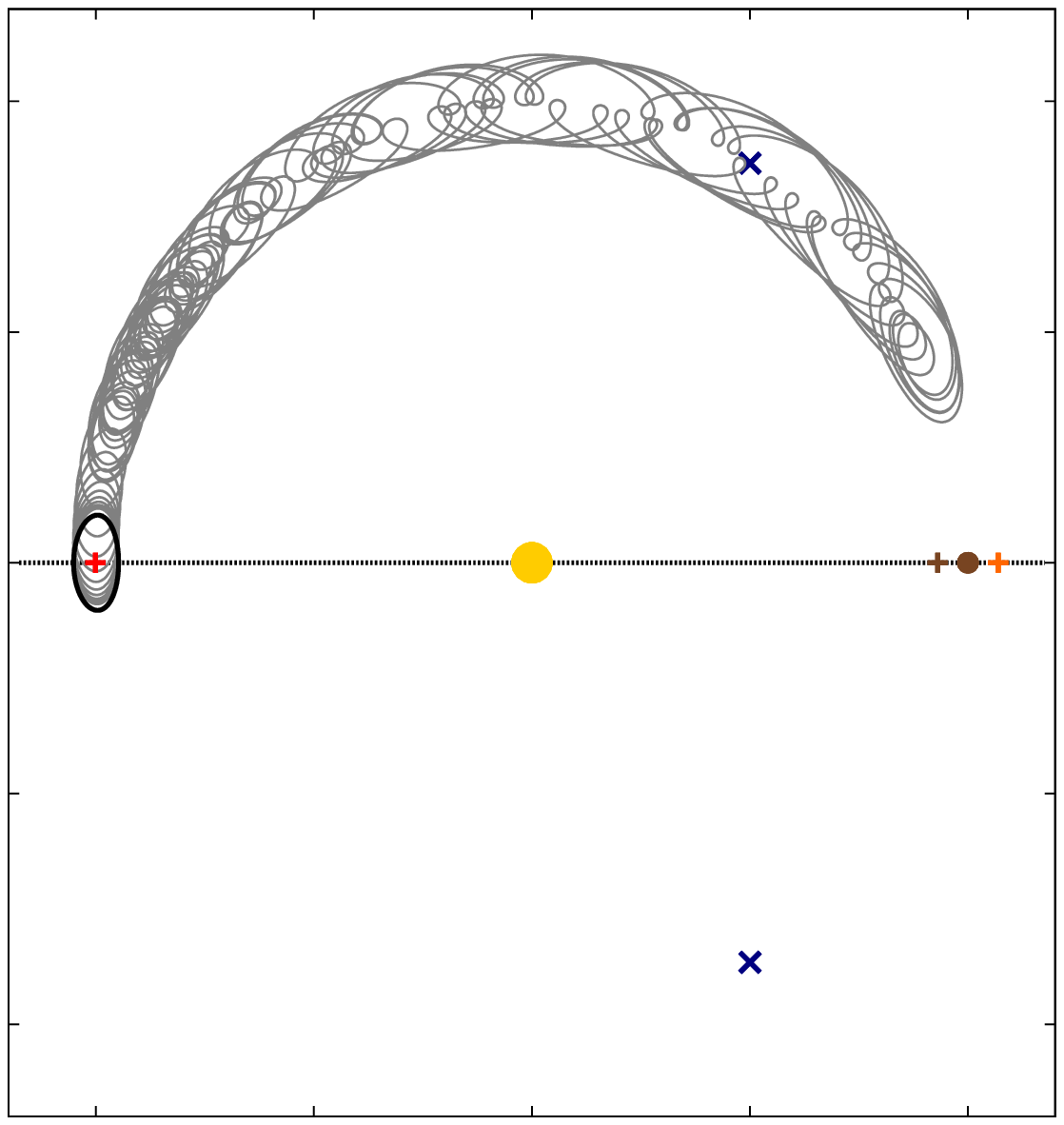}}%
    \put(0.06392778,0.114694){\color[rgb]{0,0,0}\makebox(0,0)[rt]{\lineheight{1.25}\smash{\begin{tabular}[t]{r}-1\end{tabular}}}}%
    \put(0.06742595,0.30872177){\color[rgb]{0,0,0}\makebox(0,0)[rt]{\lineheight{1.25}\smash{\begin{tabular}[t]{r}-0.5\end{tabular}}}}%
    \put(0.0676612,0.50261065){\color[rgb]{0,0,0}\makebox(0,0)[rt]{\lineheight{1.25}\smash{\begin{tabular}[t]{r}0\end{tabular}}}}%
    \put(0.06763342,0.69649954){\color[rgb]{0,0,0}\makebox(0,0)[rt]{\lineheight{1.25}\smash{\begin{tabular}[t]{r}0.5\end{tabular}}}}%
    \put(0.06268073,0.89052732){\color[rgb]{0,0,0}\makebox(0,0)[rt]{\lineheight{1.25}\smash{\begin{tabular}[t]{r}1\end{tabular}}}}%
    \put(0.15231111,0.01511067){\color[rgb]{0,0,0}\makebox(0,0)[t]{\lineheight{1.25}\smash{\begin{tabular}[t]{c}-1\end{tabular}}}}%
    \put(0.33725464,0.01511067){\color[rgb]{0,0,0}\makebox(0,0)[t]{\lineheight{1.25}\smash{\begin{tabular}[t]{c}-0.5\end{tabular}}}}%
    \put(0.52375929,0.01511067){\color[rgb]{0,0,0}\makebox(0,0)[t]{\lineheight{1.25}\smash{\begin{tabular}[t]{c}0\end{tabular}}}}%
    \put(0.70707872,0.01511067){\color[rgb]{0,0,0}\makebox(0,0)[t]{\lineheight{1.25}\smash{\begin{tabular}[t]{c}0.5\end{tabular}}}}%
    \put(0.88779682,0.01511067){\color[rgb]{0,0,0}\makebox(0,0)[t]{\lineheight{1.25}\smash{\begin{tabular}[t]{c}1\end{tabular}}}}%
    \put(0.69438568,0.17318081){\makebox(0,0)[rt]{\lineheight{1.25}\smash{\begin{tabular}[t]{r}$L_5$\end{tabular}}}}%
    \put(0.69438568,0.84401406){\makebox(0,0)[rt]{\lineheight{1.25}\smash{\begin{tabular}[t]{r}$L_4$\end{tabular}}}}%
    \put(0.86521901,0.53151405){\makebox(0,0)[rt]{\lineheight{1.25}\smash{\begin{tabular}[t]{r}$L_1$\end{tabular}}}}%
    \put(0.88381344,0.53151407){\makebox(0,0)[lt]{\lineheight{1.25}\smash{\begin{tabular}[t]{l}$L_2$\end{tabular}}}}%
    \put(0.14375936,0.52318072){\makebox(0,0)[lt]{\lineheight{1.25}\smash{\begin{tabular}[t]{l}$L_3$\end{tabular}}}}%
    \put(0.12068685,0.93245443){\makebox(0,0)[lt]{\lineheight{1.25}\smash{\begin{tabular}[t]{l}e.\end{tabular}}}}%
  \end{picture}%
\endgroup%
\\
		\def\svgwidth{0.326\textwidth}
\begingroup%
  \makeatletter%
  \providecommand\color[2][]{%
    \errmessage{(Inkscape) Color is used for the text in Inkscape, but the package 'color.sty' is not loaded}%
    \renewcommand\color[2][]{}%
  }%
  \providecommand\transparent[1]{%
    \errmessage{(Inkscape) Transparency is used (non-zero) for the text in Inkscape, but the package 'transparent.sty' is not loaded}%
    \renewcommand\transparent[1]{}%
  }%
  \providecommand\rotatebox[2]{#2}%
  \newcommand*\fsize{\dimexpr\f@size pt\relax}%
  \newcommand*\lineheight[1]{\fontsize{\fsize}{#1\fsize}\selectfont}%
  \ifx\svgwidth\undefined%
    \setlength{\unitlength}{360bp}%
    \ifx\svgscale\undefined%
      \relax%
    \else%
      \setlength{\unitlength}{\unitlength * \real{\svgscale}}%
    \fi%
  \else%
    \setlength{\unitlength}{\svgwidth}%
  \fi%
  \global\let\svgwidth\undefined%
  \global\let\svgscale\undefined%
  \makeatother%
  \begin{picture}(1,1)%
    \lineheight{1}%
    \setlength\tabcolsep{0pt}%
    \put(0,0){\includegraphics[width=\unitlength]{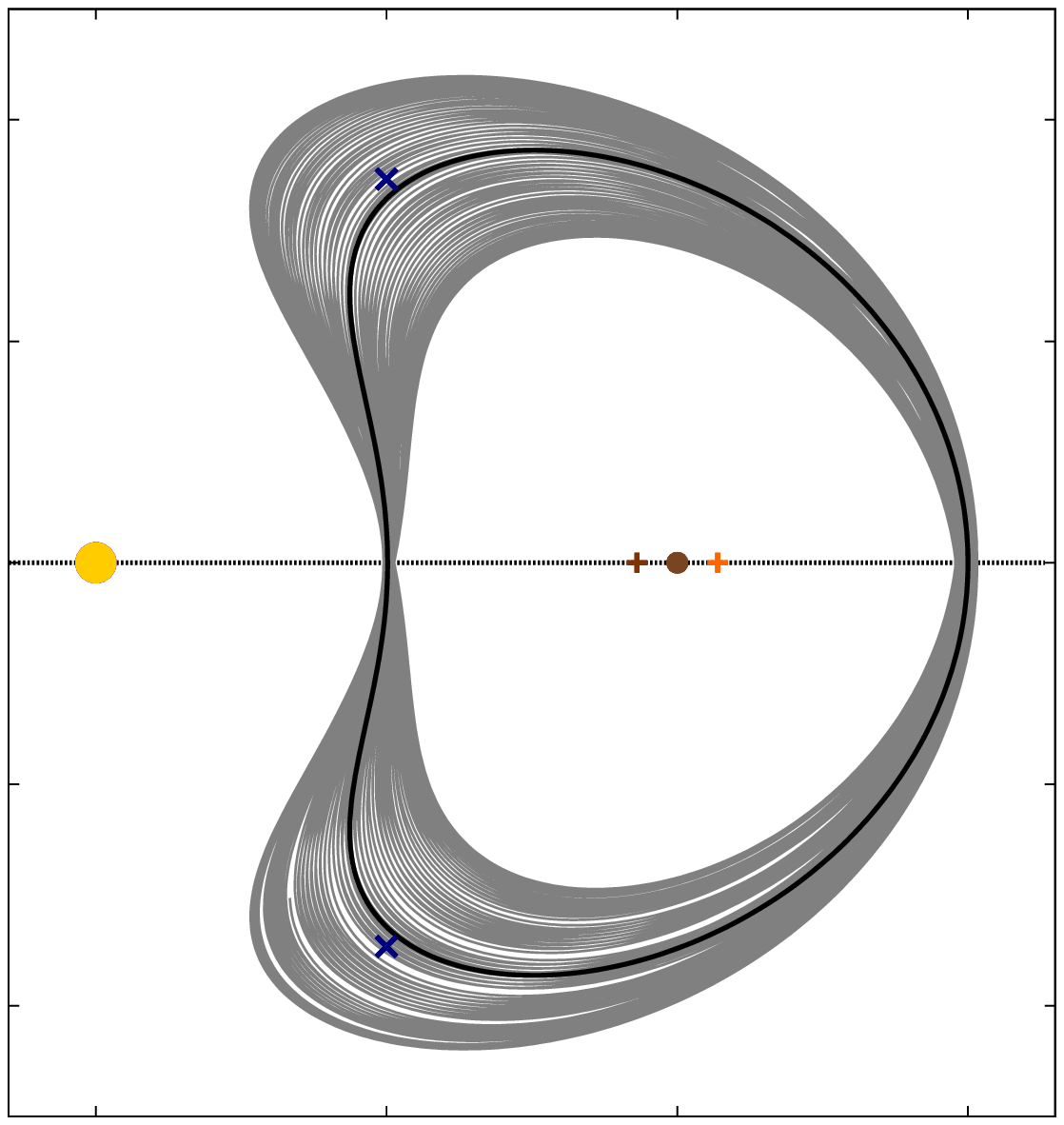}}%
    \put(0.0626908,0.13024955){\color[rgb]{0,0,0}\makebox(0,0)[rt]{\lineheight{1.25}\smash{\begin{tabular}[t]{r}-1\end{tabular}}}}%
    \put(0.06618897,0.31649955){\color[rgb]{0,0,0}\makebox(0,0)[rt]{\lineheight{1.25}\smash{\begin{tabular}[t]{r}-0.5\end{tabular}}}}%
    \put(0.0676612,0.50261065){\color[rgb]{0,0,0}\makebox(0,0)[rt]{\lineheight{1.25}\smash{\begin{tabular}[t]{r}0\end{tabular}}}}%
    \put(0.06763342,0.68872176){\color[rgb]{0,0,0}\makebox(0,0)[rt]{\lineheight{1.25}\smash{\begin{tabular}[t]{r}0.5\end{tabular}}}}%
    \put(0.06268073,0.87497177){\color[rgb]{0,0,0}\makebox(0,0)[rt]{\lineheight{1.25}\smash{\begin{tabular}[t]{r}1\end{tabular}}}}%
    \put(0.15723151,0.01511067){\color[rgb]{0,0,0}\makebox(0,0)[t]{\lineheight{1.25}\smash{\begin{tabular}[t]{c}0\end{tabular}}}}%
    \put(0.40152318,0.01511067){\color[rgb]{0,0,0}\makebox(0,0)[t]{\lineheight{1.25}\smash{\begin{tabular}[t]{c}0.5\end{tabular}}}}%
    \put(0.64349127,0.01511067){\color[rgb]{0,0,0}\makebox(0,0)[t]{\lineheight{1.25}\smash{\begin{tabular}[t]{c}1\end{tabular}}}}%
    \put(0.88955702,0.01511067){\color[rgb]{0,0,0}\makebox(0,0)[t]{\lineheight{1.25}\smash{\begin{tabular}[t]{c}1.5\end{tabular}}}}%
    \put(0.65359,0.52674288){\makebox(0,0)[lt]{\lineheight{1.25}\smash{\begin{tabular}[t]{l}$L_2$\end{tabular}}}}%
    \put(0.64171667,0.52674288){\makebox(0,0)[rt]{\lineheight{1.25}\smash{\begin{tabular}[t]{r}$L_1$\end{tabular}}}}%
    \put(0.38755001,0.83507621){\makebox(0,0)[rt]{\lineheight{1.25}\smash{\begin{tabular}[t]{r}$L_4$\end{tabular}}}}%
    \put(0.38755001,0.18924289){\makebox(0,0)[rt]{\lineheight{1.25}\smash{\begin{tabular}[t]{r}$L_5$\end{tabular}}}}%
    \put(0.11747233,0.91585287){\makebox(0,0)[lt]{\lineheight{1.25}\smash{\begin{tabular}[t]{l}b.\end{tabular}}}}%
  \end{picture}%
\endgroup%
		\def\svgwidth{0.326\textwidth}
\begingroup%
  \makeatletter%
  \providecommand\color[2][]{%
    \errmessage{(Inkscape) Color is used for the text in Inkscape, but the package 'color.sty' is not loaded}%
    \renewcommand\color[2][]{}%
  }%
  \providecommand\transparent[1]{%
    \errmessage{(Inkscape) Transparency is used (non-zero) for the text in Inkscape, but the package 'transparent.sty' is not loaded}%
    \renewcommand\transparent[1]{}%
  }%
  \providecommand\rotatebox[2]{#2}%
  \newcommand*\fsize{\dimexpr\f@size pt\relax}%
  \newcommand*\lineheight[1]{\fontsize{\fsize}{#1\fsize}\selectfont}%
  \ifx\svgwidth\undefined%
    \setlength{\unitlength}{360bp}%
    \ifx\svgscale\undefined%
      \relax%
    \else%
      \setlength{\unitlength}{\unitlength * \real{\svgscale}}%
    \fi%
  \else%
    \setlength{\unitlength}{\svgwidth}%
  \fi%
  \global\let\svgwidth\undefined%
  \global\let\svgscale\undefined%
  \makeatother%
  \begin{picture}(1,1)%
    \lineheight{1}%
    \setlength\tabcolsep{0pt}%
    \put(0,0){\includegraphics[width=\unitlength]{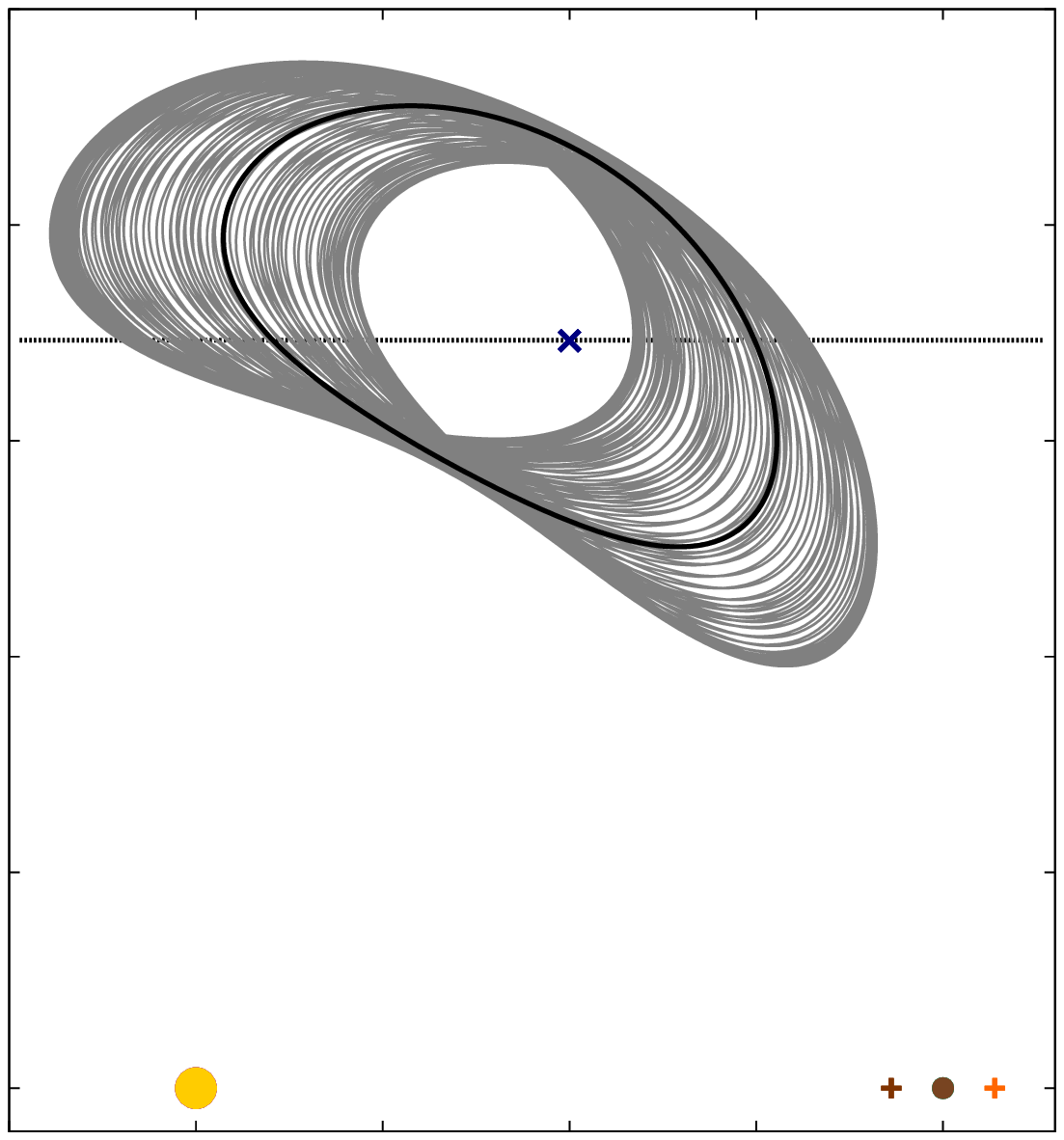}}%
    \put(0.08002231,0.07302733){\color[rgb]{0,0,0}\makebox(0,0)[rt]{\lineheight{1.25}\smash{\begin{tabular}[t]{r}0\end{tabular}}}}%
    \put(0.07998897,0.2520551){\color[rgb]{0,0,0}\makebox(0,0)[rt]{\lineheight{1.25}\smash{\begin{tabular}[t]{r}0.25\end{tabular}}}}%
    \put(0.07999453,0.43094399){\color[rgb]{0,0,0}\makebox(0,0)[rt]{\lineheight{1.25}\smash{\begin{tabular}[t]{r}0.5\end{tabular}}}}%
    \put(0.07998897,0.60997176){\color[rgb]{0,0,0}\makebox(0,0)[rt]{\lineheight{1.25}\smash{\begin{tabular}[t]{r}0.75\end{tabular}}}}%
    \put(0.07504184,0.78899955){\color[rgb]{0,0,0}\makebox(0,0)[rt]{\lineheight{1.25}\smash{\begin{tabular}[t]{r}1\end{tabular}}}}%
    \put(0.07855668,0.96802732){\color[rgb]{0,0,0}\makebox(0,0)[rt]{\lineheight{1.25}\smash{\begin{tabular}[t]{r}1.25\end{tabular}}}}%
    \put(0.0930852,0.01511067){\color[rgb]{0,0,0}\makebox(0,0)[t]{\lineheight{1.25}\smash{\begin{tabular}[t]{c}-0.25\end{tabular}}}}%
    \put(0.25112039,0.01511067){\color[rgb]{0,0,0}\makebox(0,0)[t]{\lineheight{1.25}\smash{\begin{tabular}[t]{c}0\end{tabular}}}}%
    \put(0.40596484,0.01511067){\color[rgb]{0,0,0}\makebox(0,0)[t]{\lineheight{1.25}\smash{\begin{tabular}[t]{c}0.25\end{tabular}}}}%
    \put(0.56096762,0.01511067){\color[rgb]{0,0,0}\makebox(0,0)[t]{\lineheight{1.25}\smash{\begin{tabular}[t]{c}0.5\end{tabular}}}}%
    \put(0.71582595,0.01511067){\color[rgb]{0,0,0}\makebox(0,0)[t]{\lineheight{1.25}\smash{\begin{tabular}[t]{c}0.75\end{tabular}}}}%
    \put(0.86821349,0.01511067){\color[rgb]{0,0,0}\makebox(0,0)[t]{\lineheight{1.25}\smash{\begin{tabular}[t]{c}1\end{tabular}}}}%
    \put(0.83763697,0.10377545){\makebox(0,0)[rt]{\lineheight{1.25}\smash{\begin{tabular}[t]{r}$L_1$\end{tabular}}}}%
    \put(0.90973332,0.10377562){\makebox(0,0)[t]{\lineheight{1.25}\smash{\begin{tabular}[t]{c}$L_2$\end{tabular}}}}%
    \put(0.54117697,0.7204421){\makebox(0,0)[lt]{\lineheight{1.25}\smash{\begin{tabular}[t]{l}$L_4$\end{tabular}}}}%
    \put(0.12060547,0.91585287){\makebox(0,0)[lt]{\lineheight{1.25}\smash{\begin{tabular}[t]{l}d.\end{tabular}}}}%
  \end{picture}%
\endgroup%
		\def\svgwidth{0.326\textwidth}
\begingroup%
  \makeatletter%
  \providecommand\color[2][]{%
    \errmessage{(Inkscape) Color is used for the text in Inkscape, but the package 'color.sty' is not loaded}%
    \renewcommand\color[2][]{}%
  }%
  \providecommand\transparent[1]{%
    \errmessage{(Inkscape) Transparency is used (non-zero) for the text in Inkscape, but the package 'transparent.sty' is not loaded}%
    \renewcommand\transparent[1]{}%
  }%
  \providecommand\rotatebox[2]{#2}%
  \newcommand*\fsize{\dimexpr\f@size pt\relax}%
  \newcommand*\lineheight[1]{\fontsize{\fsize}{#1\fsize}\selectfont}%
  \ifx\svgwidth\undefined%
    \setlength{\unitlength}{360bp}%
    \ifx\svgscale\undefined%
      \relax%
    \else%
      \setlength{\unitlength}{\unitlength * \real{\svgscale}}%
    \fi%
  \else%
    \setlength{\unitlength}{\svgwidth}%
  \fi%
  \global\let\svgwidth\undefined%
  \global\let\svgscale\undefined%
  \makeatother%
  \begin{picture}(1,1)%
    \lineheight{1}%
    \setlength\tabcolsep{0pt}%
    \put(0,0){\includegraphics[width=\unitlength]{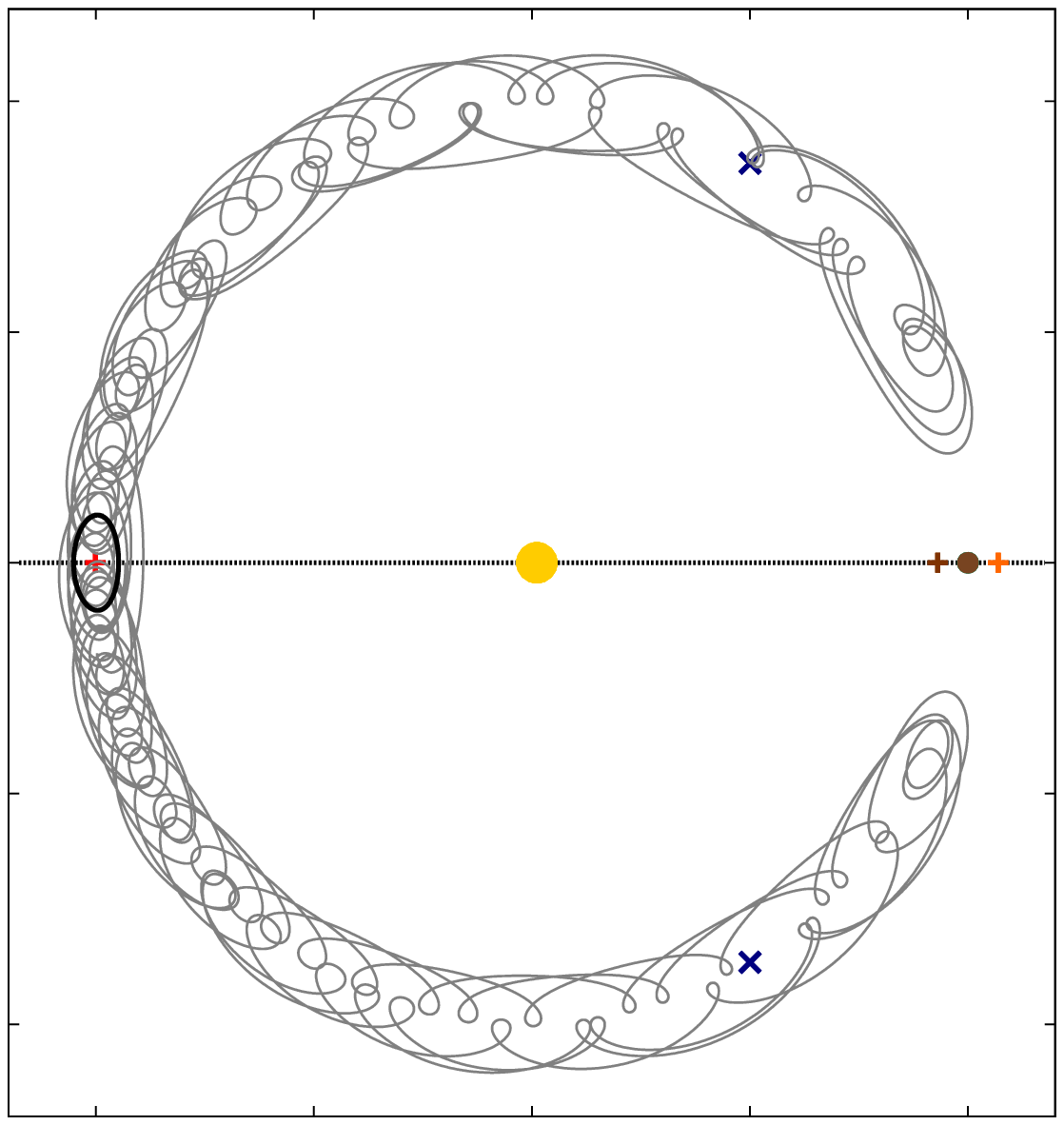}}%
    \put(0.06392778,0.114694){\color[rgb]{0,0,0}\makebox(0,0)[rt]{\lineheight{1.25}\smash{\begin{tabular}[t]{r}-1\end{tabular}}}}%
    \put(0.06742595,0.30872177){\color[rgb]{0,0,0}\makebox(0,0)[rt]{\lineheight{1.25}\smash{\begin{tabular}[t]{r}-0.5\end{tabular}}}}%
    \put(0.0676612,0.50261065){\color[rgb]{0,0,0}\makebox(0,0)[rt]{\lineheight{1.25}\smash{\begin{tabular}[t]{r}0\end{tabular}}}}%
    \put(0.06763342,0.69649954){\color[rgb]{0,0,0}\makebox(0,0)[rt]{\lineheight{1.25}\smash{\begin{tabular}[t]{r}0.5\end{tabular}}}}%
    \put(0.06268073,0.89052732){\color[rgb]{0,0,0}\makebox(0,0)[rt]{\lineheight{1.25}\smash{\begin{tabular}[t]{r}1\end{tabular}}}}%
    \put(0.15231111,0.01511067){\color[rgb]{0,0,0}\makebox(0,0)[t]{\lineheight{1.25}\smash{\begin{tabular}[t]{c}-1\end{tabular}}}}%
    \put(0.33725464,0.01511067){\color[rgb]{0,0,0}\makebox(0,0)[t]{\lineheight{1.25}\smash{\begin{tabular}[t]{c}-0.5\end{tabular}}}}%
    \put(0.52375929,0.01511067){\color[rgb]{0,0,0}\makebox(0,0)[t]{\lineheight{1.25}\smash{\begin{tabular}[t]{c}0\end{tabular}}}}%
    \put(0.70707872,0.01511067){\color[rgb]{0,0,0}\makebox(0,0)[t]{\lineheight{1.25}\smash{\begin{tabular}[t]{c}0.5\end{tabular}}}}%
    \put(0.88779682,0.01511067){\color[rgb]{0,0,0}\makebox(0,0)[t]{\lineheight{1.25}\smash{\begin{tabular}[t]{c}1\end{tabular}}}}%
    \put(0.6972222,0.17652969){\makebox(0,0)[rt]{\lineheight{1.25}\smash{\begin{tabular}[t]{r}$L_5$\end{tabular}}}}%
    \put(0.6972222,0.84736301){\makebox(0,0)[rt]{\lineheight{1.25}\smash{\begin{tabular}[t]{r}$L_4$\end{tabular}}}}%
    \put(0.8597222,0.53069635){\makebox(0,0)[rt]{\lineheight{1.25}\smash{\begin{tabular}[t]{r}$L_1$\end{tabular}}}}%
    \put(0.91280893,0.53486302){\makebox(0,0)[t]{\lineheight{1.25}\smash{\begin{tabular}[t]{c}$L_2$\end{tabular}}}}%
    \put(0.14242892,0.52236301){\makebox(0,0)[lt]{\lineheight{1.25}\smash{\begin{tabular}[t]{l}$L_3$\end{tabular}}}}%
    \put(0.12223307,0.91544597){\makebox(0,0)[lt]{\lineheight{1.25}\smash{\begin{tabular}[t]{l}f.\end{tabular}}}}%
  \end{picture}%
\endgroup%
		\caption{Dynamics located in the neighborhood of 
			(a., b.) the family $f$, 
			(c.) the long-periodic family $\sL_4^l$, 
			(d.) the short-periodic family $\sL_4^s$, and
			  (e., f.) the Lyapunov family $\sL_3$.  
			Considering a periodic orbit that belongs to each family (black curve) 
			whose cross-section is denoted ${(X_0, Y_0, \dot{X}_0, \dot{Y}_0)}$
			a trajectory located in its vicinity,
				that is, with an initial condition $(X,Y, \dot{X}, \dot{Y}) = (X_0,Y_0, \dot{X}_0, \dot{Y}_0) + \cO(\eps)$
			is propagated after $100$ revolutions of Jupiter. 
		The  dynamics observed are : 
			(a.) the ‘‘satellized" retrograde satellite orbits, 
			(b.) the quasi-satellite motion, 
			(c., d., e.) the tadpole motion, and
			(f.) the horseshoe motion.}


\label{fig:2}

	\end{center}
\end{figure*}			

		We recall that the configuration space of the circular-planar case 
			coincides with the orbital plane of the planet.
		In the following, the motion of the particle will be described in terms of polar coordinates 
		with  ${\phi = \arg(\bR)}$ 
		that illustrates the relative motion between the planet and the particle and ${R =\norm{\bR}}$.
		Most of the results mentioned below can be found with different notations 
		in the book of Szebehely \cite{1967Sz}.
		
		First of all, the five Lagrange fixed points, denoted $L_j$ for ${j=1,2,3,4,5}$, 
			are the unique equilibria of the restricted three-body problem in the circular case.
		$L_1$ and $L_2$ belong to the Sun-planet axis, in
		\bes
			{\phi_{j} = 0\degre}, \quad {R_{j} = 1 + (-1)^j \times\left(\frac{\eps}{3}\right)^{1/3}  + \cO(\eps^{2/3})},
		\ees
			that is, from either side of the planet.
		Moreover, they embody the diameter of the Hill's sphere of the planet, 
			that is the region of the configuration space 
			inside which the gravitational influence of the planet dominates with respect to the one of the Sun.
		$L_3$ is also located on the Sun-planet axis, in
		\bes
			\phi_{3}=180\degre, \quad R_3 =  1 -  \frac{7}{12}\eps + \cO(\eps^2).
		\ees
		$L_4$ and $L_5$ are the Lagrange configurations such that
			the particle lie at the vertex of an equilateral triangle formed with the Sun and the planet,
			that is, in ${\phi_{j} = (-1)^j \times60\degre}$ and ${R_{j}= 1}$.		
			
		For ${j=4,5}$ and $\eps$ small enough\footnote{
		More precisely, $L_4$ and $L_5$ are elliptic fixed points 
		for ${\eps <\eps_*}$ with ${\eps_* \simeq 0,038}$.},
		$L_j$ is an elliptic equilibrium where two one-parameter families of periodic orbits stem from.
				They are tangential to each center eigenspace of the equilibrium point.
		Being the two center eigenspaces associated with 
			frequencies respectively in $\cO(1)$ and $\cO(\sqrt{\eps})$,
		these families are generally denoted as short-periodic $\sL_j^s$ and long-periodic $\sL_j^l$,
		in correspondence to their associated timescale in the neighborhood of the equilibrium.
		$L_1$, $L_2$ and $L_3$ are unstable for all $\eps>0$ 
		and each equilibrium possesses one center eigenspace.
		The same reasoning applies and provides three one-parameter families of periodic orbits
		generally known as the Lyapunov families $\sL_1$, $\sL_2$ and $\sL_3$.
		Only $\sL_3$ will be discussed in the following.	

		The Poincar\'e map is the classical way to compute periodic orbits.
		For $\sL_j^s$ and $\sL_j^l$, suitable sections are given by 
		${\Sigma = \{ Y = R_j \sin\phi_j,\, \dot{Y}>0 \}}$ and ${\Sigma\cap\{ X>0\}}$, 
		which require three free parameters (e.g., the energy, $X$ and $\dot{X}$) in order to locate the crossing.
		We recall that the Lyapunov trajectories are symmetrical with respect to the Sun-planet axis and 
			cross the $X$-axis in ${\dot{X} = 0}$.
		Thus, a natural section, that requires only two parameters (e.g., the energy and $X$), 
		is given by ${\Sigma_0 = \{ Y = 0, \, \dot{Y}<0, \,\dot{X} = 0 \}}$.
		Then, a fixed point method is generally performed from a suitable initial guess 
		that makes the method convergent.
		For that purpose, a first approximation of a crossing is obtained 
			by the resolution of the linearized system associated with the equilibrium $L_j$
		and a continuation method is implemented.					

		Figure \ref{fig:1} displays some periodic orbits, 
			computed in the case of a Sun-Jupiter like system (${\eps = 1/1000}$)
			by varying $X$ along the section.
			
		Since ${X> -R_3}$ increases, 
			the size of a trajectory that belongs to $\sL_3$ (Fig.~\ref{fig:1}a) increases 
			and its shape no longer looks like to an ellipse centered on $L_3$.
		More precisely, $\phi$ and $R$ oscillate respectively about $180\degre$ and 1,
			whose respective amplitude increases with $X$ 
			and reaches large values close to $180\degre$ and 1.
		Moreover, the ``guiding center" of each periodic trajectory
			(the approximate position around which the trajectory oscillates) 
		remains $L_3$.
		Decreasing ${X< 1/2}$, the shape of the trajectories of $\sL_4^s$ (Fig.~\ref{fig:1}b) 
				has a quite similar evolution to the one of $\sL_3$. 
		However, two main differences exist: the shape is not symmetric,
			and the guiding center shifts from $L_4$ toward $L_3$ along the circle ${R = 1}$.
		The same behavior is observed symmetrically for $\sL_5^s$.
		We point out that, for a given energy, 
			$\sL_4^s$ and $\sL_5^s$ merge together with $\sL_3$.
		This result was found by Deprit et al. \cite{1967DeJaPa} for an Earth-Moon like system (${\eps=1/81}$) 
		in the circular-planar case.
		The features of $\sL_4^l$  (Fig.~\ref{fig:1}c) are different.
		Indeed, as long as ${X<1/2}$ decreases, 
				the size of a trajectory increases,
				while its shape changes and looks like a tadpole, with 
				the head centered on $L_4$ and the tail that extends toward $L_3$.
		In other words, by decreasing $X$, $R$ oscillates about 1 with an amplitude that increases 
		but remains much smaller than 1, while $\phi$ encompasses $60\degre$ with increasing oscillations 
		included in the range $]0\degre,180\degre[$.

		Other families of periodic orbits exist,
			and several classifications have been realized 
			(see \cite{1933St,1967Sz,1997He}).
		Among them, the family $f$ is especially remarkable:
			it is a one-parameter family of symmetrical periodic orbits
			whose motion in the synodic reference frame looks like the one of a retrograde satellite of the planet,
			and that extends from an infinitesimal neighborhood of the planet 
			(i.e., inside its Hill's sphere)		
			to the collision with the Sun (i.e., far beyond the Hill's sphere of the planet).
		Its computation is similar to the one of the Lyapunov families. 
		However, since it does not originate from a Lagrange fixed point,
			the initial guess of the method is given by the two following limit cases (see \cite{1974Be}):
		\begin{itemize}
			\item	$\dot{Y} 	= -X + \sqrt{\frac{2-X}{X}}$ 	for $\eps\simeq 0$,
			\item $\dot{Y} 	= -(X-1) - \sqrt{\frac{\eps}{X-1}}$ for $X\simeq 1$.
		\end{itemize}

		Figure \ref{fig:1}d depicts some trajectories of the family $f$. 
		By varying ${X>1}$, their shape has the same evolution to the one of $\sL_3$.
		More precisely,
			the family $f$ seems the symmetrical family of $\sL_3$ with respect to the $Y$-axis,
		characterized by $\phi$ that oscillates about zero and thus a guiding center located on the planet. 		

		The stability character of a periodic orbit 
			can be deduced from the monodromy matrix.
		For $\eps$ small enough\footnote{More precisely,
			the whole family $f$ is stable for $\eps< \eps_*$ with $\eps_*< 0.0477$ (see \cite{1970HeGu}).},
			the family $f$ is normally elliptic 
			except in two particular orbits 
			that split the neighborhood of the family in three different domains 
			(see \cite{2017PoRoVi} for more details). 
			One belongs to the Hill's sphere and corresponds to the ‘‘satellized" retrograde satellite orbits.
			The two others stand for the quasi-satellite orbits, also known as distant retrograde orbits
			(DRO).
		Examples of ‘‘satellized" retrograde satellite and quasi-satellite orbits, 
			computed during 100 revolutions of Jupiter,
		are depicted in Fig.~\ref{fig:2}a-b.			
		Both of the families $\sL_j^l$ and $\sL_j^s$ are normally elliptic close to the equilibrium. 
		The tadpole-shaped trajectories depicted in Fig.~\ref{fig:2}c and Fig.~\ref{fig:2}d,
		start in the neighborhood of periodic orbits
		that belongs to $\sL_4^l$ and $\sL_4^s$, respectively.
		A part of $\sL_3$ near the equilibrium is normally hyperbolic  and 
		two types of dynamics can be observed in its neighborhood:
			tadpole-shaped orbits with a large amplitude (Fig.~\ref{fig:2}e), and
			horseshoe-shaped orbits that encompass the three fixed points $L_3$, $L_4$ and $L_5$
			(Fig.~\ref{fig:2}f).
		The horseshoe-shaped orbit is characterized 
		by $R$ that oscillates about 1 with an amplitude smaller than 1,
		and $\phi$ that features very large oscillations centered on $180\degre$.

		To summarize the situation, we described four types of dynamics --
			‘‘satellized" retrograde satellite, quasi-satellite, tadpole motion and horseshoe motion --
			that starts in the vicinity of periodic orbits that belong to the six families mentioned above.
		These dynamics are related by the same features, that is, $R$ that oscillates about 1
		and $\phi$ that does not circulates but oscillates around a given value.
		A natural issue is to understand how these dynamics are organized in the phase space
			of the restricted three-body problem,
			and especially, if some boundaries  can be identified.
		Nevertheless, the four dimensions of the phase space 
			make difficult  the achievement of this goal.
	
		A way to overcome this difficulty is given by a suitable perturbative treatment
			that focuses on the families of periodic orbits.
		First of all, let us recall that the particle and the planet are considered in mean-motion resonance, 
		and especially in $p$:$q$ mean-motion resonance,
		if they complete respectively $p$ and $q$ revolutions around the Sun in the same time.		
		According to the Poincar\'e classification  (see \cite{2012Ch}),
			a periodic orbit of the second or the third ‘‘sort" (also translated as ‘‘kind")			
			is the continuation, from the limit case $\eps =0$,
			of a heliocentric Kepler orbit in mean-motion resonance with the planet.
		For instance, 
		 	the families $\sL_3$, $\sL_j^s$ as well as the part of the family $f$ that stands outside the Hill's sphere 
				are the continuation of Kepler orbits in 1:1 mean-motion resonance 
				(see the book of H\'enon \cite{1997He} for complete details on the periodic orbit classification).
		Hence, a perturbative treatment that considers $\eps$ as a small parameter, 
			and focuses on a small enough neighborhood of a given mean-motion resonance
			provides another way to approach some families of periodic orbits 
			and thus to understand the corresponding dynamics.
		This is the underlying idea associated with the averaged problem that is considered in this work
			and that we recall in the following section.	


	\subsection{Perturbative treatment of a mean-motion resonance: the averaged problem}
\label{sec:pert_ap}				

		From now on, 
			we go back to the general case of the restricted three-body problem.
		If we consider $\eps$ as a small parameter,
			the Hamiltonian function given in the heliocentric reference frame, Eq.~\eqref{eq:Helio_Ham},
			can be split in two terms, 
			namely, $\cH = \cH_\rK + \cH_\rP$ such that
		\be
			\begin{aligned}
			\cH_\rK(\br, \brd)  	&= \frac{1}{2}\norm{\brd}^2 - \frac{1}{\norm{\br}},\\
			\cH_\rP(\br, \lam') 	&= -\frac{\eps}{\norm{\br- \br'(\lam')}} 
							+ \frac{\eps}{\norm{\br}} 
							+ \eps\frac{\br \bigcdot \br'(\lam')}{\norm{\br'(\lam')}^3}.
			\end{aligned}
\label{eq:HamPert}
		\ee
		$\cH_\rK$ corresponds to the unperturbed Kepler motion of the particle,
		more precisely the motion around a fixed center of mass 1,
		while
		$\cH_\rP$ models the perturbations that depend on $\eps$:
			the gravitational influence of the planet,
			the acceleration of the heliocentric frame,
			and a term associated with our choice of the Kepler problem.

		A closed-solution of $\cH_\rK$ describes an ellipse 
			whose shape, orientation and position at a time $t$ are given by the orbital elements 
		${(a, e, I, \Om,  \om, v(t))}$.
		We recall that the position at a time $t$ can also be described by the mean anomaly $M(t)$, 
			a fictitious angle, linear with respect to the time and 
			whose rate of variation -- generally known as mean motion -- reads ${\dot{M}(a) = 1/\sqrt{a}^3}$ in the units adopted here.
		Instead of using the orbital elements, 
			the Poincar\'e complex variables are adopted in order to preserve the symplectic geometry of the problem.
		In the following,
			the angles
			${\varpi = \Omega + \omega}$ and  
			${\lam = M + \varpi}$
		denote respectively 
			the longitude of the periaster and 
			the mean longitude.
		The symplectic transformation associated with the Poincar\'e variables reads 	
			$$\Upsh: (\br, \brd, \lam', \Xih) \mapsto (\lam, \Lam, \xt, x, \yt, y, \lam', \Xih)$$
			with
			\begin{eqnarray*}\nonumber
			\Lam &=& \sqrt{a}, \\\label{eq:Poinc_transf}
			x &=& \sqrt{\Lam(1-\sqrt{1-e^2})}\exp i\varpi,\\\nonumber
			y &=& \sqrt{\Lam\sqrt{1-e^2}(1-\cos I)}\exp i\Om, 
\end{eqnarray*}
		that are respectively conjugated to $\lam$, $\xt = -i\xb$ and $\yt = -i\yb$. 
		We specify that 
			$x$ and $y$ derive from the angular momentum in the heliocentric reference frame 
			as follows: 
	\begin{eqnarray*}
			\norm{\brL}\circ\Upsh &=& \Lam - \modu{x}^2, \\
			\brL_3\circ\Upsh &=& \Lam - \modu{x}^2 -\modu{y}^2.   
	\end{eqnarray*}
		Moreover, $x\sqrt{2/\Lam}$ and $y\sqrt{8/\Lam}$ are equivalent to $e \exp(i\varpi)$ and $I \exp(i\Om)$
			for quasi-circular and quasi-planar orbits.

\begin{sloppypar}	
		In the extended phase space, 
			the integrable motion is given by the Hamiltonian ${\Xih + \Hh_\rK}$ with 
			${\Hh_\rK(\Lam) = -1/(2\Lam^2)}$.
		Being the planet and the particle coupled but independent, 
			the solutions describe two ellipses
			whose respective mean motions are equal to ${\dot{\lam}(\Lam) = 1/\Lam^3}$ and ${\dot{\lam}' = 1}$.
		In other words, the solutions of the problem correspond to quasi-periodic orbits with two frequencies.
\end{sloppypar}

		Since the frequencies are commensurable they can be periodic. 
		In such a case, the planet and the particle are considered in mean-motion resonance.
		Studying the restricted three-body problem in this perturbative framework consists in 
		understanding how the perturbation ${\Hh_\rP =\cH_\rP\circ\Upsh}$ transforms the unperturbed phase space. 
		More precisely, the analysis can be split in two disjoint situations:
		\begin{itemize}
		\item at a suitable distance to mean-motion resonances
			via a secular model built in order to study the persistence of the quasi-periodic solutions;
		\item or, on the contrary, in a neighborhood of a mean-motion resonance
			with the help of special variables and an adapted averaging process.
		\end{itemize}


		\subsubsection{The resonant variables}

		For $p$ and $q$ two coprime positive integers, 
			let us consider a neighborhood of the $p$:$q$ mean-motion resonance.		
		An unperturbed solution is associated with the $p$:$q$ mean-motion resonance if
			the semi-major axis of the particle is equal to ${\tilde{a} = (q/p)^{2/3}}$.
		In what follows, $\tilde{a}$ will be defined as the semi-major axis of the ‘‘exact" mean-motion resonance.

 		The symplectic transformation 
		$$\Upsc : (\theta, u, \xt, x, \yt, y, \lam', \Xi) \mapsto (\lam, \Lam, \xt, x, \yt, y, \lam', \Xih)$$
		with
	\be
	\begin{aligned}
			\theta 	&= \lam - \frac{p}{q}\lam', \\
			u		&= \Lam - \sqrt{\tilde{a}}, \\
			\Xih 		&= \Xi -  \frac{p}{q}u,
\label{eq:res_transf}
	\end{aligned}
		\ee
		introduces 
			the resonant angle $\theta$ that characterizes the commensurability,
			and $u$, its conjugated action, whose modulus measures the distance to 
			the ‘‘exact" mean-motion resonance.
		We recall that $\theta$ is not a physical angle and thus it is difficult to represent.
		Nevertheless, for quasi-circular and quasi-planar orbits, 
			the angular separation $\phi$ between the particle and the planet 
			is equivalent to ${\theta + (p-q)q^{-1}\lam'}$.

		Notice that the resonant angle $-q(p-q)^{-1}\theta$ 
			is generally used in the literature 
			in order to take advantage of 
			the properties of the leading harmonic of the Hamiltonian
			(see, e.g., \cite{1994Mo,2002Mo}).
		This choice is arbitrary, 
			and does not affect the dynamics of the solutions but changes their representation.
		Likewise, in the framework of a $p$:$q$ retrograde mean-motion resonance
			(sometimes denoted as a $-p$:$q$ mean-motion resonance),
			the considered resonant angle is neither $\theta$  nor proportional to $\theta$
			(see, e.g. \cite{2013MoNa,2020Si}).
		In such a case, that has become an important topic in recent years,
			the particle orbits around the Sun 
			in the opposite direction to the one of the planet, 
			and the canonical variables usually used in order to describe the motion are not the Poincar\'e variables
			introduced by the transformation $\Upsc$.
		However, being the study equivalent to 
			the  one of trajectories in (prograde) mean-motion resonance with inclinations $\modu{I}> \pi/2$,
			the following discussions and results remain valid for a retrograde mean-motion resonance. 
		Only the representation of the dynamics will change. 
	
		In the resonant variables given by Eq.~\eqref{eq:res_transf}, the integrable Hamiltonian reads ${\Xi + H_\rK}$ with 
		\bes
			H_\rK(u) = -\frac{1}{2(\sqrt{\tilde{a}} + u)^2} - \frac{p}{q}u.
		\ees
		$H_\rK$ highlights that  $\theta$ is constant for ${u=0}$,
			while it circulates for ${\modu{u} > 0}$.
		More precisely, the angular variables evolve at different rates: 
			$\lam'$ is a ‘‘fast" angle with a frequency $1$,
			 $\theta$ undergoes ‘‘slow" drift in $\cO(u)$ 
			 while $(\varpi, \Omega)$ are fixed.
		Consequently, for a small enough $\modu{u}$, 
			the timescales of the integrable problem are separated.
		In the full problem that reads ${\Xi + H}$ with
		\bes
			H = H_\rK + H_\rP \qtext{and}
			H_\rP = \Hh_\rP\circ\Upsc,
		\ees
		all the variables might vary and the motion is very tricky to understand.
		However, for $\eps$ and $\modu{u}$ small enough, 
			the timescales separation between the ‘‘fast" and ‘‘slow" degrees of freedom still remains.
		A classical way to exploit this feature is to replace the original problem
			by another one in which the fast oscillations have been removed.
		For that purpose, an averaging over the period of revolution of the planet is performed.
		This process defines the averaged problem.
								 
			
		\subsubsection{The averaged problem}
\label{sec:AP}

		The averaged Hamiltonian reads
		${\Hb = H_\rK  + \Hb_\rP}$ 
		with
		\be
		\begin{aligned}
			&\Hb_\rP(\theta, u, \xt, x, \yt, y) \\
			&\phantom{\Hb_\rP}= \frac{1}{2\pi}\int_0^{2\pi} H_\rP(\theta, u,  \xt, x, \yt, y, \lam')\rd \lam'.
\label{eq:Pert_Moy}
		\end{aligned}
		\ee
		Since $\Hb$ does not depend on $\lam'$, $\Xi$ is a first integral that can be dropped.
		Hence, only three degrees of freedom are required in order to explore the averaged phase space:  
			the resonant variables $(\theta, u)$, 
			and $(\xt,x)$, $(\yt, y)$, that are
			respectively, the Poincar\'e variables associated with eccentricity and inclination.
	
		There exist at least two classical techniques of computation of the averaged problem. 
		The analytical one is	based on the expansion of the Hamiltonian 
				in power series of eccentricity and inclination (see, e.g., \cite{2013RoPo}). 
		In spite of its efficiency for quasi-circular and quasi-coplanar orbits, 
			reaching higher values of eccentricity or inclination requires high order expansions
			which generate very heavy expressions.
		Also worth mentioning the asymmetric expansion developed by
			Ferraz-Mello and Sato \cite{1989FeSa} 
			in order to deal with highly eccentric trajectories in mean-motion resonance.
		The other technique consists on a numerical evaluation of the integral of Eq.~\eqref{eq:Pert_Moy} and its derivatives.
		It is a powerful tool since it deals with the Hamiltonian in its exact form
			which allows to explore the phase space for all values of eccentricity lower than one
			and all values of inclination.
	
\begin{sloppypar}	
		Following the idea of Poincar\'e \cite{1902Po}, 
		Schubart \cite{1964Sc,1968Sc,1978Sc} developed a numerical averaging procedure
		for the Hamiltonian in canonical resonant variables.
		Moons \cite{1994Mo} extended the method of Schubart 
				and provided an algorithm 
				that allows to compute the equations of motion
				of the averaged problem, and thus to construct an integrator 
				for trajectories in $p$:$q$ mean-motion resonance, for $p\neq q$.
		This algorithm has been adapted by Nesvorn\'{y} et al. \cite{2002NeThFe}
			in order to deal with the $1$:$1$ mean-motion resonance.
		In either case, the numerical averaging can be implemented as follows.		
		Let $f$ an auxiliary function that depends on $(\theta, u , \xt, x, \yt, y, E, E')$
			where $E$ and $E'$ are respectively the eccentric anomaly of the particle and the planet.
		The averaging of $f$ over $\lam'$ being calculated for fixed $(\theta, u, \xt,x,\yt,y)$,
			the Kepler equation implies that: ${\rd \lam' = L(u,x,E)\rd E}$ with ${L(u,x,E)= qp^{-1}(1-e(u,x)\cos E)}$.
		Moreover, since 
			\bes 
				\theta = E - e(u, x)\sin E + \varpi(x) - pq^{-1}(E' - e'\sin E')
			\ees
			then $E'$ can be expressed in terms of $(\theta, u, x, E)$ and $e'$.
		Finally, the averaging of $2\pi f$, that reads
		\bes
		\begin{aligned}
			&2\pi \overline{f}(\theta, u, \xt,x,\yt,y)\\
			  &=\int^{2\pi}_0
				f(\theta, u , \xt, x, \yt, y, E, E'(\theta, u,x,E))L(u,x,E)\rd E,
		\end{aligned}
		\ees	
		is computed by discretizing the variable $E$ as ${E_k= k2\pi/N}$ with ${100 \leq N \leq 300}$ 
		(see \cite{1964Sc} for more details). 
\end{sloppypar}

		In the averaged problem, the phase space to explore is 6-dimensional.
		However, and similarly to the classical approach, 
			the dimension can be reduced in the framework of the circular case ($e'=0$).


\begin{table*}[h!]	
\caption{Comparison of the features of the restricted three-body problem in the averaged phase space
		and in the synodic reference frame.
		$\sC$, $\Hb$ and $K$  denotes respectively the Jacobi constant, the averaged Hamiltonian
		and the quantity conserved in the averaged problem in the circular case.
		The notations ‘‘($N$ d.o.f)" and ‘‘(Non-aut.)" stand respectively for ‘‘$N$ degrees of freedom" and ‘‘non-autonomous".}
		\label{tab:1bis}
\begin{center}
\begin{tabular}{ c l l l l }
\hline\noalign{\smallskip}
			& {\bf General (3 d.o.f.)} 	& {\bf Circular (3 d.o.f.)} 	& {\bf Planar (2 d.o.f.)} & {\bf Circular-Planar (2 d.o.f.)}\\
			\noalign{\smallskip}\hline\noalign{\smallskip}
{\bf Synodic RF} 	& (Non-aut.)  			& $\sC$		& (Non-aut.)& $\sC$\\
\noalign{\smallskip}\hline\noalign{\smallskip}
 {\bf Averaged Pb} 	& $\Hb$	& $\Hb$, $K$ 	&  $\Hb$ & $\Hb$, $K$ 	\\ 
 \noalign{\smallskip}\hline 
\end{tabular}
\end{center}
\label{tab:1}
\end{table*}


		\subsubsection{Reduction in the circular case ($e'=0$)}

		First of all, 
			we recall that the perturbation $\cH_P$ is analytical outside the collision manifold 
			and thus can be expanded
				in power series of eccentricity and inclination.
		In the Poincar\'e complex variables, the expansion reads
		\bes
			\sum_{(l,\lt,m,\mt, k,k')\in\sD} f_{l,\lt,m,\mt}^{k,k'}(u)x^l\xt^\lt y^m \yt^\mt  e^{ i(k\lam + k'\lam') }
		\ees
		where the integers occurring in these summations satisfy the relations
		$$
			\sD = \left\{\quad\begin{aligned}
				&(l,\lt, m, \mt, k, k')\in \NN^4\times\ZZ^2\\
				&\text{s.t.}\quad m - \mt =2j, \quad j\in\ZZ,\\
				&\text{and}\quad l + m -(\lt + \mt) + k + k' = 0
			\end{aligned}\quad\right\}
		$$
		known as D'Alembert rules.
		These relations are the result of the invariance of the Hamiltonian $\cH$ 
			under the action of symmetry groups:
				the orthogonal symmetry with respect to the orbital plane of the planet,
				and the group of rotations SO(2) around the vertical axis. 
		 In the resonant variables, since the expansion of $H_\rP$ reads 	
		\bes
			\sum_{(l,\lt,m,\mt, k,k')\in\sD} f_{l,\lt,m,\mt}^{k,k'}(u)x^l\xt^\lt y^m \yt^\mt  e^{i\left(k\theta + (pk+qk')q^{-1}\lam'\right)},
		\ees
		the integers occuring in the expansion of the integral of Eq.~\eqref{eq:Pert_Moy} satisfy the relations
		$$
			\sDb = \sD\cap
				\left\{\quad \begin{aligned}
				&(l,\lt, m, \mt, k, k')\in \NN^4\times\ZZ^2 \\
				&\text{s.t.}\quad kp + k'q = 0
				\end{aligned}\quad
				\right\}.
		$$
		In other words, the angular part of the averaged Hamiltonian 
		depends on the linear combination of only two angles: 
		a ‘‘modified" resonant angle ${\theta - (q-p)q^{-1}\varpi}$
			and the argument of periaster ${\omega=\varpi-\Omega}$.
		Since the averaged Hamiltonian is invariant under the rotations of a third angle,	
			the symplectic geometry imposes the following quantity 
		\bes
			K= \modu{x}^2 + \modu{y}^2 + (p-q)q^{-1}u
		\ees
		to be a first integral.	
		We point out that this  property
			can also be derived from the Jacobi constant, Eq.~\eqref{eq:Jacobi}.
		Indeed, for a given $c$, such that ${\sC(\bR, \bRd) = c}$,
		the composition of transformations  ${\Ups_{\rS\rF}\circ\Upsh\circ\Upsc}$ in resonant variables 
			provides the following expression of the Jacobi constant:
		\be
		c =  2\sqrt{\tilde{a}}+ \eps -2(H  +K).
\label{eq:Jacobi_C}
		\ee		
		Thus, the average of the Jacobi constant over $\lam'$  introduces the averaged Hamiltonian $\Hb$, such as
		 $c =  2\sqrt{\tilde{a}}+ \eps -2(\Hb  +K)$,
		that is conserved in the averaged problem and implies
			that $K$ is  also a first integral of the averaged problem.
	
		Without revealing details on the symplectic transformation that takes advantage of $K$, 
			we outline that 
		${q(p-q)^{-1}K}$, $\modu{x}^2$ and $\modu{y}^2$
			respectively conjugated with 
				$\theta$, ${-\varpi + q(q-p)^{-1}\theta}$ and ${-\Omega + q(q-p)^{-1}\theta}$,
			are action-angle variables that can be used.
		Since these previous variables are singular for the $1$:$1$ mean-motion resonance,
			the canonical variables  that can be adopted  are 
				$u$, $K$ and $\modu{y}^2$, respectively conjugated to $\theta$, $-\varpi$ and $\omega$.
		
		Being the degree of freedom associated with $K$ separable to the other two, 
			a reduction is possible.
		By fixing a value $K$, seen as a parameter, and eliminating its conjugated cyclic angle,
			one degree of freedom is removed.
		Consequently, the averaged phase space can be described by a 1-parameter family of reduced Hamiltonians 
		 	with two degrees of freedom.
		We point out that in the circular-planar case, 
			the number of degrees can be reduced to one.
		Hence, for a fixed $K$, 
			the ‘‘reduced" averaged Hamiltonian is integrable
			and the description of the phase portrait obtained for various values of $K$
			allows to understand the global dynamics of the mean-motion resonance.

		\subsubsection{Some conclusions about the averaged problem}

		In Table \ref{tab:1}, we summarize the respective features of the averaged problem
			with respect to the classical approach in the synodic reference frame. 
		First of all, the averaged problem has the advantage to describe the solutions
			in terms of orbital elements (or variables close to these ones),
			and thus profits of the symplectic geometry of the problem
			which allows to reduce by one unit the dimension of the phase space to explore in any case.
		The algorithms of Moons \cite{1994Mo} and Nesvorn\'{y} et al. \cite{2002NeThFe} 
		are easy to implement to this end.
		Furthermore, it gives a complete understanding of the resonant dynamics in the circular-planar case.		

		However, the averaged problem possesses also some important drawbacks.
		First of all, since $H$ has been replaced by $\Hb$ in order to remove the fast oscillations,
			it does not correspond to the original problem but approximates it 
			with an accuracy that depends on the size of $\eps$.
		Besides, according to the remark of Schubart \cite{1964Sc}, 
			it has been shown by Robutel and Pousse \cite{2013RoPo} and Pousse et al. \cite{2017PoRoVi},
			that the averaged problem fails to describe trajectories
			that feature close encounters with the planet.
		In such a case, the ‘‘distance" between the averaged Hamiltonian and the original one is important,
			and the results given by the averaged problem may not be reliable.
		The clarification of the accuracy as well as the limit of validity is a serious issue.
		We devote the next section to that purpose. 
		

	\section{On the validity of the averaged problem}	
\label{sec:validity}

		\subsection{Notations}

		Before going further, let us introduce some useful notations associated with the Hamiltonian formalism.
		
		\begin{sloppypar}
		In the following, since it will be necessary to switch from the resonant variables to the heliocentric coordinates,
			we denote	 $$\Ups:(\theta,u,\xt, x, \yt, y, \lam', \Xi) \mapsto (\br, \brd, \lam', \Xih)$$
			the composition of transformations  ${\Upsh\circ\Upsc}$.
 		These two sets of variables preserve the symplectic form, that is,
		\begin{eqnarray*}
				\sum_{i=1}^3			\rd \br_i &\wedge& \rd \brd_i 
							+ 	\rd \lam'\wedge \rd \Xih\\
			&=&					\rd \theta \wedge \rd u 
							+ 	\rd \xt \wedge \rd x 
							+ 	\rd \yt \wedge \rd y 
							+ 	\rd \lam'\wedge \rd \Xi.
				\end{eqnarray*}
		Hence, the Lie derivative of an auxilliary function ${\cG(\br, \brd, \lam', \Xih)}$
			along the Hamiltonian flow of a given function ${\cF(\br, \brd, \lam', \Xih)}$
			reads:
			\bes
			\begin{aligned}
				\cL_\cF \cG 
					&	= \dron{\cG}{\br} \bigcdot \dron{\cF}{\brd} 		- \dron{\cF}{\br} \bigcdot \dron{g}{\brd} 
						+ \dron{\cG}{\lam'}\dron{\cF}{\Xih}			-\dron{\cF}{\lam'}\dron{\cG}{\Xih}\\
						&= 	\dron{g}{\theta}\dron{f}{u}	 	- \dron{f}{\theta}\dron{g}{u} 
						+ 	\dron{g}{\xt}\dron{f}{x} 	 	- \dron{f}{\xt}\dron{g}{x} 
						+ 	\dron{g}{\yt}\dron{f}{y} 	 	\\
						&\phantom{=}- \dron{f}{\yt}\dron{g}{y} 
						+ 	\dron{g}{\lam'}\dron{f}{\Xi}		-\dron{f}{\lam'}\dron{g}{\Xi}\\
						&= \cL_f g
			\end{aligned}
			\ees
			with ${f = \cF\circ\Ups}$ and ${g = \cG\circ\Ups}$.
		Finally, $\Phi_t^h(\bX_0)$ denotes the Hamiltonian flow at a time $t$, 
		generated by an auxiliary function $h(\bX)$ 
		that crosses $\bX_0$ at $t=0$.		
		\end{sloppypar}

				
		\subsection{The averaging process}

		According to the perturbation theory, 
			the averaging process coincides with the existence of a symplectic transformation $\Upsb$,
			close to  the identity,
			which maps the original Hamiltonian ${\Xi + H}$ to
			${\Xi 
					+ \Hb
					+ H_*}$,
		where $H_*$ is a remainder that is supposed to be small with respect to $\Hb_\rP$ 
		and thus neglected in the averaged problem.		
		$\Upsb$ is computed with 
			the time-one map of the Hamiltonian flow generated by some auxiliary function $S$, 
			that is, ${\Upsb = \Phi_1^S = \exp \cL_{S}}$,
			which satisfies
		\bes
			\cL_S\Xi = -\dron{S}{\lam'} =   \Hb_\rP - H_\rP.
		\ees
		In this paper, we choose 
		\bes
		\begin{aligned}
			&S(\theta, u, \xt, x,\yt,y,\lam') \\
			&\phantom{S }= \frac{1}{2\pi} \int_0^{2\pi} s (H_\rP - \Hb_\rP)_{(\theta, u, \xt,x,\yt,y,\lam' + s)} \rd s.
		\end{aligned}
		\ees	
		Based on the previous assumptions, the remainder of the averaging process reads
		\be
		H_* =  	 (\exp\cL_S - \Iden)H +  (\exp\cL_S - \cL_S - \rI\rd)\Xi.
\label{eq:remainder}
		\ee
		$H_*$ can be neglected if and only if 
			it is a perturbation of higher order with respect to $\Hb_P$.
		However, since $H_*$ depends on the derivatives of $H_\rP$ and $S$ 
			that increase as long as the planet and the particle are getting closer,
			then $\modu{H_*}$ and $\modu{\Hb_\rP}$ can increase simultaneously 
			according to the distance to the singularity 
			and can be at least of the same order.
		In such a case,
			the hierarchy between the perturbative terms is not ensured and
			the approximation provided by the averaged Hamiltonian $\Hb$ 
			might not reflect properly the dynamics of the restricted three-body problem.
		In other words, in the neighborhood of the collision manifold can exist an ‘‘exclusion zone"
		inside which the solutions of the restricted three-body problem 
		fall outside the scope of the averaged Hamiltonian.
		
		The following section is devoted to the characterization of this exclusion zone 
			through a quantitative treatment of the averaging process.
	
	
		\subsection{Quantitative treatment of the averaging process}		
\label{sec:quantitative}

		We first introduce a domain and a norm on the extended phase space 
			that will allow us to compute quantitative estimates.

		For given $\rho>0$, $\sigma>0$, $\Delta>0$, $\Deltat>0$, small enough,
			and a given $\kappa>0$, independent of the previous quantities,
			we define the following domain of the extended phase space:
		\bes
		\begin{aligned} 
			&\fD_\kappa=\\ 
				&\left\{\,
				\begin{aligned}
			&(\theta, u, \xt, x, \yt, y, \lam', \Xi) 
						\in \TT
						\times\RR
						\times\CC^4
						\times \TT 
						\times\RR\\
					&\mbox{s.t.}\quad \modu{u}\leq \kappa\rho, \quad
					\max(\modu{x}, \modu{\xt}) \leq \tilde{r}_{\sigma/\kappa},\\
					&\min_{\lam'\in\TT}\left(\norm{\br\circ\Ups - \br'}\right)\geq \Delta/\kappa,
					\quad\min_{\lam'\in\TT}\norm{\br\circ\Ups} \geq \Deltat/\kappa
				\end{aligned}\,\right\}
		\end{aligned}
		\ees
		with ${\tilde{r}_{\sigma/\kappa}= \tilde{a}^{1/4}(1-\sigma/\kappa)}$
		and being $\tilde{a}$ the resonant semi-major axis $\tilde{a}= (q/p)^{2/3}$.
		Hence, we consider a neighborhood of the $p$:$q$ mean-motion resonance
		which very excludes high eccentricities
		(${\modu{x} \simeq \tilde{a}^{1/4}}$)
		as well as sets of elements associated with the crossing of the
			spheres of radius  $\Delta/\kappa$ and $\Deltat/\kappa$, 
			respectively centered on the planet and the Sun in the heliocentric reference frame.	

		In this development, we are not interested in the situations of close encounters with the Sun, 
			that occur for a very high eccentricity.
		They are avoided by considering $\Deltat$ and $\sigma$
				as arbitrarily fixed small numbers independent of $\eps$, $\rho$ and $\Delta$.

		The estimates will be computed through the supremum norm on $\fD_\kappa$, 
			denoted  $\norm{\,\cdot\,}_{\kappa}$ such that
		\bes
			\norm{\bpf}_{\kappa}= \max_{i\leq n } \sup_{\fD_\kappa} \modu{\bpf_i}
		\ees
 		where $\bpf=(\bpf_i)_{i\leq n}$ is a $n$-dimensional vector field 
			that depends on the resonant variables ${(\theta, u, \xt, x, \yt, y, \lam', \Xi)}$.
		Since, we do not attempt to obtain estimates with particularly sharp constants,
			all constants have been suppressed and replaced by the P\"oschel's notation, that is,
		\bes
			x \leqp y, \quad x \pleq y, \qtext{and} x \eqp y 
		\ees
		to indicate respectively that
		\bes
			x < Cy, \quad Cx <y, \qtext{and} x = Cy
		\ees
		with some constant $C\geq 1$ independent of $\eps$, $\rho$ and $\Delta$.

		In this setting,  the size of the functions involved in the averaging process can be estimated.
		Hence, we state the following:
\begin{lemma}
\label{lem:HKHPS}
		\begin{sloppypar}
		For ${\rho>0}$, ${\Delta>0}$ and ${\eps>0}$, small enough quantities, that is
		\bes
			\rho \pleq 1, \quad \Delta \pleq 1, \quad \eps \pleq 1,
		\ees
		the Hamiltonian of the restricted three-body problem ${\Xi +H}$,
		the averaged Hamiltonian ${\Xi  + \Hb}$, 
		and
		the symplectic transformation $\Ups$,
		are analytic on the collisionless domain $\fD_2$.
\end{sloppypar}

\begin{sloppypar}		
		Consequently,
			$H_\rK$, $H_\rP$, $\Hb_\rP$ and $\Ups$ are bounded
			together with their partial derivatives 
			with respect to $\theta$, $u$, $\xt$, $x$, $\yt$ and $y$.
		More precisely, for ${n\geq 1}$ and ${(W_i)_{i\leq n}\in\{\theta, u, \xt, x, \yt, y\}}$,
			the following thresholds are satisfied on the smaller domain $\fD_{3/2}$:
		\bes
		\begin{aligned}
		&\begin{array}{lll}
			\norm{H_\rK}_{3/2} \eqp 1,&\quad
			\norm{H_\rP}_{3/2} \leqp \frac{\eps}{\Delta},&\quad
			\norm{\Hb_\rP}_{3/2} \leqp \frac{\eps}{\Delta},\\
			\norm{H_\rK'}_{3/2} \leqp \rho,&\quad
			\norm{H_\rK''}_{3/2} \leqp \rho,&\quad
			\norm{H_\rK'''}_{3/2} \leqp 1,\\
		\end{array}\\
			&\norm{\frac{\partial^n H_\rP}{\partial W_1 \ldots \partial W_n}}_{3/2} \leqp \frac{\eps}{\Delta^{n+1}},\quad
			\norm{\frac{\partial^n(\br\circ\Ups)}{\partial{W_1}\ldots\partial{W_n}}}_{3/2} \leqp 1.			
		\end{aligned}
		\ees
		and
		\be
			\rho^2 \leqp \norm{H_{\rK} - H_{\rK}(0)}_{3/2} \leqp \rho^2
\label{eq:Lem1_HK_born}
		\ee

\end{sloppypar}
\end{lemma}
		
		The previous lemma allows to state an averaging theorem where quantitative estimates
			on the averaging process are computed.
\begin{theorem}
		Assuming $\eps$, $\rho$ and $\Delta$ small enough such that		
		\be
			\eps\pleq \rho\Delta^2 \qtext{and} \eps \pleq \Delta^3,
\label{eq:theo1_cond}
		\ee 
		there exists a symplectic transformation close to the identity, denoted as
		$$
			\Upsb: \left\{
			\begin{array}{ccc}
				{\fD_{4/3}} 	&\rightarrow &\fD_{3/2}\\
				(\und{\theta}, \und{u}, \und{\xt}, \und{x}, \und{\yt}, \und{y}, \lam',\und{\Xi}) 
					&\mapsto &
				(\theta, u, \xt, x, \yt, y, \lam', \Xi)	,
			\end{array}
				\right.$$		
		with
		\bes
		\begin{gathered}
				\norm{\und{\Xi} - \Xi}_{{4/3}} \leqp \frac{\eps}{\Delta},\quad
				\norm{\und{W} - W}_{{4/3}} \leqp \frac{\eps}{\Delta^2}\\
		\end{gathered}
		\ees
		for $W \in\{\theta, u, \xt, x, \yt, y\}$,	
		such that, in the ‘‘averaged" resonant variables	
			${(\und{\theta}, \und{u}, \und{\xt}, \und{x}, \und{\yt}, \und{y}, \lam',\und{\Xi})}$, 
			the Hamiltonian reads:		
		\bes
			(\Xi + H)\circ\Upsb
			 = \und{\Xi} + \Hb + H_*
		\ees
		where  $H_*$ is the remainder of the averaging process.
	
		Furthermore, on the domain $\fD_{3/2}$, 
			$H_*$  
			together with its partial derivatives with respect to $\theta$, $u$, $\xt$, $x$, $\yt$ and $y$ 
			are bounded  and satisfy the following thresholds:
		\bes
		\begin{aligned}
			&\norm{H_*}_{4/3} 	\leq \frac{\eps}{\Delta}\eta \quad
				\qtext{with} \eta \eqp \left( \frac{\eps}{\Delta^3} + \frac{\rho}{\Delta}\right),\\
			&\norm{\frac{\partial H_*}{\partial W}}_{4/3} \leqp\frac{\eps}{\Delta^2}( \eta + \rho) 
		\quad 
			\text{for $W\in\{\theta, u, \xt, x, \yt, y\}$.}
		\end{aligned}
		\ees
\label{theo:averaging}
\end{theorem}

		Lemma \ref{lem:HKHPS} and Theorem \ref{theo:averaging} 
		provide the estimates that allow
			to compare how $\modu{\Hb_\rP}$ and $\modu{H_*}$ increase
		as long as the planet and the particle are getting closer.
		In order to clarify our reasonings,
		we first relate the upper bound of the distance to the resonance to $\eps$ and $\Delta$
		by choosing 
		\be 
			\rho \eqp \sqrt{\frac{\eps}{\Delta}}
\label{eq:rho}
		\ee such that the two terms in $\eta$ depend on the same quantity.
		Hence, we have 
		\be
			\eta \eqp \sqrt{\frac{\eps}{\Delta^3}}\left(1 + \sqrt{\frac{\eps}{\Delta^3}}\right)
\label{eq:eta}
		\ee
		which imposes the lower bound ${\Delta\geqp  \eps^{1/3} }$ 
		in order to get decreasing perturbations in the ‘‘averaged" resonant variables.
		More precisely, we recover the size of the Hill's sphere of the planet.
		We recall that $\Delta$ denotes 
			the minimal mutual distance, 
			that is, the minimal distance between the particle and the planet,
			which is allowed in the considered domain $\fD_\kappa$.		
		As a consequence, if we relate $\Delta$ to $\eps$ as follows:
		\bes
		\begin{gathered}
			\Delta = \rN_\eps R_\rH
			\qtext{with}
			\rN_{\eps} =\eps^{-\alpha},\quad
			0< \alpha \leq 1/3,
		\end{gathered}
		\ees
		being $R_\rH= \left(\frac{\eps}{3}\right)^{1/3}$ the Hill's radius of the planet,
		then Theorem \ref{theo:averaging} ensures that
			the domain $\fD_{4/3}$ stands outside the exclusion zone of the averaged problem 
		for $\eps$ small enough and $\rN_{\eps}>1$, 
		that is, for a minimal mutual distance larger than the Hill's radius of the planet.
	
		In spite of this feature, Theorem \ref{theo:averaging} does not establish that 
			a given solution of the averaged problem that starts inside $\fD_{4/3}$,
			does not escape and does not cross the exclusion zone of the averaged phase space.
		To this end, a careful analyze of the behavior of the averaged solutions has to be led
			for each type of dynamics in mean-motion resonance.
		Nevertheless, assuming that the solution remains inside $\fD_{4/3}$ until a certain amount of time,
			a theorem of stability over finite times
			can be proven in the restricted three-body problem.
			
		Before stating the theorem, let us denote the solution governed by the averaged Hamiltonian ${\Xi + \Hb}$,
			that starts in ${\bX_0\in\fD_{1}}$
			and remains inside this domain up to a given time ${\cT_1>0}$, as
		\bes
		{\und{\bX}(t) = (\und{\bW}(t), \lam'(t), \und{\Xi}(t))}
		\ees
		with
			\bes{\und{\bW}(t)=(
			\und{\theta}(t), 
			\und{u}(t), 
			\und{\xt}(t), 
			\und{x}(t), 
			\und{\yt}(t)) }
			\qtext{and}{\lam'(t) = t}.\ees
		For ${\modu{t}\leq \cT_1}$ and ${\rN_{\eps} >1}$,
			$\und{\bX}(t)$ approximates the solution of the restricted three-body problem that starts in $\bX_0$.
		In the resonant variables, the solution governed by ${\Xi + H}$ can be written as ${(\bW(t), \lam'(t), \Xi(t))}$
		with
		$$
			\begin{aligned}
			\theta(t) 	&= \und{\theta}(t) 	+ \delta_1(t),\\
			\xt(t)		&= \und{\xt}(t)		+ \delta_3(t),\\
			\yt(t)		&= \und{\yt}(t)		+ \delta_5(t),\\
			\end{aligned}\quad
			\begin{aligned}
			u(t)		&= \und{u}(t)		+ \delta_2(t),\\
			x(t)		&= \und{x}(t)		+ \delta_4(t),\\
			y(t)		&= \und{y}(t)		+ \delta_6(t),\\
			\end{aligned}
		$$
		being  ${(\delta_i(t))_{i\leq 6}}$ the functions
		that denote the error in the approximate solution given by the averaged problem.		

\begin{theorem}
\label{theo:stability}
		With the previous notations, the errors in the approximate solution satisfy the following upper bound:
		\bes
			\modu{\delta_i} \leqp \frac{\eps^{1/3}}{\rN_\eps^{2} } 
			\qtext{for $\modu{t}\leq \min\left(\cT, \cT_1\right)$ and $\cT = 2\pi\sqrt{\rN_\eps^{3}}$.}
		\ees
 \end{theorem}		
		In the following reasonings, it is assumed that ${\cT \leq \cT_1}$.
		First of all, in the limit case given by ${\rN_{\eps}= \eps^{-1/3}}$,
		that is, ${\Delta = 1}$ and ${\rho \eqp \sqrt{\eps}}$, 
			Theorem \ref{theo:stability} ensures that, up to a finite time in $\cO(1/\sqrt{\eps})$,
			the approximate solution given by the averaged problem remains inside a neighborhood 
			of order $\cO(\eps)$
			of the solution obtained in the synodic reference frame.
		Hence, we obtain the results
			for which the particle is considered distant enough from the planet.
			
		The other limit case, that is, for  ${\rN_{\eps} \simeq 1}$, 
			the particle can approach the edge of the Hill's sphere
			while the distance to the resonance can reach the order  $\cO(\eps^{1/3})$.
		Even though
			the gravitational influence of the planet is not dominant,
			it can be strong enough with respect to the one of the Sun.
		Hence, we can only ensure that 
			the accuracy of the approximate solution will not exceed a quantity of order  $\cO(\eps^{1/3})$ 
			for one or few periods of revolution of the planet.		
		In such a case, 
			the solution of the averaged problem might not be reliable
			in order to approach the one obtained in the synodic reference frame.
			
		By increasing ${\rN_\eps>1}$, 
			the minimal mutual distance moves away from the Hill's sphere, 
			the gravitational effect of the planet becomes weaker with respect to the one of the Sun,
			and the upper bound on the distance to the mean-motion resonance decreases as 
			${\rho \eqp \frac{\eps^{1/3}}{\sqrt{\rN_{\eps}}}}$.
		Thus, the approximate solution becomes more accurate
		with an upper bound on the error that decreases as ${\rN_{\eps}^{-2}}$ 
			and a time of stability that increases as ${\sqrt{\rN_{\eps}^3}}$.
		Consequently,
			multiplying by a factor $5$ the minimal mutual distance
			divides the error by a factor $25$ and multiplies the time of stability by a factor $5^{3/2}\simeq 10$.		
		In a more practical way, for a given number ${n>0}$ of revolutions of the planet, 
		with ${n \leq \cO(1/\sqrt{\eps})}$,
			this results provides a domain of initial conditions
			such that the approximate solution given by the averaged problem is reliable 
			in order to approximate the one obtained in the synodic reference frame.
		The error in the approximate solution being ${\eps^{1/3}n^{-4/3}}$, 
		the smaller the planet mass ratio is, the greater the accuracy would be.
				
		\subsection{Discussion}
		
	The proofs of Lemma \ref{lem:HKHPS}, Theorem \ref{theo:averaging} and Theorem \ref{theo:stability} 
	are given in the Appendix \ref{sec:appendix}.
	
	The key ingredient of the proof of Lemma \ref{lem:HKHPS} is our definition 
	of the collisionless domains $\fD_\kappa$ 
	given in terms of heliocentric coordinates instead of resonant variables
	in order to exclude a neighborhood of the collision manifolds.
	Since the Poincar\'e complex variables prevent singularities associated with 
	the eccentricity or inclination equal to zero,
		the transformation $\Ups$ is analytic and can be bounded independently to $\eps$, $\rho$ and $\Delta$.
	Hence, the estimates on $H_\rK$, $H_\rP$ and $\Hb_\rP$ are deduced directly from $\cH_\rK$ and $\cH_\rP$.
	
	The proof of Theorem  \ref{theo:averaging} has two parts.
	We first characterize the conditions that allow to ensure that the transformation of averaging $\Ups$
		is close to identity and maps the domain $\fD_{4/3}$ in the domain $\fD_{3/2}$
		inside which the estimates are computed in Lemma \ref{lem:HKHPS}.
	In the second part, we estimate the remainder $H_*$, Eq.~\eqref{eq:remainder},
	and its associated vector field
		by using the Taylor expansions at zero and first order
		combined with the estimates of the Lemma \ref{lem:HKHPS}.
		
	Finally, Theorem \ref{theo:stability} is a direct application of the classical strategy to prove stability over finite times
	(see \cite{1989Ar}).
	For that purpose, we compare the vector field of the approximation given by the averaged problem 
		with the one of a solution of the original problem.
	Assuming that the two solutions remain in a given neighborhood up to a time $\cT$,
	we can choose $\cT$ such that the order on the errors in the approximation is of the same order
	as the one of the transformation in ‘‘averaged" resonant variables.
		
	\smallskip
		
	The validity of the averaged problem was discussed by Robutel and Pousse
	\cite{2013RoPo}, Robutel et al. \cite{2016RoNiPo} and Pousse et al. \cite{2017PoRoVi} in the framework of the 1:1 mean-motion resonance.
	Moreover, it is a key ingredient of the proof given by Niederman et al. \cite{2020NiPoRo} 
	on the existence of the horseshoe-shaped trajectories
	followed by the two Saturn's moons, Janus and Epimetheus.
	Since the semi-major axes of the two small bodies  are almost the same,
	the issue generated by periodical close encounters is manifest.
	For instance, in \cite{2013RoPo} and \cite{2017PoRoVi}, 
		the authors highlight,
		through a frequency analysis,
		that, when an initial condition tends to the singularity of mutual collision, 
		the approximation given by the averaged problem has fundamental frequencies 
		that increase and tend to infinity.
	This phenomenon is inconsistent with the hypothesis of timescales separation
	required  by the averaged problem.
	For that reason, an arbitrary criterion\footnote{
		A solution of the averaged problem
		was considered outside the exclusion zone, 
		if the modulus of their fundamental frequencies are lower than $\dot{\lam}'/4$
		where $\dot{\lam}'=1$ denotes the frequency of averaging.}
		on the frequencies was introduced in \cite{2017PoRoVi} 
		in order to localize the exclusion zone in the averaged phase space.
	
	The work of Robutel et al. \cite{2016RoNiPo} provides a rigorous treatment of the averaging process
		for resonant dynamics in the planar planetary three-body problem.
	In the framework of quasi-circular co-orbital trajectories
	it gives quantitative estimates
		on the remainder generated by the averaging process 
		as well as its vector fields
		in order to state a theorem of stability over finite times.
	The proof uses a complex domain of holomorphy
	inside which estimates are computed through the Cauchy inequality.	
	The results of the present paper are based on the same idea but applied in the general case 
	(eccentric and spatial trajectories)
		of a generic $p$:$q$ mean-motion resonance of the restricted three-body problem.
	Even though the technique of complexifying used in \cite{2016RoNiPo} 
	is very efficient in the case of quasi-circular and quasi-planar orbits,
		the definition of the minimal mutual distance $\Delta$ in terms of resonant variables
		is really difficult in the general case.
	That is why, we have chosen the direct computation of estimates by taking advantage of the form of 
	$H_\rK$, as well as the one of the $\cH_\rP$, that only depends on $\br$ and $\lam'$.


	\section{The co-orbital motion in the circular-planar case}	
\label{sec:coorb}

\begin{sloppypar}
		In this section, we focus on the co-orbital motion (1:1 mean-motion resonance)	in the circular-planar case.
		More precisely, in the framework of the averaged problem, 
			we intend to approach the six families of periodic orbits described in Sect.~\ref{sec:rem_sol}
			(the short-periodic $\sL_j^s$ and long-periodic $\sL_j^l$ for $j=4,5$,
			the Lyapunov family $\sL_3$,
			and the family $f$)	
			as well as the dynamics observed in their neighborhood
			(the tadpole motion, the horseshoe motion, the quasi-satellite motion, and the ‘‘satellized" retrograde satellite motion),
			and identify the limit of validity of the corresponding solutions 
			by applying the Theorem \ref{theo:averaging} and Theorem \ref{theo:stability}
			developed in Sect.~\ref{sec:validity}.
\end{sloppypar}	
\begin{sloppypar}		

		First of all, we introduce the resonant variables and apply the properties stated in Sect.~\ref{sec:pert_ap}
			to the case of the 1:1 mean-motion resonance.
		The resonant degree of freedom is described by 
			the angle  ${\theta = \lam -\lam'}$
			and the action ${u=\sqrt{a} - 1}$ that measures the distance to the exact mean-motion resonance
			given by the semi-major axis $\tilde{a}=1$.
		In the circular-planar case, the secular variations of the orbits are provided by $(\xt, x)$
			for which ${K= \modu{x}^2}$ is a first integral of the averaged problem.
		For a fixed $K$, seen as a parameter, the reduced averaged Hamiltonian, denoted as follows
		\bes
		\Hb^K(\theta,u) = \Hb(\theta, u, \xt(K), x(K)),
		\ees
			is integrable with one degree of freedom, 
			and thus
			allows to understand the co-orbital dynamics through a phase portrait.
		
		Instead of using $K$, we introduce the parameter $e_0$ such as
		\bes
			K = 1 - \sqrt{1- e_0^2}.
		\ees
		Hence, $e_0$ defines the eccentricity of a trajectory 
			that crosses the orbit of the planet, 
			that is, at the exact mean-motion resonance $u=0$.
		Furthermore, since 
		\bes 
			e = \sqrt{1 - \left(1 - \frac{K}{(1+u)}\right)^2} = e_0\left(1 + \cO(u)\right),
		\ees 
			$e_0$ approximates the eccentricity of the trajectories 
			that belong to a given phase portrait.
		Notice that $e_0$ is also connected to the Jacobi constant through Eq.~\eqref{eq:Jacobi_C}.
		More precisely, Theorem \ref{theo:averaging} and Eq.~\eqref{eq:eta}
		imply that inside the domain of validity of the averaged problem, denoted $\fD_1(\Delta)$ in the previous section,
		the Jacobi constant $\sC(\bR, \dot{\bR}) = c$ reads:
		\bes
		\begin{aligned}
			 c 	&= - 2\Hb 	+ 2\sqrt{1- e_0^2} 	+ \cO\left(\sqrt{\frac{\eps^3}{\Delta^5}}\right).
		\end{aligned}
		\ees
		Consequently,  Eq.~\eqref{eq:rho} and the estimates of Lemma \ref{lem:HKHPS}
			ensure the following relation between the two quantities: 
		\bes
		\begin{aligned}
			 c 	&=  1 	+ 2\sqrt{1- e_0^2} 	+ \cO\left(\frac{\eps}{\Delta}\right)
		\end{aligned}
		\ees

\end{sloppypar}
	
		Before going further,
			we will see in the next section 
			how a trajectory of a given phase portrait is related to the solutions of the averaged problem.
		Besides, in order to bridge a gap between the classical and the perturbative approaches,
			we will detail how a solution of the averaged problem describes the motion of a particle 
			in the synodic reference frame.
		These correspondences were described in \cite{2017PoRoVi},
			but the relationship with the synodic reference frame needed to understand the dynamics were lacking.
		
		
		\subsection{Reading a phase portrait}
\label{sec:reading}
\begin{sloppypar}		
 	
		For a given value of ${e_0\geq 0}$ and a given initial condition ${(\theta_i,u_i)}$, 
			a trajectory that belongs to the corresponding phase portrait 
			is generally a periodic solution but can also be an equilibrium of $\Hb^{K(e_0)}$.
		If we denote  its frequency $\nu$,
			a periodic trajectory can be written as
		\bes
		\begin{aligned}
			\theta(t) 	&= \theta_i + F_1(\nu t),\\
			u(t) 		&= u_i +  F_2(\nu t), 
		\end{aligned}
		\ees
		where the functions $F_j$ are $2\pi$-periodic such that ${F_j(0) = 0}$.			

		The dynamics of the angle ${\varpi=\arg(x)}$ is required 
			in order to relate this trajectory to the solutions of the averaged problem.	
		Since
		\bes
			\dot{\varpi}(t) = -\dron{}{K}\Hb^{K(e_0)}(\theta(t),u(t))
		\ees	
		is $2\pi/\nu$-periodic, 
		there exists $F_3$, a $2\pi$-periodic function  with mean zero and ${F_3(0)=0}$,
			such that for all ${\varpi_i\in\TT}$,
		\bes
			\varpi(t) = \varpi_i + gt + F_3(\nu t),
		\ees
		where
		\bes
			g = -\frac{\nu}{2\pi}\int_0^{2\pi} \dron{}{K}\Hb^{K(e_0)}(\theta(t), u(t))\rd t
		\ees
		is the secular precession frequency of $\varpi$.			
		In other words,  a solution of the averaged problem that starts in ${(\theta_i, u_i, \xt_i, x_i)}$ 
		with 
		\bes 
			e_0 = \sqrt{1 - \left(1 - \modu{x_i}^2\right)^2}, \quad \varpi_i = \arg(x_i)
		\ees 
		can generally be written as
		\be
		\begin{aligned}
			\theta(t) 	&= \theta_i + F_1(\nu t),\\
			u(t) 		&= u_i +  F_2(\nu t), \\
		\end{aligned}\quad
		\begin{aligned}
			x(t)       	&=  x_i\exp i(gt + F_3(\nu t)),\\
			\xt(t) 		& = \xt_i/ \exp i(gt + F_3(\nu t)),
		\end{aligned}
\label{eq:Sol_AP}
		\ee
		and a periodic trajectory of a given phase portrait generally corresponds		 
			to a set of quasi-periodic solutions
				parametrized by $\varpi_i\in\TT$,
				whose fundamental frequencies are given by $\nu$ and $g$.
		The same reasoning applies for an equilibrium of $\Hb^{K(e_0)}$:
			it corresponds to a set of periodic solutions of the averaged problem,
			parametrized by $\varpi_i\in\TT$,
			and that can be written as
			\be
		\begin{aligned}
			\theta(t) 	&= \theta_i,\\
			u(t) 		&= u_i, \\
		\end{aligned}\qquad
		\begin{aligned}
			x(t)       	&=  x_i\exp i(gt),\\
			\xt(t) 		& = \xt_i/ \exp i(gt).
		\end{aligned}
\label{eq:SolFP_AP}
		\ee
		Nevertheless, $\varpi$ being ignorable when the osculating ellipses are circles ($e_0= 0$),
		 	the solutions have the same features in the averaged problem as in the phase portrait of $\Hb^{K=0}$.
\end{sloppypar}

		Theorem \ref{theo:stability} ensures that a given solution of the averaged problem 
			that lies outside the Hill's sphere of the planet,
			approximates for a finite time the motion of a particle 
			that starts at the same initial condition.			
		Hence, in the Poincar\'e complex variables, 
			the motion of a particle
			that crosses
		$(\lam_i, \Lam_i, \xt_i, x_i)$ with $\lam_i = \theta_i$ and $\Lam_i = 1+ u_i$ at $t=0$,
			can be approximated for a finite time by Eq.~\eqref{eq:Sol_AP} or  Eq.~\eqref{eq:SolFP_AP} such that
		\bes
			(\lam(t), \Lam(t), \xt(t), x(t)) 
		\ees
		with 
		\bes
			\lam(t) = t + \theta(t), \quad \Lam(t) = 1 + u(t).
		\ees
		In terms of orbital elements, the approximation of the variations reads
		\bes
		\begin{array}{llllllllll}
			a(t) 		&=&1 + 2u_i			& 	&		&+& 2F_1(\nu t) + \cO_2(u(t)),\\
			e(t)		&=& e_0   			&	&		&+&e_0\cO(u(t)),		 \\
			\varpi(t) 	&=& \varpi_i 			&+ 	&gt		&+ &F_3(\nu t)\\
			M(t) 		&=&   \theta_i - \varpi_i 	&+ 	&(1-g)t	&+&[F_1-F_3](\nu t)
		\end{array}
\label{eq:Sol_EO}
		\ees		
		Hence, the semi-major axis and the eccentricity experience a slow oscillation of frequency $\nu$
		and of amplitude of the order $\cO(u)$,
			respectively about 1 and $e_0$.
		The variations of the longitude of the periaster correspond to the composition of a secular drift of frequency $g$
			with an oscillation of frequency $\nu$.
		Finally, the motion of the mean longitude is given by a fast drift of frequency $1-g$ 
		composed with a slow oscillation of frequency $\nu$. 

\begin{sloppypar}
		We recall that  the orbital elements are related to the polar coordinates $(\phi,R) = (\arg(\bR), \norm{\bR})$ 
			of the synodic reference frame
			 as follows:
\begin{eqnarray}\nonumber
			\phi &=& \theta + v-M,\\\label{eq:Prop_EO}
			 \quad R &=&   a(1 -e\cos E)\\\nonumber
				v &=& M + G_1(e, M), \quad E =M + G_2(e,M).
\end{eqnarray}
		The functions $G_j$,  
			that satisfy $G_j(e, M) =\cO(e)$,
			derive from the Kepler equation ${M = E-e\sin E}$ and the difference between the true 
			and eccentric anomaly ${\tan (v/2) = \sqrt{\frac{1+e}{1-e}}\tan (E/2)}$.
\end{sloppypar}
	
\begin{sloppypar}		
		In the synodic reference frame, the approximate motion of the particle  can be written as
		\bes
		\begin{array}{lllllllll}
			\phi(t)	&=&  \theta_i& + &G_1(e_0, M(t))  &  +& F_1(\nu t)  &+& \cO_2(u(t), e_0)	\\		
			R(t) 		&=&	1 + 2u_i& - & e_0\cos(M(t)) &+& 2F_2(\nu t) &+&\cO_2(u(t), e_0)		
			\end{array}
		\ees
		with 	${M(t) =   \theta_i - \varpi_i + (1-g)t + [F_1-F_3](\nu t)}$.
		As a consequence, a periodic trajectory of a given phase portrait generally provides 
			a quasi-periodic approximation
			of the motion whose fundamental frequencies are $1-g$ and $\nu$.
		More precisely, the motion is characterized by $R$ 
			that oscillates about  1 with an amplitude of the order ${\cO(e_0)+ \cO(u(t))}$,
			while $\phi$ is the sum of the periodic oscillation generated by $\theta(t)$
			with the quasi-periodic oscillations generated by $G_1$
				whose amplitude is of the order $\cO(e_0)$.
		The same reasoning applies for an equilibrium of $\Hb^{K(e_0)}$:
			it provides a set of periodic trajectories of frequency $1-g$ whose motion follows:
		\bes
		\begin{array}{lllllll}
			\phi(t)	&=&  \theta_i& 	+ &	G_1\left(e_0,  M(t))\right) 
						&+& \cO_2(u_i, e_0),	\\		
			R(t) 		& =& 1 + 2u_i&		 - &	e_0\cos(M(t)) 
						&+&\cO_2(u_i, e_0),
		\end{array}
		\ees
		with ${M(t) = \theta_i - \varpi_i + (1 - g)t}$.
		Hence, the motion is characterized by an oscillation of frequency $1-g$ 
		around a guiding center located in ${(\phi, R) = \left(\theta_i, 1 + \cO(u_i)\right)}$ 
		and whose amplitude is of the order $\cO(e_0)$.
		Finally, for ${e_0=0}$, the motion of the particle is approximated by
		\be
		\begin{array}{lllllll}
			\phi(t)	&=&  \theta_i &+ &F_1(\nu t),\\
			R(t)		&=& 1 + 2 u_i& + &2F_2(\nu t) &+& \cO_2(u(t)),
		\end{array}
\label{eq:approx_0}
		\ee
		and a trajectory of the phase portrait provides an equilibrium  
		in the synodic reference frame
		or a periodic trajectory of frequency $\nu$ 
		characterized by $R$ that oscillates around 1 with an amplitude of the order $\cO(u)$,
			that is, much smaller than 1. 
\end{sloppypar}
	

				\subsection{Phase portraits for a Sun-Jupiter like system ($\eps = 1/1000$)}
\begin{figure*}
	\begin{center}
		\tiny
		\def\svgwidth{0.75\textwidth}
\begingroup%
  \makeatletter%
  \providecommand\color[2][]{%
    \errmessage{(Inkscape) Color is used for the text in Inkscape, but the package 'color.sty' is not loaded}%
    \renewcommand\color[2][]{}%
  }%
  \providecommand\transparent[1]{%
    \errmessage{(Inkscape) Transparency is used (non-zero) for the text in Inkscape, but the package 'transparent.sty' is not loaded}%
    \renewcommand\transparent[1]{}%
  }%
  \providecommand\rotatebox[2]{#2}%
  \newcommand*\fsize{\dimexpr\f@size pt\relax}%
  \newcommand*\lineheight[1]{\fontsize{\fsize}{#1\fsize}\selectfont}%
  \ifx\svgwidth\undefined%
    \setlength{\unitlength}{598.885849bp}%
    \ifx\svgscale\undefined%
      \relax%
    \else%
      \setlength{\unitlength}{\unitlength * \real{\svgscale}}%
    \fi%
  \else%
    \setlength{\unitlength}{\svgwidth}%
  \fi%
  \global\let\svgwidth\undefined%
  \global\let\svgscale\undefined%
  \makeatother%
  \begin{picture}(1,0.48089302)%
    \lineheight{1}%
    \setlength\tabcolsep{0pt}%
    \put(0,0){\includegraphics[width=\unitlength]{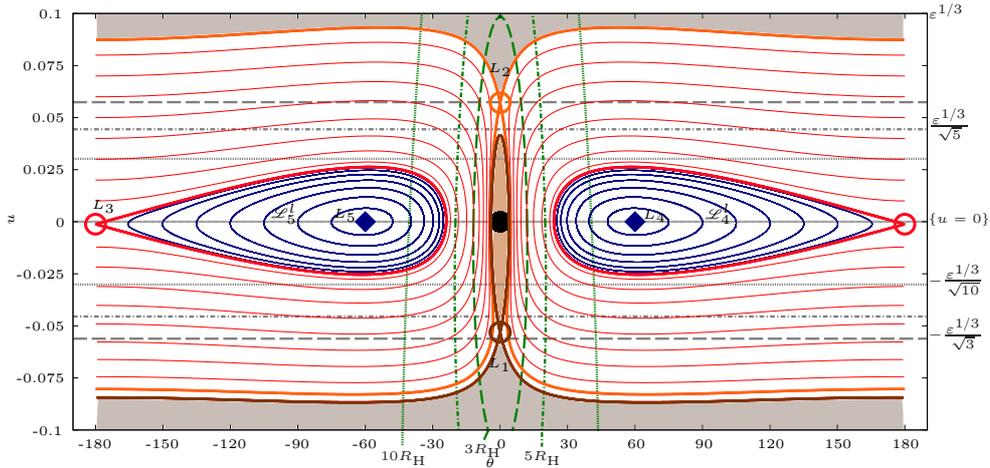}}%
    \put(0.41052725,0.01209075){\makebox(0,0)[t]{\lineheight{1.25}\smash{\begin{tabular}[t]{c}$10R_\rH$\end{tabular}}}}%
    \put(0.55078726,0.01209075){\makebox(0,0)[t]{\lineheight{1.25}\smash{\begin{tabular}[t]{c}$5R_\rH$\end{tabular}}}}%
    \put(0.49067584,0.01960498){\makebox(0,0)[t]{\lineheight{1.25}\smash{\begin{tabular}[t]{c}$3R_\rH$\end{tabular}}}}%
    \put(0.93970953,0.19118838){\makebox(0,0)[lt]{\lineheight{1.25}\smash{\begin{tabular}[t]{l}$-\frac{\eps^{1/3}}{\sqrt{10}}$\end{tabular}}}}%
    \put(0.93970953,0.46169068){\makebox(0,0)[lt]{\lineheight{1.25}\smash{\begin{tabular}[t]{l}$\eps^{1/3}$\\\end{tabular}}}}%
    \put(0.93970953,0.34898139){\makebox(0,0)[lt]{\lineheight{1.25}\smash{\begin{tabular}[t]{l}$\frac{\eps^{1/3}}{\sqrt{5}}$\end{tabular}}}}%
    \put(0.93970953,0.13608606){\makebox(0,0)[lt]{\lineheight{1.25}\smash{\begin{tabular}[t]{l}$-\frac{\eps^{1/3}}{\sqrt{3}}$\end{tabular}}}}%
    \put(0.93970953,0.25380465){\makebox(0,0)[lt]{\lineheight{1.25}\smash{\begin{tabular}[t]{l}$\{u=0\}$\end{tabular}}}}%
    \put(0.06653355,0.03822074){\color[rgb]{0,0,0}\makebox(0,0)[rt]{\lineheight{1.25}\smash{\begin{tabular}[t]{r}-0.1\end{tabular}}}}%
    \put(0.06865974,0.09115236){\color[rgb]{0,0,0}\makebox(0,0)[rt]{\lineheight{1.25}\smash{\begin{tabular}[t]{r}-0.075\end{tabular}}}}%
    \put(0.06866308,0.14408398){\color[rgb]{0,0,0}\makebox(0,0)[rt]{\lineheight{1.25}\smash{\begin{tabular}[t]{r}-0.05\end{tabular}}}}%
    \put(0.06865974,0.1970156){\color[rgb]{0,0,0}\makebox(0,0)[rt]{\lineheight{1.25}\smash{\begin{tabular}[t]{r}-0.025\end{tabular}}}}%
    \put(0.06880783,0.24994722){\color[rgb]{0,0,0}\makebox(0,0)[rt]{\lineheight{1.25}\smash{\begin{tabular}[t]{r}0\end{tabular}}}}%
    \put(0.06878445,0.30287885){\color[rgb]{0,0,0}\makebox(0,0)[rt]{\lineheight{1.25}\smash{\begin{tabular}[t]{r}0.025\end{tabular}}}}%
    \put(0.06878779,0.35581047){\color[rgb]{0,0,0}\makebox(0,0)[rt]{\lineheight{1.25}\smash{\begin{tabular}[t]{r}0.05\end{tabular}}}}%
    \put(0.06878445,0.4087421){\color[rgb]{0,0,0}\makebox(0,0)[rt]{\lineheight{1.25}\smash{\begin{tabular}[t]{r}0.075\end{tabular}}}}%
    \put(0.06665826,0.46167372){\color[rgb]{0,0,0}\makebox(0,0)[rt]{\lineheight{1.25}\smash{\begin{tabular}[t]{r}0.1\end{tabular}}}}%
    \put(0.09938234,0.02494609){\color[rgb]{0,0,0}\makebox(0,0)[t]{\lineheight{1.25}\smash{\begin{tabular}[t]{c}-180\end{tabular}}}}%
    \put(0.16734186,0.02494609){\color[rgb]{0,0,0}\makebox(0,0)[t]{\lineheight{1.25}\smash{\begin{tabular}[t]{c}-150\end{tabular}}}}%
    \put(0.23538488,0.02494609){\color[rgb]{0,0,0}\makebox(0,0)[t]{\lineheight{1.25}\smash{\begin{tabular}[t]{c}-120\end{tabular}}}}%
    \put(0.30334607,0.02494609){\color[rgb]{0,0,0}\makebox(0,0)[t]{\lineheight{1.25}\smash{\begin{tabular}[t]{c}-90\end{tabular}}}}%
    \put(0.37138909,0.02494609){\color[rgb]{0,0,0}\makebox(0,0)[t]{\lineheight{1.25}\smash{\begin{tabular}[t]{c}-60\end{tabular}}}}%
    \put(0.43934862,0.02494609){\color[rgb]{0,0,0}\makebox(0,0)[t]{\lineheight{1.25}\smash{\begin{tabular}[t]{c}-30\end{tabular}}}}%
    \put(0.50928461,0.02494609){\color[rgb]{0,0,0}\makebox(0,0)[t]{\lineheight{1.25}\smash{\begin{tabular}[t]{c}0\end{tabular}}}}%
    \put(0.57729138,0.02494609){\color[rgb]{0,0,0}\makebox(0,0)[t]{\lineheight{1.25}\smash{\begin{tabular}[t]{c}30\end{tabular}}}}%
    \put(0.64524308,0.02494609){\color[rgb]{0,0,0}\makebox(0,0)[t]{\lineheight{1.25}\smash{\begin{tabular}[t]{c}60\end{tabular}}}}%
    \put(0.71321565,0.02494609){\color[rgb]{0,0,0}\makebox(0,0)[t]{\lineheight{1.25}\smash{\begin{tabular}[t]{c}90\end{tabular}}}}%
    \put(0.78085586,0.02494609){\color[rgb]{0,0,0}\makebox(0,0)[t]{\lineheight{1.25}\smash{\begin{tabular}[t]{c}120\end{tabular}}}}%
    \put(0.84881539,0.02494609){\color[rgb]{0,0,0}\makebox(0,0)[t]{\lineheight{1.25}\smash{\begin{tabular}[t]{c}150\end{tabular}}}}%
    \put(0.91685841,0.02494609){\color[rgb]{0,0,0}\makebox(0,0)[t]{\lineheight{1.25}\smash{\begin{tabular}[t]{c}180\end{tabular}}}}%
    \put(0.01805045,0.25125826){\color[rgb]{0,0,0}\rotatebox{90}{\makebox(0,0)[lt]{\lineheight{1.25}\smash{\begin{tabular}[t]{l}$u$\end{tabular}}}}}%
    \put(0.49304053,0.00507586){\color[rgb]{0,0,0}\makebox(0,0)[lt]{\lineheight{1.25}\smash{\begin{tabular}[t]{l}$\theta$\end{tabular}}}}%
    \put(0.36313362,0.25856681){\color[rgb]{0,0,0}\makebox(0,0)[rt]{\lineheight{1.25}\smash{\begin{tabular}[t]{r}$L_5$\end{tabular}}}}%
    \put(0.65178541,0.25808009){\color[rgb]{0,0,0}\makebox(0,0)[lt]{\lineheight{1.25}\smash{\begin{tabular}[t]{l}$L_4$\end{tabular}}}}%
    \put(0.49622392,0.4071505){\color[rgb]{0,0,0}\makebox(0,0)[lt]{\lineheight{1.25}\smash{\begin{tabular}[t]{l}$L_2$\end{tabular}}}}%
    \put(0.51904752,0.10672679){\color[rgb]{0,0,0}\makebox(0,0)[rt]{\lineheight{1.25}\smash{\begin{tabular}[t]{r}$L_1$\end{tabular}}}}%
    \put(0.09638275,0.2675732){\color[rgb]{0,0,0}\makebox(0,0)[lt]{\lineheight{1.25}\smash{\begin{tabular}[t]{l}$L_3$\end{tabular}}}}%
    \put(0.30302202,0.25856681){\color[rgb]{0,0,0}\makebox(0,0)[rt]{\lineheight{1.25}\smash{\begin{tabular}[t]{r}$\sL_5^l$\end{tabular}}}}%
    \put(0.71440168,0.25808009){\color[rgb]{0,0,0}\makebox(0,0)[lt]{\lineheight{1.25}\smash{\begin{tabular}[t]{l}$\sL_4^l$\end{tabular}}}}%
  \end{picture}%
\endgroup%
		\caption{Phase portrait of a particle in quasi-circular motion ($e_0=0$) 
			for a Sun-Jupiter like system ($\eps = 1/1000$).  
		The black dot denotes the singularity associated with the collision with Jupiter,
		while the brown circle, 
			orange circle, 
			red circle 
			and the two blue diamonds
		correspond respectively to 
			$L_1$, 
			$L_2$, 
			$L_3$ 
			and $L_j$ for $j=4,5$.
		The separatrices that originate from 
			$L_1$, 
			$L_2$ 
			and $L_3$,
		represented respectively by 
			brown, 
			orange 
			and red thick curves, 
			divide the phase portrait in six regions. 
		Although they are very close to each other, 
		the separatrices of $L_1$ do not coincide with those that originate from $L_2$.
		The beige domain, 
			centered on the singularity,
			embodies the Hill's sphere of the planet, 
			inside which the averaged Hamiltonian does not reflect properly
		the dynamics of the restricted three-body problem.
		The upper and lower grey domains 
			lay outside the co-orbital resonance ($\theta$ circulates).
		The blue and red trajectories 
			are level curves associated with 
				tadpole-shaped 
				and horseshoe-shaped periodic orbits, respectively.
		More precisely, the blue level curves belong to the families $\sL_4^l$ and $\sL_5^l$.
		Finally, the green curves exhibit the elements 
			associated with a minimal mutual distance $\Delta = NR_\rH$, 
			while the grey lines depict $\{\modu{u} = \eps^{1/3}/\sqrt{N}\}$.
		For a given $N$, they bound the domain $\fD_1(\Delta)$ 
		inside which Therorem \ref{theo:stability} is applied.}
\label{fig:3}
	\end{center}
\end{figure*}

\begin{figure*}[!h]
	\begin{center}
		\tiny
		\def\svgwidth{0.45\textwidth}
\begingroup%
  \makeatletter%
  \providecommand\color[2][]{%
    \errmessage{(Inkscape) Color is used for the text in Inkscape, but the package 'color.sty' is not loaded}%
    \renewcommand\color[2][]{}%
  }%
  \providecommand\transparent[1]{%
    \errmessage{(Inkscape) Transparency is used (non-zero) for the text in Inkscape, but the package 'transparent.sty' is not loaded}%
    \renewcommand\transparent[1]{}%
  }%
  \providecommand\rotatebox[2]{#2}%
  \newcommand*\fsize{\dimexpr\f@size pt\relax}%
  \newcommand*\lineheight[1]{\fontsize{\fsize}{#1\fsize}\selectfont}%
  \ifx\svgwidth\undefined%
    \setlength{\unitlength}{360.07850647bp}%
    \ifx\svgscale\undefined%
      \relax%
    \else%
      \setlength{\unitlength}{\unitlength * \real{\svgscale}}%
    \fi%
  \else%
    \setlength{\unitlength}{\svgwidth}%
  \fi%
  \global\let\svgwidth\undefined%
  \global\let\svgscale\undefined%
  \makeatother%
  \begin{picture}(1,0.62546385)%
    \lineheight{1}%
    \setlength\tabcolsep{0pt}%
    \put(0,0){\includegraphics[width=\unitlength]{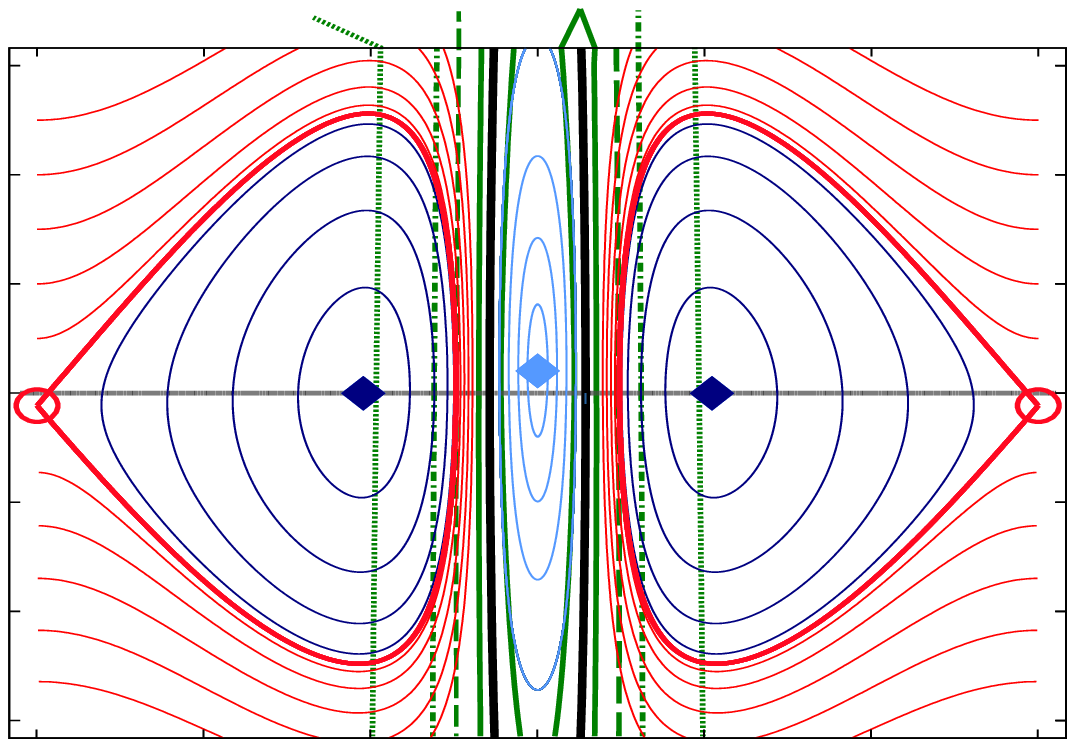}}%
    \put(0.60349078,0.60728456){\makebox(0,0)[lt]{\lineheight{1.25}\smash{\begin{tabular}[t]{l}$5R_\rH$\end{tabular}}}}%
    \put(0.55350101,0.60728456){\makebox(0,0)[lt]{\lineheight{1.25}\smash{\begin{tabular}[t]{l}$R_\rH$\end{tabular}}}}%
    \put(0.47411259,0.60728456){\makebox(0,0)[t]{\lineheight{1.25}\smash{\begin{tabular}[t]{c}$3R_\rH$\end{tabular}}}}%
    \put(0.37542516,0.60728456){\makebox(0,0)[rt]{\lineheight{1.25}\smash{\begin{tabular}[t]{r}$10R_\rH$\end{tabular}}}}%
    \put(0.1019005,0.02767649){\color[rgb]{0,0,0}\makebox(0,0)[rt]{\lineheight{1.25}\smash{\begin{tabular}[t]{r}-0.03\end{tabular}}}}%
    \put(0.1019439,0.11498081){\color[rgb]{0,0,0}\makebox(0,0)[rt]{\lineheight{1.25}\smash{\begin{tabular}[t]{r}-0.02\end{tabular}}}}%
    \put(0.09837402,0.20228513){\color[rgb]{0,0,0}\makebox(0,0)[rt]{\lineheight{1.25}\smash{\begin{tabular}[t]{r}-0.01\end{tabular}}}}%
    \put(0.10216299,0.28958945){\color[rgb]{0,0,0}\makebox(0,0)[rt]{\lineheight{1.25}\smash{\begin{tabular}[t]{r}0\end{tabular}}}}%
    \put(0.09858149,0.37689378){\color[rgb]{0,0,0}\makebox(0,0)[rt]{\lineheight{1.25}\smash{\begin{tabular}[t]{r}0.01\end{tabular}}}}%
    \put(0.10215136,0.46419812){\color[rgb]{0,0,0}\makebox(0,0)[rt]{\lineheight{1.25}\smash{\begin{tabular}[t]{r}0.02\end{tabular}}}}%
    \put(0.10210796,0.55150244){\color[rgb]{0,0,0}\makebox(0,0)[rt]{\lineheight{1.25}\smash{\begin{tabular}[t]{r}0.03\end{tabular}}}}%
    \put(0.13747033,0.58288422){\color[rgb]{0,0,0}\makebox(0,0)[t]{\lineheight{1.25}\smash{\begin{tabular}[t]{c}-180\end{tabular}}}}%
    \put(0.27094255,0.58288422){\color[rgb]{0,0,0}\makebox(0,0)[t]{\lineheight{1.25}\smash{\begin{tabular}[t]{c}-120\end{tabular}}}}%
    \put(0.40441756,0.58288422){\color[rgb]{0,0,0}\makebox(0,0)[t]{\lineheight{1.25}\smash{\begin{tabular}[t]{c}-60\end{tabular}}}}%
    \put(0.54103887,0.58288422){\color[rgb]{0,0,0}\makebox(0,0)[t]{\lineheight{1.25}\smash{\begin{tabular}[t]{c}0\end{tabular}}}}%
    \put(0.67443777,0.58288422){\color[rgb]{0,0,0}\makebox(0,0)[t]{\lineheight{1.25}\smash{\begin{tabular}[t]{c}60\end{tabular}}}}%
    \put(0.80726159,0.58288422){\color[rgb]{0,0,0}\makebox(0,0)[t]{\lineheight{1.25}\smash{\begin{tabular}[t]{c}120\end{tabular}}}}%
    \put(0.94073386,0.58288422){\color[rgb]{0,0,0}\makebox(0,0)[t]{\lineheight{1.25}\smash{\begin{tabular}[t]{c}180\end{tabular}}}}%
    \put(0.03008556,0.29129461){\color[rgb]{0,0,0}\rotatebox{90}{\makebox(0,0)[lt]{\lineheight{1.25}\smash{\begin{tabular}[t]{l}$u$\end{tabular}}}}}%
    \put(0.54105069,0.60872804){\color[rgb]{0,0,0}\makebox(0,0)[t]{\lineheight{1.25}\smash{\begin{tabular}[t]{c}$\theta$\end{tabular}}}}%
    \put(0.15201822,0.5369481){\color[rgb]{0,0,0}\makebox(0,0)[lt]{\lineheight{1.25}\smash{\begin{tabular}[t]{l}a.\end{tabular}}}}%
    \put(0.5013046,0.3398832){\color[rgb]{0,0,0}\makebox(0,0)[lt]{\lineheight{1.25}\smash{\begin{tabular}[t]{l}$f(e_0)$\end{tabular}}}}%
    \put(0.64238789,0.2486698){\color[rgb]{0,0,0}\makebox(0,0)[lt]{\lineheight{1.25}\smash{\begin{tabular}[t]{l}$\sL_4^s(e_0)$\end{tabular}}}}%
    \put(0.46137163,0.2486698){\color[rgb]{0,0,0}\makebox(0,0)[rt]{\lineheight{1.25}\smash{\begin{tabular}[t]{r}$\sL_5^s(e_0)$\end{tabular}}}}%
    \put(0.15905604,0.28124598){\color[rgb]{0,0,0}\makebox(0,0)[lt]{\lineheight{1.25}\smash{\begin{tabular}[t]{l}$\sL_3(e_0)$\end{tabular}}}}%
  \end{picture}%
\endgroup%
		\def\svgwidth{0.45\textwidth}
\begingroup%
  \makeatletter%
  \providecommand\color[2][]{%
    \errmessage{(Inkscape) Color is used for the text in Inkscape, but the package 'color.sty' is not loaded}%
    \renewcommand\color[2][]{}%
  }%
  \providecommand\transparent[1]{%
    \errmessage{(Inkscape) Transparency is used (non-zero) for the text in Inkscape, but the package 'transparent.sty' is not loaded}%
    \renewcommand\transparent[1]{}%
  }%
  \providecommand\rotatebox[2]{#2}%
  \newcommand*\fsize{\dimexpr\f@size pt\relax}%
  \newcommand*\lineheight[1]{\fontsize{\fsize}{#1\fsize}\selectfont}%
  \ifx\svgwidth\undefined%
    \setlength{\unitlength}{359.91066742bp}%
    \ifx\svgscale\undefined%
      \relax%
    \else%
      \setlength{\unitlength}{\unitlength * \real{\svgscale}}%
    \fi%
  \else%
    \setlength{\unitlength}{\svgwidth}%
  \fi%
  \global\let\svgwidth\undefined%
  \global\let\svgscale\undefined%
  \makeatother%
  \begin{picture}(1,0.62612456)%
    \lineheight{1}%
    \setlength\tabcolsep{0pt}%
    \put(0,0){\includegraphics[width=\unitlength]{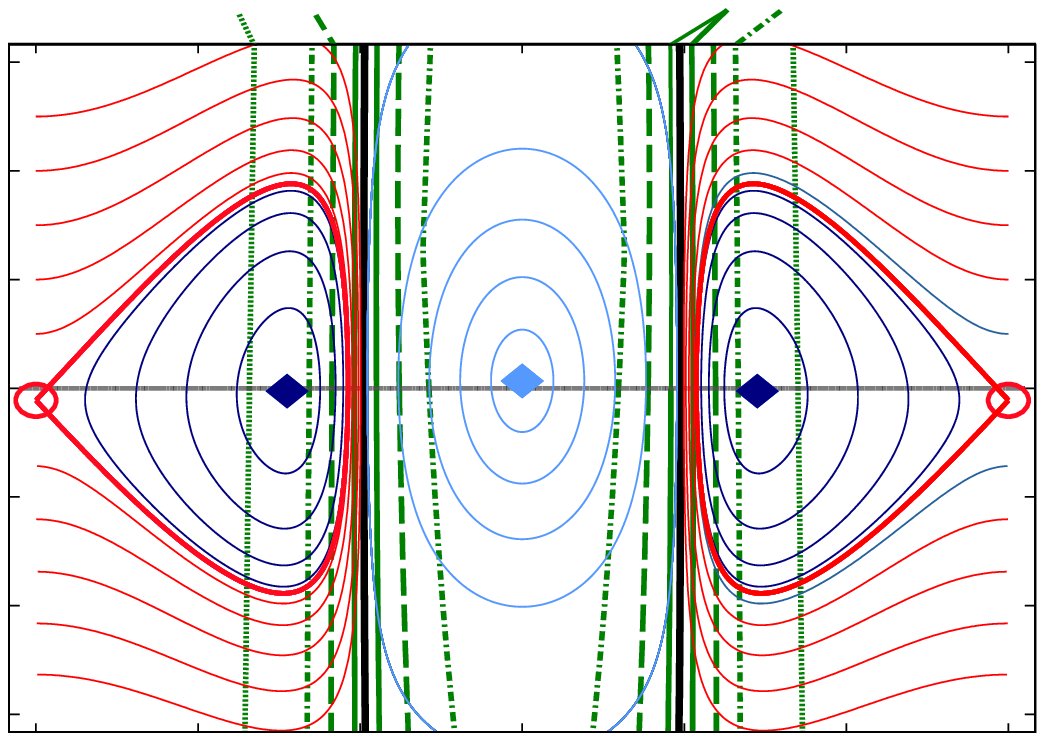}}%
    \put(0.8653062,0.02975969){\color[rgb]{0,0,0}\makebox(0,0)[lt]{\lineheight{1.25}\smash{\begin{tabular}[t]{l}-0.03\end{tabular}}}}%
    \put(0.8653062,0.11687292){\color[rgb]{0,0,0}\makebox(0,0)[lt]{\lineheight{1.25}\smash{\begin{tabular}[t]{l}-0.02\end{tabular}}}}%
    \put(0.8653062,0.20398618){\color[rgb]{0,0,0}\makebox(0,0)[lt]{\lineheight{1.25}\smash{\begin{tabular}[t]{l}-0.01\end{tabular}}}}%
    \put(0.87141632,0.29109941){\color[rgb]{0,0,0}\makebox(0,0)[lt]{\lineheight{1.25}\smash{\begin{tabular}[t]{l}0\end{tabular}}}}%
    \put(0.87141632,0.37821264){\color[rgb]{0,0,0}\makebox(0,0)[lt]{\lineheight{1.25}\smash{\begin{tabular}[t]{l}0.01\end{tabular}}}}%
    \put(0.87141632,0.46532589){\color[rgb]{0,0,0}\makebox(0,0)[lt]{\lineheight{1.25}\smash{\begin{tabular}[t]{l}0.02\end{tabular}}}}%
    \put(0.87141632,0.55243912){\color[rgb]{0,0,0}\makebox(0,0)[lt]{\lineheight{1.25}\smash{\begin{tabular}[t]{l}0.03\end{tabular}}}}%
    \put(0.05262483,0.58375222){\color[rgb]{0,0,0}\makebox(0,0)[t]{\lineheight{1.25}\smash{\begin{tabular}[t]{c}-180\end{tabular}}}}%
    \put(0.18226753,0.58375222){\color[rgb]{0,0,0}\makebox(0,0)[t]{\lineheight{1.25}\smash{\begin{tabular}[t]{c}-120\end{tabular}}}}%
    \put(0.31191301,0.58375222){\color[rgb]{0,0,0}\makebox(0,0)[t]{\lineheight{1.25}\smash{\begin{tabular}[t]{c}-60\end{tabular}}}}%
    \put(0.44470624,0.58375222){\color[rgb]{0,0,0}\makebox(0,0)[t]{\lineheight{1.25}\smash{\begin{tabular}[t]{c}0\end{tabular}}}}%
    \put(0.57427556,0.58375222){\color[rgb]{0,0,0}\makebox(0,0)[t]{\lineheight{1.25}\smash{\begin{tabular}[t]{c}60\end{tabular}}}}%
    \put(0.7032696,0.58375222){\color[rgb]{0,0,0}\makebox(0,0)[t]{\lineheight{1.25}\smash{\begin{tabular}[t]{c}120\end{tabular}}}}%
    \put(0.83291232,0.58375222){\color[rgb]{0,0,0}\makebox(0,0)[t]{\lineheight{1.25}\smash{\begin{tabular}[t]{c}180\end{tabular}}}}%
    \put(0.95884943,0.29280083){\color[rgb]{0,0,0}\rotatebox{90}{\makebox(0,0)[lt]{\lineheight{1.25}\smash{\begin{tabular}[t]{l}$u$\end{tabular}}}}}%
    \put(0.44055038,0.60953947){\color[rgb]{0,0,0}\makebox(0,0)[t]{\lineheight{1.25}\smash{\begin{tabular}[t]{c}$\theta$\end{tabular}}}}%
    \put(0.81381559,0.53760031){\color[rgb]{0,0,0}\makebox(0,0)[rt]{\lineheight{1.25}\smash{\begin{tabular}[t]{r}b.\end{tabular}}}}%
    \put(0.28541127,0.60781578){\color[rgb]{0,0,0}\makebox(0,0)[rt]{\lineheight{1.25}\smash{\begin{tabular}[t]{r}$3R_\rH$\end{tabular}}}}%
    \put(0.22289578,0.60781578){\color[rgb]{0,0,0}\makebox(0,0)[rt]{\lineheight{1.25}\smash{\begin{tabular}[t]{r}$10R_\rH$\end{tabular}}}}%
    \put(0.5832374,0.60781578){\color[rgb]{0,0,0}\makebox(0,0)[lt]{\lineheight{1.25}\smash{\begin{tabular}[t]{l}$R_\rH$\end{tabular}}}}%
    \put(0.63324975,0.60781578){\color[rgb]{0,0,0}\makebox(0,0)[lt]{\lineheight{1.25}\smash{\begin{tabular}[t]{l}$5R_\rH$\end{tabular}}}}%
  \end{picture}%
\endgroup%
		\def\svgwidth{0.45\textwidth}
\begingroup%
  \makeatletter%
  \providecommand\color[2][]{%
    \errmessage{(Inkscape) Color is used for the text in Inkscape, but the package 'color.sty' is not loaded}%
    \renewcommand\color[2][]{}%
  }%
  \providecommand\transparent[1]{%
    \errmessage{(Inkscape) Transparency is used (non-zero) for the text in Inkscape, but the package 'transparent.sty' is not loaded}%
    \renewcommand\transparent[1]{}%
  }%
  \providecommand\rotatebox[2]{#2}%
  \newcommand*\fsize{\dimexpr\f@size pt\relax}%
  \newcommand*\lineheight[1]{\fontsize{\fsize}{#1\fsize}\selectfont}%
  \ifx\svgwidth\undefined%
    \setlength{\unitlength}{360.07850647bp}%
    \ifx\svgscale\undefined%
      \relax%
    \else%
      \setlength{\unitlength}{\unitlength * \real{\svgscale}}%
    \fi%
  \else%
    \setlength{\unitlength}{\svgwidth}%
  \fi%
  \global\let\svgwidth\undefined%
  \global\let\svgscale\undefined%
  \makeatother%
  \begin{picture}(1,0.62997438)%
    \lineheight{1}%
    \setlength\tabcolsep{0pt}%
    \put(0,0){\includegraphics[width=\unitlength]{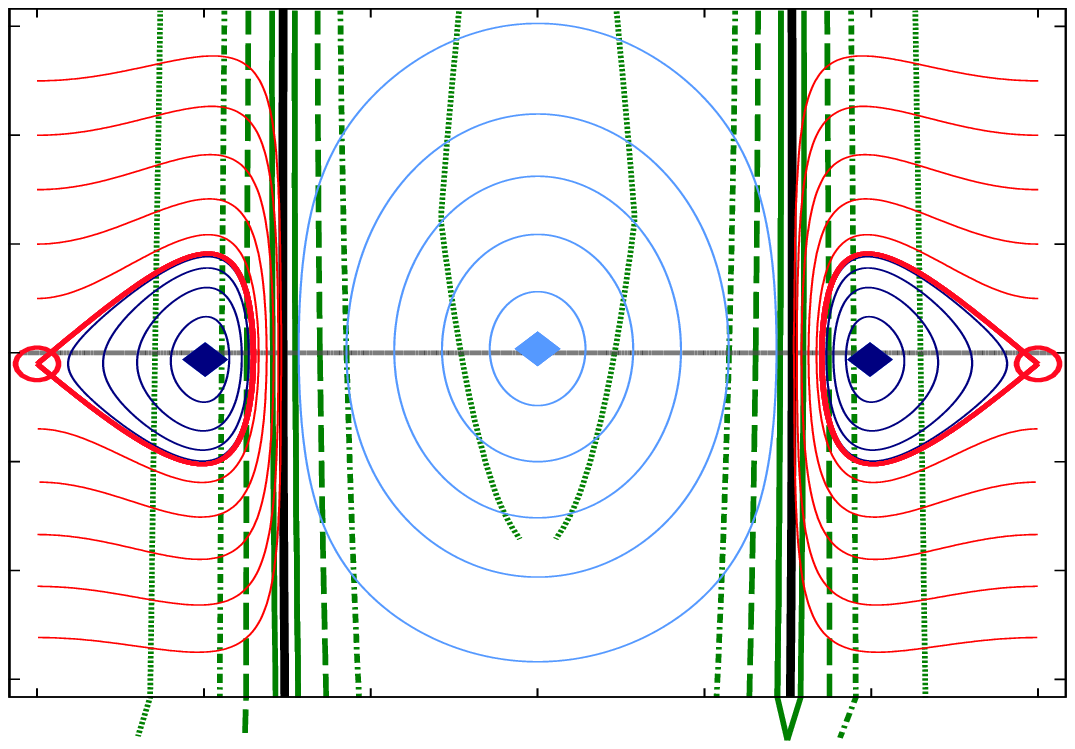}}%
    \put(0.1019005,0.06551309){\color[rgb]{0,0,0}\makebox(0,0)[rt]{\lineheight{1.25}\smash{\begin{tabular}[t]{r}-0.03\end{tabular}}}}%
    \put(0.1019439,0.15281741){\color[rgb]{0,0,0}\makebox(0,0)[rt]{\lineheight{1.25}\smash{\begin{tabular}[t]{r}-0.02\end{tabular}}}}%
    \put(0.09837402,0.24012173){\color[rgb]{0,0,0}\makebox(0,0)[rt]{\lineheight{1.25}\smash{\begin{tabular}[t]{r}-0.01\end{tabular}}}}%
    \put(0.10216299,0.32742605){\color[rgb]{0,0,0}\makebox(0,0)[rt]{\lineheight{1.25}\smash{\begin{tabular}[t]{r}0\end{tabular}}}}%
    \put(0.09858149,0.41473038){\color[rgb]{0,0,0}\makebox(0,0)[rt]{\lineheight{1.25}\smash{\begin{tabular}[t]{r}0.01\end{tabular}}}}%
    \put(0.10215136,0.50203472){\color[rgb]{0,0,0}\makebox(0,0)[rt]{\lineheight{1.25}\smash{\begin{tabular}[t]{r}0.02\end{tabular}}}}%
    \put(0.10210796,0.58933904){\color[rgb]{0,0,0}\makebox(0,0)[rt]{\lineheight{1.25}\smash{\begin{tabular}[t]{r}0.03\end{tabular}}}}%
    \put(0.03008556,0.32913121){\color[rgb]{0,0,0}\rotatebox{90}{\makebox(0,0)[lt]{\lineheight{1.25}\smash{\begin{tabular}[t]{l}$u$\end{tabular}}}}}%
    \put(0.1019005,0.06551309){\color[rgb]{0,0,0}\makebox(0,0)[rt]{\lineheight{1.25}\smash{\begin{tabular}[t]{r}-0.03\end{tabular}}}}%
    \put(0.1019439,0.15281741){\color[rgb]{0,0,0}\makebox(0,0)[rt]{\lineheight{1.25}\smash{\begin{tabular}[t]{r}-0.02\end{tabular}}}}%
    \put(0.09837402,0.24012173){\color[rgb]{0,0,0}\makebox(0,0)[rt]{\lineheight{1.25}\smash{\begin{tabular}[t]{r}-0.01\end{tabular}}}}%
    \put(0.10216299,0.32742605){\color[rgb]{0,0,0}\makebox(0,0)[rt]{\lineheight{1.25}\smash{\begin{tabular}[t]{r}0\end{tabular}}}}%
    \put(0.09858149,0.41473038){\color[rgb]{0,0,0}\makebox(0,0)[rt]{\lineheight{1.25}\smash{\begin{tabular}[t]{r}0.01\end{tabular}}}}%
    \put(0.10215136,0.50203472){\color[rgb]{0,0,0}\makebox(0,0)[rt]{\lineheight{1.25}\smash{\begin{tabular}[t]{r}0.02\end{tabular}}}}%
    \put(0.10210796,0.58933904){\color[rgb]{0,0,0}\makebox(0,0)[rt]{\lineheight{1.25}\smash{\begin{tabular}[t]{r}0.03\end{tabular}}}}%
    \put(0.03008556,0.32913121){\color[rgb]{0,0,0}\rotatebox{90}{\makebox(0,0)[lt]{\lineheight{1.25}\smash{\begin{tabular}[t]{l}$u$\end{tabular}}}}}%
    \put(0.22812047,0.00723827){\makebox(0,0)[rt]{\lineheight{1.25}\smash{\begin{tabular}[t]{r}$10R_\rH$\end{tabular}}}}%
    \put(0.33226463,0.00723827){\makebox(0,0)[rt]{\lineheight{1.25}\smash{\begin{tabular}[t]{r}$3R_\rH$\end{tabular}}}}%
    \put(0.74884026,0.00723827){\makebox(0,0)[rt]{\lineheight{1.25}\smash{\begin{tabular}[t]{r}$R_\rH$\end{tabular}}}}%
    \put(0.80716048,0.00723827){\makebox(0,0)[rt]{\lineheight{1.25}\smash{\begin{tabular}[t]{r}$5R_\rH$\end{tabular}}}}%
    \put(0.13748927,0.02784032){\color[rgb]{0,0,0}\makebox(0,0)[t]{\lineheight{1.25}\smash{\begin{tabular}[t]{c}-180\end{tabular}}}}%
    \put(0.2709292,0.02784032){\color[rgb]{0,0,0}\makebox(0,0)[t]{\lineheight{1.25}\smash{\begin{tabular}[t]{c}-120\end{tabular}}}}%
    \put(0.40437199,0.02784032){\color[rgb]{0,0,0}\makebox(0,0)[t]{\lineheight{1.25}\smash{\begin{tabular}[t]{c}-60\end{tabular}}}}%
    \put(0.54105474,0.02784032){\color[rgb]{0,0,0}\makebox(0,0)[t]{\lineheight{1.25}\smash{\begin{tabular}[t]{c}0\end{tabular}}}}%
    \put(0.67441918,0.02784032){\color[rgb]{0,0,0}\makebox(0,0)[t]{\lineheight{1.25}\smash{\begin{tabular}[t]{c}60\end{tabular}}}}%
    \put(0.80719142,0.02784032){\color[rgb]{0,0,0}\makebox(0,0)[t]{\lineheight{1.25}\smash{\begin{tabular}[t]{c}120\end{tabular}}}}%
    \put(0.94063138,0.02784032){\color[rgb]{0,0,0}\makebox(0,0)[t]{\lineheight{1.25}\smash{\begin{tabular}[t]{c}180\end{tabular}}}}%
    \put(0.53838708,0.00200953){\color[rgb]{0,0,0}\makebox(0,0)[t]{\lineheight{1.25}\smash{\begin{tabular}[t]{c}$\theta$\end{tabular}}}}%
    \put(0.15184964,0.08088163){\color[rgb]{0,0,0}\makebox(0,0)[lt]{\lineheight{1.25}\smash{\begin{tabular}[t]{l}c.\end{tabular}}}}%
  \end{picture}%
\endgroup%
		\def\svgwidth{0.45\textwidth}
\begingroup%
  \makeatletter%
  \providecommand\color[2][]{%
    \errmessage{(Inkscape) Color is used for the text in Inkscape, but the package 'color.sty' is not loaded}%
    \renewcommand\color[2][]{}%
  }%
  \providecommand\transparent[1]{%
    \errmessage{(Inkscape) Transparency is used (non-zero) for the text in Inkscape, but the package 'transparent.sty' is not loaded}%
    \renewcommand\transparent[1]{}%
  }%
  \providecommand\rotatebox[2]{#2}%
  \newcommand*\fsize{\dimexpr\f@size pt\relax}%
  \newcommand*\lineheight[1]{\fontsize{\fsize}{#1\fsize}\selectfont}%
  \ifx\svgwidth\undefined%
    \setlength{\unitlength}{360.16143036bp}%
    \ifx\svgscale\undefined%
      \relax%
    \else%
      \setlength{\unitlength}{\unitlength * \real{\svgscale}}%
    \fi%
  \else%
    \setlength{\unitlength}{\svgwidth}%
  \fi%
  \global\let\svgwidth\undefined%
  \global\let\svgscale\undefined%
  \makeatother%
  \begin{picture}(1,0.62500514)%
    \lineheight{1}%
    \setlength\tabcolsep{0pt}%
    \put(0,0){\includegraphics[width=\unitlength]{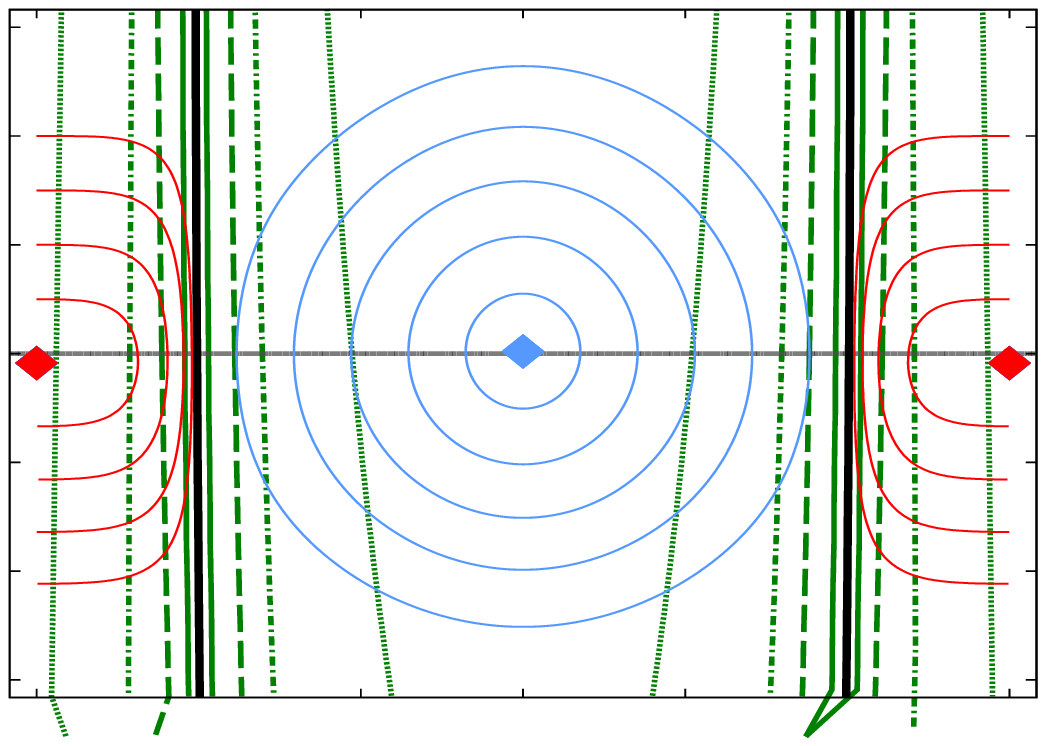}}%
    \put(0.05282965,0.02453706){\color[rgb]{0,0,0}\makebox(0,0)[t]{\lineheight{1.25}\smash{\begin{tabular}[t]{c}-180\end{tabular}}}}%
    \put(0.18241297,0.02453706){\color[rgb]{0,0,0}\makebox(0,0)[t]{\lineheight{1.25}\smash{\begin{tabular}[t]{c}-120\end{tabular}}}}%
    \put(0.31199908,0.02453706){\color[rgb]{0,0,0}\makebox(0,0)[t]{\lineheight{1.25}\smash{\begin{tabular}[t]{c}-60\end{tabular}}}}%
    \put(0.44473151,0.02453706){\color[rgb]{0,0,0}\makebox(0,0)[t]{\lineheight{1.25}\smash{\begin{tabular}[t]{c}0\end{tabular}}}}%
    \put(0.57424154,0.02453706){\color[rgb]{0,0,0}\makebox(0,0)[t]{\lineheight{1.25}\smash{\begin{tabular}[t]{c}60\end{tabular}}}}%
    \put(0.70317648,0.02453706){\color[rgb]{0,0,0}\makebox(0,0)[t]{\lineheight{1.25}\smash{\begin{tabular}[t]{c}120\end{tabular}}}}%
    \put(0.83275983,0.02453706){\color[rgb]{0,0,0}\makebox(0,0)[t]{\lineheight{1.25}\smash{\begin{tabular}[t]{c}180\end{tabular}}}}%
    \put(0.86513886,0.06378057){\color[rgb]{0,0,0}\makebox(0,0)[lt]{\lineheight{1.25}\smash{\begin{tabular}[t]{l}-0.03\end{tabular}}}}%
    \put(0.86513886,0.15080385){\color[rgb]{0,0,0}\makebox(0,0)[lt]{\lineheight{1.25}\smash{\begin{tabular}[t]{l}-0.02\end{tabular}}}}%
    \put(0.86513886,0.23782713){\color[rgb]{0,0,0}\makebox(0,0)[lt]{\lineheight{1.25}\smash{\begin{tabular}[t]{l}-0.01\end{tabular}}}}%
    \put(0.87124625,0.32485041){\color[rgb]{0,0,0}\makebox(0,0)[lt]{\lineheight{1.25}\smash{\begin{tabular}[t]{l}0\end{tabular}}}}%
    \put(0.87124625,0.4118737){\color[rgb]{0,0,0}\makebox(0,0)[lt]{\lineheight{1.25}\smash{\begin{tabular}[t]{l}0.01\end{tabular}}}}%
    \put(0.87124625,0.49889698){\color[rgb]{0,0,0}\makebox(0,0)[lt]{\lineheight{1.25}\smash{\begin{tabular}[t]{l}0.02\end{tabular}}}}%
    \put(0.87124625,0.58592026){\color[rgb]{0,0,0}\makebox(0,0)[lt]{\lineheight{1.25}\smash{\begin{tabular}[t]{l}0.03\end{tabular}}}}%
    \put(0.9586393,0.34044418){\color[rgb]{0,0,0}\rotatebox{90}{\makebox(0,0)[t]{\lineheight{1.25}\smash{\begin{tabular}[t]{c}$u$\end{tabular}}}}}%
    \put(0.44214095,-0.00128779){\color[rgb]{0,0,0}\makebox(0,0)[t]{\lineheight{1.25}\smash{\begin{tabular}[t]{c}$\theta$\end{tabular}}}}%
    \put(0.81585309,4.27817344){\color[rgb]{0,0,0}\makebox(0,0)[rt]{\lineheight{1.25}\smash{\begin{tabular}[t]{r}d.\end{tabular}}}}%
    \put(0.04577889,0.00627477){\makebox(0,0)[lt]{\lineheight{1.25}\smash{\begin{tabular}[t]{l}$10R_\rH$\end{tabular}}}}%
    \put(0.13740443,0.00627477){\makebox(0,0)[lt]{\lineheight{1.25}\smash{\begin{tabular}[t]{l}$3R_\rH$\end{tabular}}}}%
    \put(0.62468586,0.00627477){\makebox(0,0)[lt]{\lineheight{1.25}\smash{\begin{tabular}[t]{l}$R_\rH$\end{tabular}}}}%
    \put(0.73713509,0.00627477){\makebox(0,0)[lt]{\lineheight{1.25}\smash{\begin{tabular}[t]{l}$5R_\rH$\end{tabular}}}}%
    \put(0.81324876,0.0765251){\color[rgb]{0,0,0}\makebox(0,0)[rt]{\lineheight{1.25}\smash{\begin{tabular}[t]{r}d.\end{tabular}}}}%
    \put(0.07166187,0.31539544){\makebox(0,0)[lt]{\lineheight{1.25}\smash{\begin{tabular}[t]{l}$\sL_3(e_0)$\end{tabular}}}}%
  \end{picture}%
\endgroup%
		\caption{Phase portraits of the reduced Hamiltonian $\Hb^{K(e_0)}$ 
		for a Sun-Jupiter like system ($\eps=1/1000$).
		They enlarge the area bounded by $\{\modu{u} = \sqrt{\eps}\}$ in Fig.~\ref{fig:4}.
		For a, b, c and d, $e_0$ is equal to $0.15$, $0.5$, $0.75$ and  $0.925$, respectively.
		The black curves represent collision with the planet. 
		The blue, 
			sky blue 
			and red trajectories 
		are level curves associated with
			tadpole, 
			quasi-satellite 
			and horseshoe motion, respectively. 
		The red circles, 
			sky blue 
			and blue diamonds are equilibria.
		More precisely, they correspond respectively to orbits of the families of periodic orbits
			$\sL_3$, 
			$f$ 
			and $\sL^s_j$ for $j=4,5$.
		From each hyperbolic equilibrium emerges a separatrix (red thick curve) that divides 
		tadpole motion  and horseshoe motion.
		For d, the red diamond denotes an orbit that belong to the stable part of $\sL_3$.
		Finally, the green curves exhibit the elements 
			associated with a minimal mutual distance $\Delta = NR_\rH$, 
			and bound the domain $\fD_1(\Delta)$ 
		inside which Therorem \ref{theo:stability} is applied.}
\label{fig:4}
	\end{center}
\end{figure*}

\begin{sloppypar}
		Figures \ref{fig:3} and \ref{fig:4} display the phase portraits of the 
		reduced averaged Hamiltonian $\Hb^{K(e_0)}$ associated with five values of $e_0$.
		They are obtained for a Sun-Jupiter like system (${\eps=1/1000}$) 
			by implementing the algorithm of Nesvorn\'{y} et al. \cite{2002NeThFe}.			
		They are equivalent to the one showed in \cite{2002Mo,2002NeThFe}
			and extensively described in \cite{2017PoRoVi}.
		We will limit ourselves to present what will be useful to understand 
		the trajectories described in Sect.~\ref{sec:rem_sol}
		as well as to apply the Theorems stated in Sect.~\ref{sec:validity}.
		Notice that the phase portraits  are invariant by the symmetry with respect to the $u$-axis 
			${(\theta,u) \mapsto (2\pi-\theta, u)}$.
\end{sloppypar}

		In Fig.~\ref{fig:3}, $e_0$ is equal to zero and particle's motion is quasi-circular.
		Let us mention that in this case the reduced averaged Hamiltonian reads
		\bes
			\Hb^{0}(\theta, u) = -\frac{1-\eps}{2a} - u 
						+ \eps\left(
						\cos \theta - \frac{1}{\sqrt{a^2 + 1 -2a\cos\theta}}
						\right)
		\ees
		with ${a= (1+u)^2}$,
		which allows to recover some classical analytical properties (see, e.g, \cite{1977Ga,1999MuDe})
		that will be recalled in the following.
		Being the computations similar to the ones given in \cite{2013RoPo},
			the details are not given.

		First of all, the black dot at ${\theta=u=0}$ embodies the collision with the planet, where $\Hb^0$ is not defined.
		The phase portrait possesses five equilibria 
			that correspond to the Lagrange fixed points $L_j$.
		The two elliptic equilibria located in ${(\theta_j,u_j) = \left((-1)^j\times 60\degre,0\right)}$,
			stand for $L_4$ and $L_5$
		while the hyperbolic ones  in 
		\bes
		\begin{aligned}
			(\theta_3,u_3)	&= \left(180\degre, 1 - \frac{7}{6}\eps + \cO(\eps^2)\right),\\
			(\theta_j, u_j)	&=\left(0\degre, (-1)^j\times\left(\frac{\eps}{6}\right)^{1/3} + \cO(\eps^{2/3})\right)
		\end{aligned}
		\ees
		approximate respectively $L_3$ and $L_j$ for ${j=1,2}$.
		With respect to the synodic reference frame  (see Sect.~\ref{sec:rem_sol}),		
			$L_4$ are $L_5$ are recovered at their exact location
		while $L_3$  is approximated within an accuracy $\cO(\eps)$.
		For ${j=1,2}$, $L_j$ are found within an accuracy $\cO(\eps^{1/3})$
			which highlights the weakness of the averaged problem at the edge of the Hill's sphere.

		The phase portrait is characterized by six regions.

		The beige domain centered at the singularity and bounded by the separatrices 
			originating from $L_1$ and $L_2$
		seems to be the prograde satellite-like motion.
		However, 
		since the domain belongs to the Hill's sphere of the planet,
			it lies inside the exclusion zone and 
			its trajectories fall outside the scope of the averaged problem.
		Hence, the corresponding dynamics will not be analyzed in this work.
		The upper and lower grey domains,  that  lay above the separatrices of $L_1$ and $L_2$,
			illustrate the non-resonant motion
				for which  $\theta$ circulates
					(clockwise in the upper region
					and anti-clockwise in the lower one).			
		Considering that the width along the $u$-axis of the region located inside the separatrices 
		is of the order $\cO(\eps^{1/3})$,
			 Theorem \ref{theo:averaging} implies that
			 the non-resonant regions escape from the domain of validity of the averaged problem.

		The three remaining regions are the ones of the co-orbital dynamics, 
			for which $\theta$  oscillates about a given value.
		These regions are divided by the separatrix that originates from  $L_3$.
		The solutions that librate around $L_4$ and $L_5$ are tadpole-shaped
			characterized by $\modu{\theta}$ that oscillates about $60\degre$  
			such that ${\Theta_{0}<\modu{\theta}<180\degre}$ 
			with 
		\bes
		{\Theta_{0} = 2\arcsin((\sqrt{2}-1)/2) + \cO(\eps)\simeq 23.9\degre},
		\ees
		and $u$ that oscillates around zero with an amplitude that can reach $\cO(\sqrt{\eps})$.
 		According to the Sect.~\ref{sec:reading}, these trajectories are periodic 
			and possess the same features as 
			the ones of the long-periodic families $\sL_4^l$ and $\sL_5^l$.
		Outside the separatrix, the trajectories encompass $L_3$, $L_4$ and $L_5$,
			i.e., they are characterized by $\theta$ and $u$ 
			that oscillate respectively about $180\degre$ and $0$,
			with large amplitudes.
		More precisely, 
				$\theta$ and $u$ undergo large variations such that 
				${\theta\in[\Theta_{1}, 2\pi -  \Theta_{1}]}$
					with 
		\bes{0<\cO(\eps^{1/3})<\Theta_{1}< \Theta_{0}}\ees
				and ${\modu{u} \leq U_1}$ 
					with ${\cO(\sqrt{\eps})<U_{1}< \cO(\eps^{1/3})}$.
		These solutions approximates periodic horseshoe-shaped trajectories
			(see, e.g., the study of Barrab\'es and Oll\'e \cite{2006BaOl} that focuses on these solutions).
				
		With respect to Fig.~\ref{fig:3}, the phase portraits of Fig.~\ref{fig:4}a-d, 
			respectively associated with $e_0=0.15$, $0.5$, $0.75$, $0.925$, 
			enlarge the area bounded by  $\modu{u} = \sqrt{\eps}\simeq 0.31$.
			
		For $e_0>0$, the location of the singularities evolves:
			the origin becomes a regular point surrounded 
				by a set of singular points that describes a curve.
		For small $e_0$ (e.g., Fig.~\ref{fig:4}a), 
			a new domain of co-orbital motion appears inside the collision curve.
		It is centered on an elliptic equilibrium point
			located close to the origin.
		According to Sect.~\ref{sec:rem_sol} and Sect.~\ref{sec:reading},
			the elliptic equilibrium point approximates
			a periodic orbit of the family $f$.
		Hence, the periodic trajectories that librate around,
			provide quasi-periodic approximations of quasi-satellite orbits.
		Outside the collision curve, the topology does not change with respect to 
		the one depicted in Fig.~\ref{fig:3} outside the Hill's sphere:
			two elliptic equilibria close to $L_4$'s and $L_5$'s locations 
			and a separatrix emerging from an hyperbolic equilibrium close to $L_3$
		 	that divides the regions of tadpole and horseshoe motions.
		According to Sect.~\ref{sec:reading}, 
			the equilibria of the phase portraits approximate periodic trajectories  
			whose guiding center are located respectively close to $L_3$, $L_4$ and $L_5$.		
		Hence, the elliptic equilibria belong to $\sL_4^s$ and $\sL_5^s$
		while the hyperbolic one corresponds to a trajectory of  $\sL_3$.

		For higher values of $e_0$ (e.g., Fig.~\ref{fig:4}b-c), 
			the size of the quasi-satellite domain increases
			while the one of the tadpole domains shrinks 
			when the two elliptic equilibria are getting closer to the hyperbolic one.
		The evolution of the two elliptic equilibria illustrates 
			the shift of the guiding center of $\sL_4^s$ and $\sL_5^s$ toward $L_3$ 
			described in Sect.~\ref{sec:rem_sol}.
		Finally, for very high $e_0$ (e.g. Fig.~\ref{fig:4}d), 
			the tadpole domains vanished
			and remains a domain characterized by trajectories 
			that librate around an elliptic equilibrium that belongs to $\sL_3$.
		This bifurcation of $\sL_3$ occurs for $e_0\simeq 0.917$, 
			when the short-periodic families $\sL_j^s$ merge with $\sL_3$.

		For a given minimal mutual distance $\Delta = NR_\rH$ 
			with $1\leq N \leq \eps^{-1/3}$ and $R_\rH = \left(\frac{\eps}{3}\right)^{1/3}$
			that denotes the Hill's radius of the planet,
		we recall that the domain $\fD_1(\Delta)$ introduces in Sect.~\ref{sec:quantitative} 
		is defined by the resonant variables for which the distance between the planet and the particle
			is larger than $NR_\rH$,
		and $\modu{u}$ 
		which is bounded by a quantity 
			$\rho \eqp \frac{\eps^{1/3}}{\sqrt{N}}$.
		The grey lines and green curves
		that lay over the phase portraits approximate the boundaries of $\fD_1(\Delta)$
			for several value of $N=1,3,5,10$.
		The grey lines correspond to $\{\modu{u} = \frac{\eps^{1/3}}{\sqrt{N}}\}$
			while
		the green curves exhibit the elements $(\theta, u, e(e_0,u))$
			for which the minimal mutual distance is equal to $\Delta = N R_\rH$.	
		They are computed by resolving the following equation:
		\bes
			\min_{M\in\TT}(R^2 + 1 -2R\cos \phi)_{(\theta, u, e(e_0,u), M)}= N^2 R_\rH^2.
		\ees	
		with the help of Eq.~\eqref{eq:Prop_EO}.		

		In Fig.~\ref{fig:3} and Fig.~\ref{fig:4}, 
			the exclusion zone of the averaged problem,
			is depicted by the small areas centered on the collision curves
			and bounded by the continuous green curves associated with 
			the minimal mutual distance $\Delta=R_\rH$.
		They show that, contrarily to the tadpole motion, 
			some solutions in quasi-satellite and horsehoe motion intersects the Hill's sphere and 
			fall outside the scope of the averaged Hamiltonian.
		More generally, the phase portraits also show that, 
			for a  given minimal mutual distance ${\Delta = N R_\rH}$,
			a co-orbital solution which starts inside the area  ${\{\modu{u} \leq \frac{\eps^{1/3}}{\sqrt{N}}\}}$
			can cross the sphere defined by ${\Delta = N R_\rH}$
			and thus escape from the domain $\fD_1(\Delta)$
			inside which Theorem  \ref{theo:averaging} and Theorem \ref{theo:stability} are applied.
		However, a co-orbital solution which starts in the neighborhood of the section ${\{u = 0\}}$
			 at a given minimal mutual distance $\Delta$,
			 does not experience closest encounters with the Jupiter and
			remains inside $\fD_1(\Delta)$.
		As a consequence, the section $\{u=0\}$ provides a convenient way 
			to discuss about the validity and the time of stability of the solutions of the averaged problem
			without caring about times of escape from  $\fD_1(\Delta)$.
		This study is realized in the following section.		
	
		\subsection{A ‘‘map" of the  co-orbital motion in the circular-planar case}

	\begin{figure}[h!]
	\begin{center}
		\tiny
		\def\svgwidth{0.49\textwidth}
\begingroup%
  \makeatletter%
  \providecommand\color[2][]{%
    \errmessage{(Inkscape) Color is used for the text in Inkscape, but the package 'color.sty' is not loaded}%
    \renewcommand\color[2][]{}%
  }%
  \providecommand\transparent[1]{%
    \errmessage{(Inkscape) Transparency is used (non-zero) for the text in Inkscape, but the package 'transparent.sty' is not loaded}%
    \renewcommand\transparent[1]{}%
  }%
  \providecommand\rotatebox[2]{#2}%
  \newcommand*\fsize{\dimexpr\f@size pt\relax}%
  \newcommand*\lineheight[1]{\fontsize{\fsize}{#1\fsize}\selectfont}%
  \ifx\svgwidth\undefined%
    \setlength{\unitlength}{432bp}%
    \ifx\svgscale\undefined%
      \relax%
    \else%
      \setlength{\unitlength}{\unitlength * \real{\svgscale}}%
    \fi%
  \else%
    \setlength{\unitlength}{\svgwidth}%
  \fi%
  \global\let\svgwidth\undefined%
  \global\let\svgscale\undefined%
  \makeatother%
  \begin{picture}(1,0.83333333)%
    \lineheight{1}%
    \setlength\tabcolsep{0pt}%
    \put(0,0){\includegraphics[width=\unitlength]{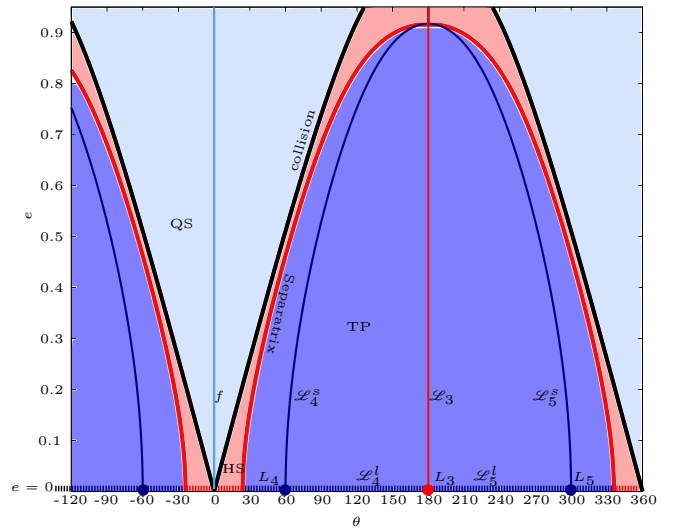}}%
    \put(0.07180714,0.13226814){\color[rgb]{0,0,0}\makebox(0,0)[rt]{\lineheight{1.25}\smash{\begin{tabular}[t]{r}0.1\end{tabular}}}}%
    \put(0.07478204,0.21166629){\color[rgb]{0,0,0}\makebox(0,0)[rt]{\lineheight{1.25}\smash{\begin{tabular}[t]{r}0.2\end{tabular}}}}%
    \put(0.07474587,0.2909487){\color[rgb]{0,0,0}\makebox(0,0)[rt]{\lineheight{1.25}\smash{\begin{tabular}[t]{r}0.3\end{tabular}}}}%
    \put(0.07493576,0.37034684){\color[rgb]{0,0,0}\makebox(0,0)[rt]{\lineheight{1.25}\smash{\begin{tabular}[t]{r}0.4\end{tabular}}}}%
    \put(0.07476396,0.44962925){\color[rgb]{0,0,0}\makebox(0,0)[rt]{\lineheight{1.25}\smash{\begin{tabular}[t]{r}0.5\end{tabular}}}}%
    \put(0.07482725,0.52902739){\color[rgb]{0,0,0}\makebox(0,0)[rt]{\lineheight{1.25}\smash{\begin{tabular}[t]{r}0.6\end{tabular}}}}%
    \put(0.07493576,0.60830981){\color[rgb]{0,0,0}\makebox(0,0)[rt]{\lineheight{1.25}\smash{\begin{tabular}[t]{r}0.7\end{tabular}}}}%
    \put(0.07474587,0.68770796){\color[rgb]{0,0,0}\makebox(0,0)[rt]{\lineheight{1.25}\smash{\begin{tabular}[t]{r}0.8\end{tabular}}}}%
    \put(0.07466449,0.76699036){\color[rgb]{0,0,0}\makebox(0,0)[rt]{\lineheight{1.25}\smash{\begin{tabular}[t]{r}0.9\end{tabular}}}}%
    \put(0.08569137,0.04152679){\color[rgb]{0,0,0}\makebox(0,0)[t]{\lineheight{1.25}\smash{\begin{tabular}[t]{c}-120\end{tabular}}}}%
    \put(0.14078627,0.04152679){\color[rgb]{0,0,0}\makebox(0,0)[t]{\lineheight{1.25}\smash{\begin{tabular}[t]{c}-90\end{tabular}}}}%
    \put(0.19587887,0.04152679){\color[rgb]{0,0,0}\makebox(0,0)[t]{\lineheight{1.25}\smash{\begin{tabular}[t]{c}-60\end{tabular}}}}%
    \put(0.25097145,0.04152679){\color[rgb]{0,0,0}\makebox(0,0)[t]{\lineheight{1.25}\smash{\begin{tabular}[t]{c}-30\end{tabular}}}}%
    \put(0.3086883,0.04152679){\color[rgb]{0,0,0}\makebox(0,0)[t]{\lineheight{1.25}\smash{\begin{tabular}[t]{c}0\end{tabular}}}}%
    \put(0.36384638,0.04152679){\color[rgb]{0,0,0}\makebox(0,0)[t]{\lineheight{1.25}\smash{\begin{tabular}[t]{c}30\end{tabular}}}}%
    \put(0.41881239,0.04152679){\color[rgb]{0,0,0}\makebox(0,0)[t]{\lineheight{1.25}\smash{\begin{tabular}[t]{c}60\end{tabular}}}}%
    \put(0.47392307,0.04152679){\color[rgb]{0,0,0}\makebox(0,0)[t]{\lineheight{1.25}\smash{\begin{tabular}[t]{c}90\end{tabular}}}}%
    \put(0.52845724,0.04152679){\color[rgb]{0,0,0}\makebox(0,0)[t]{\lineheight{1.25}\smash{\begin{tabular}[t]{c}120\end{tabular}}}}%
    \put(0.58343409,0.04152679){\color[rgb]{0,0,0}\makebox(0,0)[t]{\lineheight{1.25}\smash{\begin{tabular}[t]{c}150\end{tabular}}}}%
    \put(0.63852669,0.04152679){\color[rgb]{0,0,0}\makebox(0,0)[t]{\lineheight{1.25}\smash{\begin{tabular}[t]{c}180\end{tabular}}}}%
    \put(0.69421604,0.04152679){\color[rgb]{0,0,0}\makebox(0,0)[t]{\lineheight{1.25}\smash{\begin{tabular}[t]{c}210\end{tabular}}}}%
    \put(0.74930867,0.04152679){\color[rgb]{0,0,0}\makebox(0,0)[t]{\lineheight{1.25}\smash{\begin{tabular}[t]{c}240\end{tabular}}}}%
    \put(0.80440124,0.04152679){\color[rgb]{0,0,0}\makebox(0,0)[t]{\lineheight{1.25}\smash{\begin{tabular}[t]{c}270\end{tabular}}}}%
    \put(0.85956169,0.04152679){\color[rgb]{0,0,0}\makebox(0,0)[t]{\lineheight{1.25}\smash{\begin{tabular}[t]{c}300\end{tabular}}}}%
    \put(0.91465425,0.04152679){\color[rgb]{0,0,0}\makebox(0,0)[t]{\lineheight{1.25}\smash{\begin{tabular}[t]{c}330\end{tabular}}}}%
    \put(0.96974681,0.04152679){\color[rgb]{0,0,0}\makebox(0,0)[t]{\lineheight{1.25}\smash{\begin{tabular}[t]{c}360\end{tabular}}}}%
    \put(0.02502351,0.49623739){\color[rgb]{0,0,0}\rotatebox{90}{\makebox(0,0)[rt]{\lineheight{1.25}\smash{\begin{tabular}[t]{r}$e$\end{tabular}}}}}%
    \put(0.52643116,0.00703667){\color[rgb]{0,0,0}\makebox(0,0)[t]{\lineheight{1.25}\smash{\begin{tabular}[t]{c}$\theta$\end{tabular}}}}%
    \put(0.275085,0.47042305){\color[rgb]{0,0,0}\makebox(0,0)[rt]{\lineheight{1.25}\smash{\begin{tabular}[t]{r}QS\end{tabular}}}}%
    \put(0.30664422,0.20176866){\color[rgb]{0,0,0}\makebox(0,0)[lt]{\lineheight{1.25}\smash{\begin{tabular}[t]{l}$f$\end{tabular}}}}%
    \put(0.63814409,0.20176866){\color[rgb]{0,0,0}\makebox(0,0)[lt]{\lineheight{1.25}\smash{\begin{tabular}[t]{l}$\sL_3$\end{tabular}}}}%
    \put(0.42981076,0.20176866){\color[rgb]{0,0,0}\makebox(0,0)[lt]{\lineheight{1.25}\smash{\begin{tabular}[t]{l}$\sL_4^s$\end{tabular}}}}%
    \put(0.84149499,0.20176871){\color[rgb]{0,0,0}\makebox(0,0)[rt]{\lineheight{1.25}\smash{\begin{tabular}[t]{r}$\sL_5^s$\end{tabular}}}}%
    \put(0.64508853,0.07676866){\color[rgb]{0,0,0}\makebox(0,0)[lt]{\lineheight{1.25}\smash{\begin{tabular}[t]{l}$L_3$\end{tabular}}}}%
    \put(0.40949962,0.07676866){\color[rgb]{0,0,0}\makebox(0,0)[rt]{\lineheight{1.25}\smash{\begin{tabular}[t]{r}$L_4$\end{tabular}}}}%
    \put(0.86027622,0.0767694){\color[rgb]{0,0,0}\makebox(0,0)[lt]{\lineheight{1.25}\smash{\begin{tabular}[t]{l}$L_5$\end{tabular}}}}%
    \put(0.35519179,0.08835248){\color[rgb]{0,0,0}\makebox(0,0)[rt]{\lineheight{1.25}\smash{\begin{tabular}[t]{r}HS\end{tabular}}}}%
    \put(0.54939057,0.31070083){\color[rgb]{0,0,0}\makebox(0,0)[rt]{\lineheight{1.25}\smash{\begin{tabular}[t]{r}TP\end{tabular}}}}%
    \put(0.46287212,0.65307734){\color[rgb]{0,0,0}\rotatebox{73.591452}{\makebox(0,0)[rt]{\lineheight{1.25}\smash{\begin{tabular}[t]{r}collision\end{tabular}}}}}%
    \put(0.38860661,0.27487123){\color[rgb]{0,0,0}\rotatebox{-103.35591}{\makebox(0,0)[rt]{\lineheight{1.25}\smash{\begin{tabular}[t]{r}Separatrix\end{tabular}}}}}%
    \put(0.52553696,0.07784964){\color[rgb]{0,0,0}\makebox(0,0)[lt]{\lineheight{1.25}\smash{\begin{tabular}[t]{l}$\sL_4^l$\end{tabular}}}}%
    \put(0.74641101,0.0778497){\color[rgb]{0,0,0}\makebox(0,0)[rt]{\lineheight{1.25}\smash{\begin{tabular}[t]{r}$\sL_5^l$\end{tabular}}}}%
    \put(0.0592035,0.06083829){\color[rgb]{0,0,0}\rotatebox{-0.48961285}{\makebox(0,0)[rt]{\lineheight{1.25}\smash{\begin{tabular}[t]{r}$e=0$\end{tabular}}}}}%
  \end{picture}%
\endgroup%
		\caption{``Map" of the co-orbital motion defined by the section $\{u=0\}$.
			The black and red thick curves 
			stand respectively for the singularity of collision 
			and the crossing of the separatrices that originate from $\sL_3$ (red curve).
			They divide the map in three regions.
			The sky blue and blue regions correspond 
			to the quasi-satellite and the tadpole motion.
			They are centered respectively on the family $f$ (sky blue curve) 
			and the short periodic families $\sL_j^s$ (blue curves).
			The horseshoe region is represented in red.
			The dashed line is associated with the quasi-circular motion $(e_i=0)$ 
			for which the tadpole and horseshoe solution correspond to periodic trajectories 
			in the synodic reference frame.}
\label{fig:5}
	\end{center}
\end{figure}			
	
	\begin{figure*}[!h]
	\begin{center}
		\tiny
		\def\svgwidth{0.49\textwidth}
\begingroup%
  \makeatletter%
  \providecommand\color[2][]{%
    \errmessage{(Inkscape) Color is used for the text in Inkscape, but the package 'color.sty' is not loaded}%
    \renewcommand\color[2][]{}%
  }%
  \providecommand\transparent[1]{%
    \errmessage{(Inkscape) Transparency is used (non-zero) for the text in Inkscape, but the package 'transparent.sty' is not loaded}%
    \renewcommand\transparent[1]{}%
  }%
  \providecommand\rotatebox[2]{#2}%
  \newcommand*\fsize{\dimexpr\f@size pt\relax}%
  \newcommand*\lineheight[1]{\fontsize{\fsize}{#1\fsize}\selectfont}%
  \ifx\svgwidth\undefined%
    \setlength{\unitlength}{432bp}%
    \ifx\svgscale\undefined%
      \relax%
    \else%
      \setlength{\unitlength}{\unitlength * \real{\svgscale}}%
    \fi%
  \else%
    \setlength{\unitlength}{\svgwidth}%
  \fi%
  \global\let\svgwidth\undefined%
  \global\let\svgscale\undefined%
  \makeatother%
  \begin{picture}(1,0.83333333)%
    \lineheight{1}%
    \setlength\tabcolsep{0pt}%
    \put(0,0){\includegraphics[width=\unitlength]{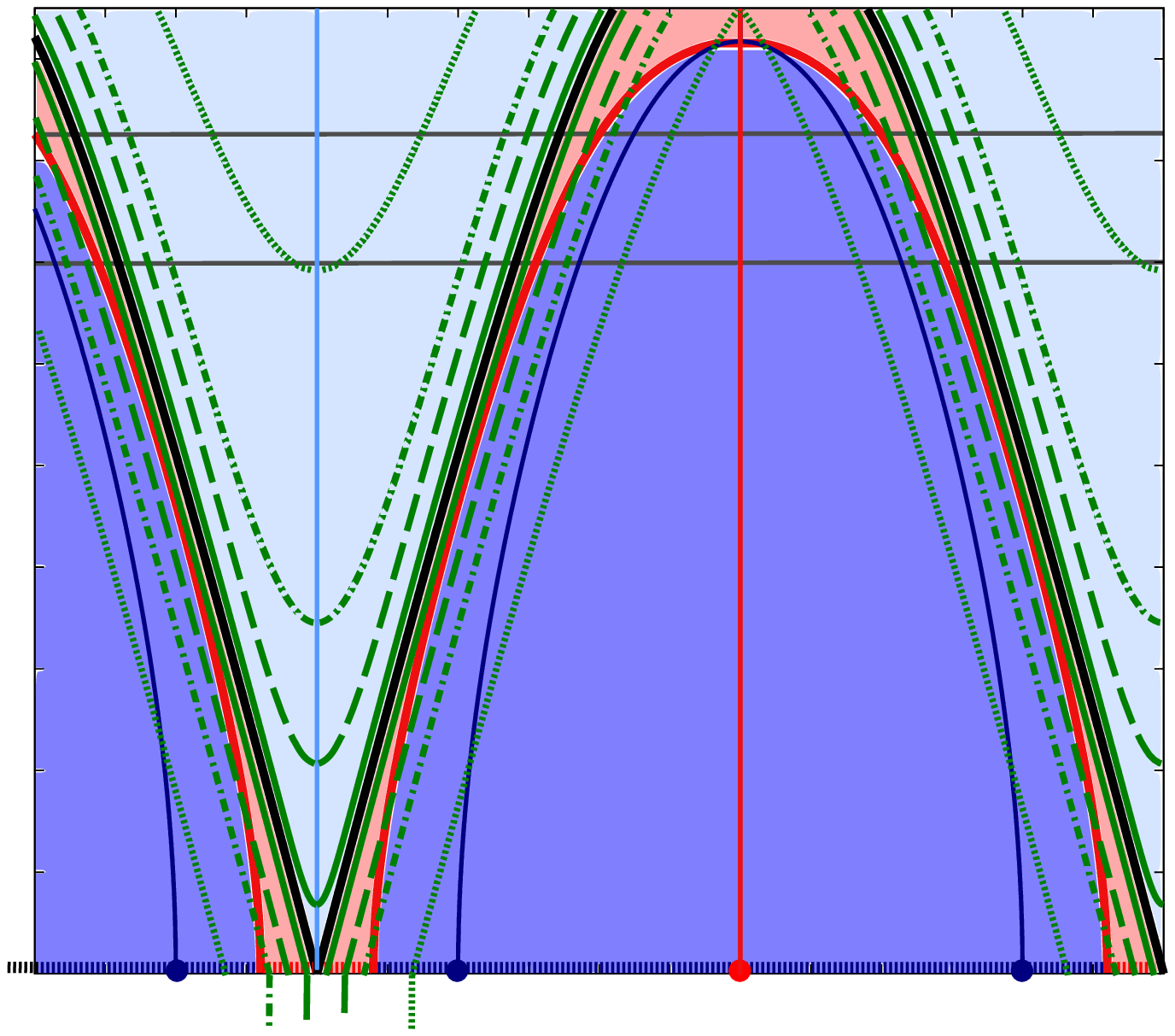}}%
    \put(0.07180714,0.13226814){\color[rgb]{0,0,0}\makebox(0,0)[rt]{\lineheight{1.25}\smash{\begin{tabular}[t]{r}0.1\end{tabular}}}}%
    \put(0.07478204,0.21166629){\color[rgb]{0,0,0}\makebox(0,0)[rt]{\lineheight{1.25}\smash{\begin{tabular}[t]{r}0.2\end{tabular}}}}%
    \put(0.07474587,0.2909487){\color[rgb]{0,0,0}\makebox(0,0)[rt]{\lineheight{1.25}\smash{\begin{tabular}[t]{r}0.3\end{tabular}}}}%
    \put(0.07493576,0.37034684){\color[rgb]{0,0,0}\makebox(0,0)[rt]{\lineheight{1.25}\smash{\begin{tabular}[t]{r}0.4\end{tabular}}}}%
    \put(0.07476396,0.44962925){\color[rgb]{0,0,0}\makebox(0,0)[rt]{\lineheight{1.25}\smash{\begin{tabular}[t]{r}0.5\end{tabular}}}}%
    \put(0.07482725,0.52902739){\color[rgb]{0,0,0}\makebox(0,0)[rt]{\lineheight{1.25}\smash{\begin{tabular}[t]{r}0.6\end{tabular}}}}%
    \put(0.07493576,0.60830981){\color[rgb]{0,0,0}\makebox(0,0)[rt]{\lineheight{1.25}\smash{\begin{tabular}[t]{r}0.7\end{tabular}}}}%
    \put(0.07474587,0.68770796){\color[rgb]{0,0,0}\makebox(0,0)[rt]{\lineheight{1.25}\smash{\begin{tabular}[t]{r}0.8\end{tabular}}}}%
    \put(0.07466449,0.76699036){\color[rgb]{0,0,0}\makebox(0,0)[rt]{\lineheight{1.25}\smash{\begin{tabular}[t]{r}0.9\end{tabular}}}}%
    \put(0.08569137,0.04152679){\color[rgb]{0,0,0}\makebox(0,0)[t]{\lineheight{1.25}\smash{\begin{tabular}[t]{c}-120\end{tabular}}}}%
    \put(0.14078627,0.04152679){\color[rgb]{0,0,0}\makebox(0,0)[t]{\lineheight{1.25}\smash{\begin{tabular}[t]{c}-90\end{tabular}}}}%
    \put(0.19587887,0.04152679){\color[rgb]{0,0,0}\makebox(0,0)[t]{\lineheight{1.25}\smash{\begin{tabular}[t]{c}-60\end{tabular}}}}%
    \put(0.25097145,0.04152679){\color[rgb]{0,0,0}\makebox(0,0)[t]{\lineheight{1.25}\smash{\begin{tabular}[t]{c}-30\end{tabular}}}}%
    \put(0.3086883,0.04152679){\color[rgb]{0,0,0}\makebox(0,0)[t]{\lineheight{1.25}\smash{\begin{tabular}[t]{c}0\end{tabular}}}}%
    \put(0.36384638,0.04152679){\color[rgb]{0,0,0}\makebox(0,0)[t]{\lineheight{1.25}\smash{\begin{tabular}[t]{c}30\end{tabular}}}}%
    \put(0.41881239,0.04152679){\color[rgb]{0,0,0}\makebox(0,0)[t]{\lineheight{1.25}\smash{\begin{tabular}[t]{c}60\end{tabular}}}}%
    \put(0.47392307,0.04152679){\color[rgb]{0,0,0}\makebox(0,0)[t]{\lineheight{1.25}\smash{\begin{tabular}[t]{c}90\end{tabular}}}}%
    \put(0.52845724,0.04152679){\color[rgb]{0,0,0}\makebox(0,0)[t]{\lineheight{1.25}\smash{\begin{tabular}[t]{c}120\end{tabular}}}}%
    \put(0.58343409,0.04152679){\color[rgb]{0,0,0}\makebox(0,0)[t]{\lineheight{1.25}\smash{\begin{tabular}[t]{c}150\end{tabular}}}}%
    \put(0.63852669,0.04152679){\color[rgb]{0,0,0}\makebox(0,0)[t]{\lineheight{1.25}\smash{\begin{tabular}[t]{c}180\end{tabular}}}}%
    \put(0.69421604,0.04152679){\color[rgb]{0,0,0}\makebox(0,0)[t]{\lineheight{1.25}\smash{\begin{tabular}[t]{c}210\end{tabular}}}}%
    \put(0.74930867,0.04152679){\color[rgb]{0,0,0}\makebox(0,0)[t]{\lineheight{1.25}\smash{\begin{tabular}[t]{c}240\end{tabular}}}}%
    \put(0.80440124,0.04152679){\color[rgb]{0,0,0}\makebox(0,0)[t]{\lineheight{1.25}\smash{\begin{tabular}[t]{c}270\end{tabular}}}}%
    \put(0.85956169,0.04152679){\color[rgb]{0,0,0}\makebox(0,0)[t]{\lineheight{1.25}\smash{\begin{tabular}[t]{c}300\end{tabular}}}}%
    \put(0.91465425,0.04152679){\color[rgb]{0,0,0}\makebox(0,0)[t]{\lineheight{1.25}\smash{\begin{tabular}[t]{c}330\end{tabular}}}}%
    \put(0.96974681,0.04152679){\color[rgb]{0,0,0}\makebox(0,0)[t]{\lineheight{1.25}\smash{\begin{tabular}[t]{c}360\end{tabular}}}}%
    \put(0.02502351,0.49623739){\color[rgb]{0,0,0}\rotatebox{90}{\makebox(0,0)[rt]{\lineheight{1.25}\smash{\begin{tabular}[t]{r}$e$\end{tabular}}}}}%
    \put(0.52643116,0.00703667){\color[rgb]{0,0,0}\makebox(0,0)[t]{\lineheight{1.25}\smash{\begin{tabular}[t]{c}$\theta$\end{tabular}}}}%
    \put(0.06298208,0.06117886){\color[rgb]{0,0,0}\rotatebox{-0.4896129}{\makebox(0,0)[rt]{\lineheight{1.25}\smash{\begin{tabular}[t]{r}$e=0$\end{tabular}}}}}%
    \put(0.9660888,0.72288589){\makebox(0,0)[rt]{\lineheight{1.25}\smash{\begin{tabular}[t]{r}Saturn\end{tabular}}}}%
    \put(0.91400547,0.62219145){\makebox(0,0)[rt]{\lineheight{1.25}\smash{\begin{tabular}[t]{r}Mars\end{tabular}}}}%
    \put(0.26017749,0.66150531){\color[rgb]{0,0,0}\rotatebox{-61.541679}{\makebox(0,0)[rt]{\lineheight{1.25}\smash{\begin{tabular}[t]{r}$10R_\rH$\end{tabular}}}}}%
    \put(0.26165908,0.42572707){\color[rgb]{0,0,0}\rotatebox{-69.040207}{\makebox(0,0)[rt]{\lineheight{1.25}\smash{\begin{tabular}[t]{r}$5R_\rH$\end{tabular}}}}}%
    \put(0.28596453,0.26600527){\color[rgb]{0,0,0}\rotatebox{-69.040207}{\makebox(0,0)[rt]{\lineheight{1.25}\smash{\begin{tabular}[t]{r}$3R_\rH$\end{tabular}}}}}%
    \put(0.29431787,0.1614759){\color[rgb]{0,0,0}\rotatebox{-75.106295}{\makebox(0,0)[rt]{\lineheight{1.25}\smash{\begin{tabular}[t]{r}$R_\rH$\end{tabular}}}}}%
    \put(0.27549029,0.01237063){\color[rgb]{0,0,0}\rotatebox{-0.228311}{\makebox(0,0)[lt]{\lineheight{1.25}\smash{\begin{tabular}[t]{l}$R_\rH$\end{tabular}}}}}%
    \put(0.31905886,0.01237589){\color[rgb]{0,0,0}\rotatebox{-0.228311}{\makebox(0,0)[lt]{\lineheight{1.25}\smash{\begin{tabular}[t]{l}$3R_\rH$\end{tabular}}}}}%
    \put(0.27137133,0.0162029){\color[rgb]{0,0,0}\rotatebox{-0.22831071}{\makebox(0,0)[rt]{\lineheight{1.25}\smash{\begin{tabular}[t]{r}$5R_\rH$\end{tabular}}}}}%
    \put(0.38207141,0.0163826){\color[rgb]{0,0,0}\rotatebox{-0.228311}{\makebox(0,0)[lt]{\lineheight{1.25}\smash{\begin{tabular}[t]{l}$10R_\rH$\end{tabular}}}}}%
  \end{picture}%
\endgroup%
		\def\svgwidth{0.49\textwidth}
\begingroup%
  \makeatletter%
  \providecommand\color[2][]{%
    \errmessage{(Inkscape) Color is used for the text in Inkscape, but the package 'color.sty' is not loaded}%
    \renewcommand\color[2][]{}%
  }%
  \providecommand\transparent[1]{%
    \errmessage{(Inkscape) Transparency is used (non-zero) for the text in Inkscape, but the package 'transparent.sty' is not loaded}%
    \renewcommand\transparent[1]{}%
  }%
  \providecommand\rotatebox[2]{#2}%
  \newcommand*\fsize{\dimexpr\f@size pt\relax}%
  \newcommand*\lineheight[1]{\fontsize{\fsize}{#1\fsize}\selectfont}%
  \ifx\svgwidth\undefined%
    \setlength{\unitlength}{432bp}%
    \ifx\svgscale\undefined%
      \relax%
    \else%
      \setlength{\unitlength}{\unitlength * \real{\svgscale}}%
    \fi%
  \else%
    \setlength{\unitlength}{\svgwidth}%
  \fi%
  \global\let\svgwidth\undefined%
  \global\let\svgscale\undefined%
  \makeatother%
  \begin{picture}(1,0.83333333)%
    \lineheight{1}%
    \setlength\tabcolsep{0pt}%
    \put(0,0){\includegraphics[width=\unitlength]{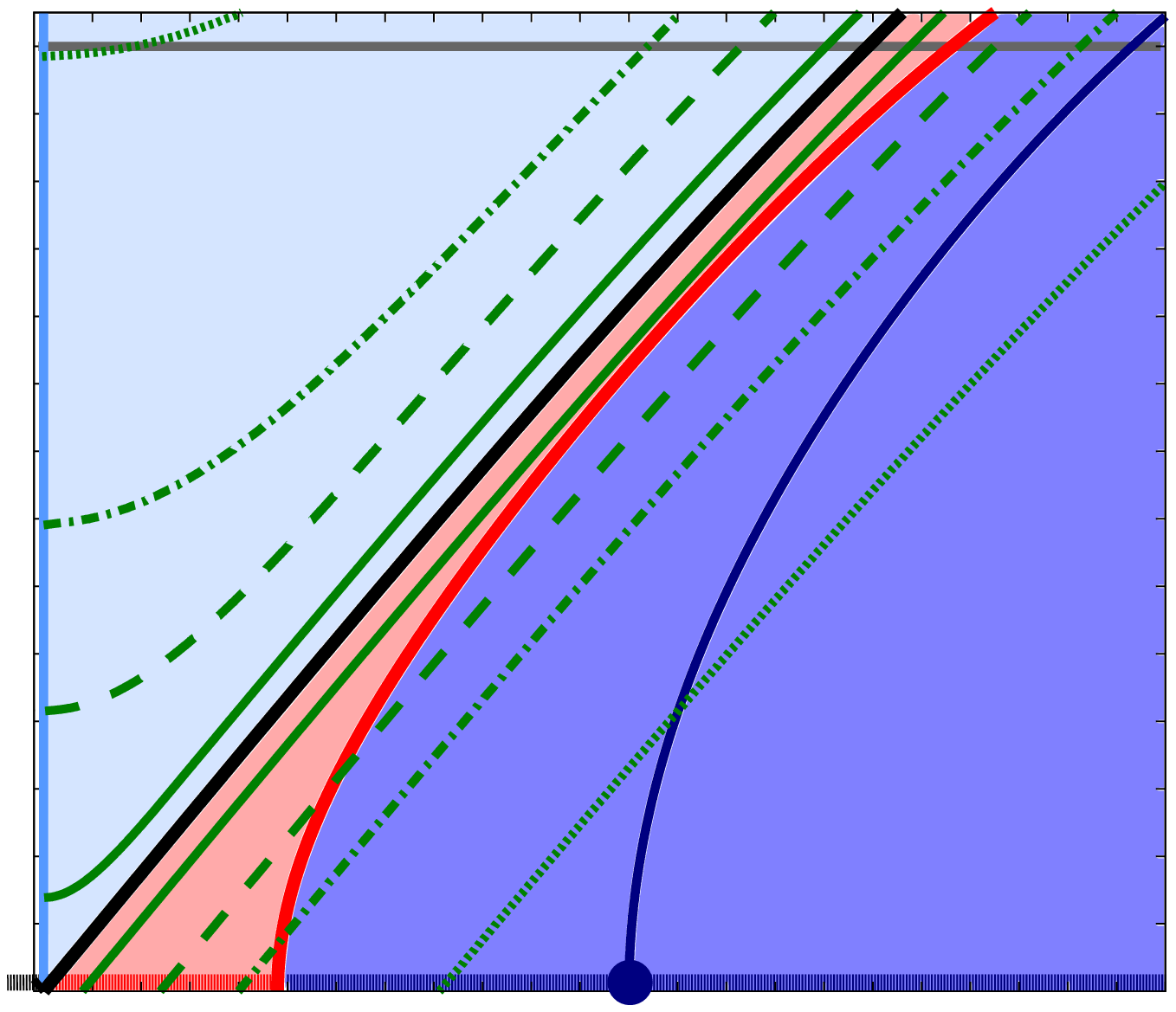}}%
    \put(0.08506026,0.10495333){\color[rgb]{0,0,0}\makebox(0,0)[rt]{\lineheight{1.25}\smash{\begin{tabular}[t]{r}0.05\end{tabular}}}}%
    \put(0.08210807,0.15692092){\color[rgb]{0,0,0}\makebox(0,0)[rt]{\lineheight{1.25}\smash{\begin{tabular}[t]{r}0.1\end{tabular}}}}%
    \put(0.08506026,0.20888851){\color[rgb]{0,0,0}\makebox(0,0)[rt]{\lineheight{1.25}\smash{\begin{tabular}[t]{r}0.15\end{tabular}}}}%
    \put(0.08508297,0.2608561){\color[rgb]{0,0,0}\makebox(0,0)[rt]{\lineheight{1.25}\smash{\begin{tabular}[t]{r}0.2\end{tabular}}}}%
    \put(0.08506026,0.31293944){\color[rgb]{0,0,0}\makebox(0,0)[rt]{\lineheight{1.25}\smash{\begin{tabular}[t]{r}0.25\end{tabular}}}}%
    \put(0.0850468,0.36490703){\color[rgb]{0,0,0}\makebox(0,0)[rt]{\lineheight{1.25}\smash{\begin{tabular}[t]{r}0.3\end{tabular}}}}%
    \put(0.08506026,0.41687462){\color[rgb]{0,0,0}\makebox(0,0)[rt]{\lineheight{1.25}\smash{\begin{tabular}[t]{r}0.35\end{tabular}}}}%
    \put(0.08523669,0.46884221){\color[rgb]{0,0,0}\makebox(0,0)[rt]{\lineheight{1.25}\smash{\begin{tabular}[t]{r}0.4\end{tabular}}}}%
    \put(0.08506026,0.5208098){\color[rgb]{0,0,0}\makebox(0,0)[rt]{\lineheight{1.25}\smash{\begin{tabular}[t]{r}0.45\end{tabular}}}}%
    \put(0.08506488,0.57277739){\color[rgb]{0,0,0}\makebox(0,0)[rt]{\lineheight{1.25}\smash{\begin{tabular}[t]{r}0.5\end{tabular}}}}%
    \put(0.08506026,0.62474499){\color[rgb]{0,0,0}\makebox(0,0)[rt]{\lineheight{1.25}\smash{\begin{tabular}[t]{r}0.55\end{tabular}}}}%
    \put(0.08512818,0.67671258){\color[rgb]{0,0,0}\makebox(0,0)[rt]{\lineheight{1.25}\smash{\begin{tabular}[t]{r}0.6\end{tabular}}}}%
    \put(0.08506026,0.72868018){\color[rgb]{0,0,0}\makebox(0,0)[rt]{\lineheight{1.25}\smash{\begin{tabular}[t]{r}0.65\end{tabular}}}}%
    \put(0.08523669,0.78064777){\color[rgb]{0,0,0}\makebox(0,0)[rt]{\lineheight{1.25}\smash{\begin{tabular}[t]{r}0.7\end{tabular}}}}%
    \put(0.106142,0.03458296){\color[rgb]{0,0,0}\makebox(0,0)[t]{\lineheight{1.25}\smash{\begin{tabular}[t]{c}0\end{tabular}}}}%
    \put(0.18065863,0.03458296){\color[rgb]{0,0,0}\makebox(0,0)[t]{\lineheight{1.25}\smash{\begin{tabular}[t]{c}10\end{tabular}}}}%
    \put(0.25625543,0.03458296){\color[rgb]{0,0,0}\makebox(0,0)[t]{\lineheight{1.25}\smash{\begin{tabular}[t]{c}20\end{tabular}}}}%
    \put(0.33143898,0.03458296){\color[rgb]{0,0,0}\makebox(0,0)[t]{\lineheight{1.25}\smash{\begin{tabular}[t]{c}30\end{tabular}}}}%
    \put(0.40654116,0.03458296){\color[rgb]{0,0,0}\makebox(0,0)[t]{\lineheight{1.25}\smash{\begin{tabular}[t]{c}40\end{tabular}}}}%
    \put(0.4815936,0.03458296){\color[rgb]{0,0,0}\makebox(0,0)[t]{\lineheight{1.25}\smash{\begin{tabular}[t]{c}50\end{tabular}}}}%
    \put(0.55665961,0.03458296){\color[rgb]{0,0,0}\makebox(0,0)[t]{\lineheight{1.25}\smash{\begin{tabular}[t]{c}60\end{tabular}}}}%
    \put(0.63166865,0.03458296){\color[rgb]{0,0,0}\makebox(0,0)[t]{\lineheight{1.25}\smash{\begin{tabular}[t]{c}70\end{tabular}}}}%
    \put(0.70682508,0.03458296){\color[rgb]{0,0,0}\makebox(0,0)[t]{\lineheight{1.25}\smash{\begin{tabular}[t]{c}80\end{tabular}}}}%
    \put(0.78190917,0.03458296){\color[rgb]{0,0,0}\makebox(0,0)[t]{\lineheight{1.25}\smash{\begin{tabular}[t]{c}90\end{tabular}}}}%
    \put(0.85646649,0.03458296){\color[rgb]{0,0,0}\makebox(0,0)[t]{\lineheight{1.25}\smash{\begin{tabular}[t]{c}100\end{tabular}}}}%
    \put(0.93158223,0.03458296){\color[rgb]{0,0,0}\makebox(0,0)[t]{\lineheight{1.25}\smash{\begin{tabular}[t]{c}110\end{tabular}}}}%
    \put(0.02502351,0.37980323){\color[rgb]{0,0,0}\rotatebox{90}{\makebox(0,0)[lt]{\lineheight{1.25}\smash{\begin{tabular}[t]{l}$e$\end{tabular}}}}}%
    \put(0.53163949,0.00703667){\color[rgb]{0,0,0}\makebox(0,0)[t]{\lineheight{1.25}\smash{\begin{tabular}[t]{c}$\theta$\end{tabular}}}}%
    \put(0.12783586,0.78636593){\rotatebox{11.071346}{\makebox(0,0)[lt]{\lineheight{1.25}\smash{\begin{tabular}[t]{l}$10R_\rH$\end{tabular}}}}}%
    \put(0.526268,0.7570324){\rotatebox{45.618525}{\makebox(0,0)[rt]{\lineheight{1.25}\smash{\begin{tabular}[t]{r}$5R_\rH$\end{tabular}}}}}%
    \put(0.59918479,0.75356005){\rotatebox{45.618525}{\makebox(0,0)[rt]{\lineheight{1.25}\smash{\begin{tabular}[t]{r}$3R_\rH$\end{tabular}}}}}%
    \put(0.6616849,0.75355995){\rotatebox{45.618525}{\makebox(0,0)[rt]{\lineheight{1.25}\smash{\begin{tabular}[t]{r}$R_\rH$\end{tabular}}}}}%
    \put(0.26891721,0.08845421){\rotatebox{49.439721}{\makebox(0,0)[lt]{\lineheight{1.25}\smash{\begin{tabular}[t]{l}$5R_\rH$\end{tabular}}}}}%
    \put(0.421696,0.08845477){\rotatebox{49.439721}{\makebox(0,0)[lt]{\lineheight{1.25}\smash{\begin{tabular}[t]{l}$10R_\rH$\end{tabular}}}}}%
    \put(0.20988929,0.08845429){\rotatebox{49.439721}{\makebox(0,0)[lt]{\lineheight{1.25}\smash{\begin{tabular}[t]{l}$3R_\rH$\end{tabular}}}}}%
    \put(0.15086138,0.08845437){\rotatebox{49.439721}{\makebox(0,0)[lt]{\lineheight{1.25}\smash{\begin{tabular}[t]{l}$R_\rH$\end{tabular}}}}}%
    \put(0.07464359,0.06328667){\color[rgb]{0,0,0}\makebox(0,0)[rt]{\lineheight{1.25}\smash{\begin{tabular}[t]{r}$e=0$\end{tabular}}}}%
    \put(0.34589901,0.7945287){\makebox(0,0)[lt]{\lineheight{1.25}\smash{\begin{tabular}[t]{l}Mars\end{tabular}}}}%
  \end{picture}%
\endgroup%
		\caption{(Left panel) Same figure as Fig.~\ref{fig:5}. 
			The dashed green curves correspond to the element $(\theta_i,e_i)$
			for which the minimal mutual distance $\Delta$ is equal to $\rN_\eps R_\rH$ 
			where $R_\rH$ denotes the Hill's radius.
		The grey lines correspond to the initial condition 
			for which the trajectory crosses the orbit of Mars
		and Saturn. (Right panel) Enlargement in the neighborhood of the collision curve.}
\label{fig:6}
	\end{center}
\end{figure*}			

		First of all, we summarize the situation.
		 Six families of periodic orbits 
			($f$, $\sL_3$, $\sL_j^s$ and $\sL_j^l$) 
		and three types of trajectories 
			(tadpole, horseshoe and quasi-satellite motion)
		described in Sect.~\ref{sec:rem_sol}
		have been recovered in the averaged problem close to the exact mean-motion resonance $u=0$.
		Notice that the domain of ‘‘satellized" retrograde satellite mentioned in Sect.~\ref{sec:rem_sol} is missing
			since it is located in the neighborhood of the family $f$ 
			that belongs to the Hill's sphere (see \cite{2017PoRoVi} for more details).

		Each domain of co-orbital motion
			 extends quasi-symmetrically with respect to the exact mean-motion resonance $u=0$
			 and is neatly defined by 
				the collision curves 
				or the separatrices that originate from the hyperbolic equilibria associated with $\sL_3$.	
		In this section, we construct a ‘‘map" of the co-orbital motion in the circular-case,
			that is, a representation of  the section $\{u=0\}$
			which can be used in order to discuss about the stability of the solutions
			as well as to compute co-orbital trajectories in the synodic reference frame.
		Two parameters are required to identify a solution of the averaged problem
			that belongs to the section $\{u=0\}$:
			the resonant angle $\theta_i$ and 
			the eccentricity of the orbit $e_i$, 
				(we recall that $e_i = e_0$ when $u=0$).		
		Hence, 
			we compute the evolution of the  cross sections of the boundaries of each domain
			(separatrix and collision curves) 
			by varying the eccentricity $e_i$.
		To that end,  we consider the following reduction of the reduced averaged Hamiltonian:
		\bes
			\Hb_0^{K(e_0)}(\theta, u) = -\frac{3}{2}u^2 + \Hb_P(\theta, 0, \xt(K), x(K))
		\ees
		which is derived from
			the Taylor expansions of $\Hb_\rP$ and $H_\rK$, 
				respectively at zero and second order in ${u=0}$.
		This reduction is reliable in the vicinity of the section ${\{u=0\}}$ and 
		has the advantage to be independent of $\eps$,
			under the following rescaling: ${\eps^{-1}\Hb_0(\sqrt{\eps}u, \theta, \xt, x)}$.
		Hence, a ‘‘map" of the section ${\{u=0\}}$ computed through $\Hb_0^{K(e_0)}$,
			is invariant under the variation of $\eps$.

		Fig.~\ref{fig:5} displays the map of the co-orbital motion in the circular-planar case.
		The black thick curves illustrate the collision with the planet.
		They can be approximated by $\modu{\theta_i}= 2 e_i \times 180\degre/\pi$ up to high eccentricities.	
		They bound the sky blue region associated with the quasi-satellite motion.
		The red thick curves depict the crossings of the separatrices that originate from $\sL_3$.
		It divides the blue and red regions of tadpole and horseshoe motion.
		Let us mention that for $e_i=0$, 
			only two domains of co-orbital motion exist: 
			$L_4$, $L_5$ and the long-periodic families $\sL_4^l$ and  $\sL_5^l$ form the tadpole region
			while the other one corresponds to the periodic horseshoe-shaped trajectories.
		The left panel of the Fig.~\ref{fig:6} displays the map of the co-orbital motion with the 
			elements $(\theta_i, e_i)$ for which the minimal mutual distance $\Delta$ is equal to ${NR_\rH}$,
			for a Sun-Jupiter like system (${\eps=1/1000}$).
		The two additional grey lines represent the elements $(\theta_i, e_i)$
			for which the particle crosses the orbit of Mars and Saturn.
		Hence, they suggest the maximal value of eccentricity 
		for which the solutions of the restricted three-body problem are reliable
			in order to describe the real motion in the Solar System.
		The right panel of Fig.~\ref{fig:6}  is an enlargement of the map on the region that surround the collision curve.
		
\begin{sloppypar}
		According to Theorem \ref{theo:averaging} and Theorem \ref{theo:stability}, 
			for a given number ${1<N\leq \eps^{-1/3}}$, and a given type of co-orbital dynamics,
			we can define a set of elements $(\theta_i, e_i)$ 
		which satisfy a mutual distance greater than $\Delta = NR_\rH$,
		and for which the time of stability of the solutions of the averaged problem is at least $\sqrt{N}^{3}$ 
		revolutions of Jupiter.
		In the synodic reference frame, 
			the couple $(\theta_i, e_i)$ provides a set of initial conditions,
				parametrized by $\varpi_i\in\TT$ 
				and that can be written as
		\be
		\begin{aligned}
			\phi_i 		&= \theta_i 					+ G_1(e_i, \theta_i - \varpi_i)\\
			R_i 			&= 1 - e_i \cos\Big(\theta_i - \varpi_i 	+ G_2(e_i, \theta_i - \varpi_i)\Big)\\
			\dot{\bR}_i 	&= \left(R_i - \frac{1}{\sqrt{1-e_i^2}}\right)
							\begin{pmatrix}\sin \phi_i\\ 
								-\cos \phi_i
							\end{pmatrix}\\
						&\phantom{=} + \frac{e_i}{\sqrt{1-e_i^2}}
							\begin{pmatrix}
								\sin \varpi_i\\ 
								-\cos \varpi_i
							\end{pmatrix}
		\end{aligned}
\label{eq:CI}
		\ee
		Hence, Theorem \ref{theo:stability} ensures that
			 an initial condition, given by Eq.~\eqref{eq:CI},
				provides a co-orbital trajectory of the same type, at least for a finite time.
		In other words, 
			transitions to another co-orbital motion or escapes from the 1:1 mean-motion resonance
			can not occur
				at least during a time $\cT=2\pi \sqrt{N^3}$.
		For instance, for $N$ equal to $3$, $5$ and $10$, 
			it ensures a time of stability corresponding approximately  to 5, 10 and 30 revolutions of Jupiter, 
			that is, more than 50, 125 and 350 years.
\end{sloppypar}

		As mentioned at the end of the previous section,
			the quasi-satellite and the horseshoe domains intersect the exclusion zone of the averaged problem.
		More precisely, for high values of $e_i$, 
			the quasi-satellite motion dominates the map
			and the size of the intersection between the quasi-satellite domain 
			and the exclusion zone is small relatively to the whole domain.
		By decreasing $e_i$, since the quasi-satellite domain shrinks with the collision curve,
			the relative size of the intersection increases until a critical value 
			for which the exclusion zone contains all the quasi-satellite orbits.
		In the case of a Sun-Jupiter like system, this critical value occurs for $e_i \simeq 0.07$.

		Notice that Pousse et al. \cite{2017PoRoVi} suggested, 
			through a frequency analysis of the family $f$, 
			 a critical value of $e_i\simeq 0.18$ for the quasi-satellite orbits.
		This value was given by an arbitrary criterion
			for which a solution of the averaged problem
			is considered outside the exclusion zone, 
			if the modulus of their fundamental frequencies $\nu$ and $g$ are lower than $\dot{\lam}'/4$,
			where $\dot{\lam}'=1$ denotes the frequency of averaging.
		Theorem \ref{theo:averaging} provides a lower value of eccentricity and thus a larger domain of validity
			of the averaged Hamiltonian for the quasi-satellite motion.
		However, based on the results in \cite{2017PoRoVi}, 
			the quasi-satellite region that surround the Hill's sphere
		is probably overlaped by secondary resonances,
			that is, the resonant structures generated by commensurabilities
			between frequencies $\nu$, $g$ and $\dot{\lam}'=1$,
			and especially between $\nu$ and $1-g$ due to the D'Alembert rules (see Sect.~\ref{sec:AP}).
		In particular, 
			it has been shown in this area (see, e.g., \cite{2017PoRoVi}) that  
		the neighborhood of the family $f$
			is divided in three disjoint regions by two critical orbits of the family $f$ associated with 
			the commensurabilty $3\nu = 1-g$.
		As a consequence, a global study of the frequencies in the averaged problem
			will probably reveal the resonant structures that destabilizes the quasi-satellite region that surround the Hill's sphere,
			and may also highlight some islands of quasi-satellite solutions for which the time of stability is larger 
			than the one given by Theorem \ref{theo:stability}. 
			
		The horseshoe motion exists outside the exclusion zone for low and very high eccentricities.
		More precisely, by increasing $e_i$, the size of the intersection between the horseshoe domain and
			the exclusion zone increases until a critical value $e_i\simeq 0.4$ for which
			all the horseshoe-shaped trajectories cross the section $\{u=0\}$ inside the Hill's sphere.
		Furthermore, there exists another critical value $e_i \simeq 0.6$
			for which a part of the horseshoe domain goes outside the exclusion zone.
		Then, for increasing $e_i$, the relative size of the intersection between the horseshoe domain and
			the exclusion zone decreases.
			
		Most of the solutions in horseshoe motion experiences closed encounters with Jupiter 
		(less than 5 Hill's radius),
			and thus have a relatively small time of stability according to Theorem \ref{theo:stability}.
		Similarly to the quasi-satellite region located at the edge of the Hill's sphere, 
			the horseshoe region is probably overlaped by secondary resonances
			which destabilize the domain.
		A global frequency analysis of the region may reveal these resonant structures.	
			
	\subsection{Conclusions}

		In this paper, we showed that the averaged problem provides another approach 
			in order to study some families of periodic orbits of the restricted three-body problem.
		More precisely, we proved that it is a valid approximation  in a particular area of the phase space 
			that focuses on mean-motion resonances.
		Through a rigorous treatment, we characterized  the domain of validity of the averaged problem		
		(Theo.~\ref{theo:averaging})
			 and proved that it is a  reliable approximation 
			 as long as the considered trajectories lay outside the Hill's sphere of the planet.
		A new result of stability over finite times has also been proved (Theo.~\ref{theo:stability}).
		As a consequence, we provided a rigorous justification 
			of the relevance of the  averaged problem
			to study some specific solutions of the restricted three-body problem.
	
\begin{figure}[!h]
	\begin{center}
		\tiny
		\def\svgwidth{0.49\textwidth}
\begingroup%
  \makeatletter%
  \providecommand\color[2][]{%
    \errmessage{(Inkscape) Color is used for the text in Inkscape, but the package 'color.sty' is not loaded}%
    \renewcommand\color[2][]{}%
  }%
  \providecommand\transparent[1]{%
    \errmessage{(Inkscape) Transparency is used (non-zero) for the text in Inkscape, but the package 'transparent.sty' is not loaded}%
    \renewcommand\transparent[1]{}%
  }%
  \providecommand\rotatebox[2]{#2}%
  \newcommand*\fsize{\dimexpr\f@size pt\relax}%
  \newcommand*\lineheight[1]{\fontsize{\fsize}{#1\fsize}\selectfont}%
  \ifx\svgwidth\undefined%
    \setlength{\unitlength}{432bp}%
    \ifx\svgscale\undefined%
      \relax%
    \else%
      \setlength{\unitlength}{\unitlength * \real{\svgscale}}%
    \fi%
  \else%
    \setlength{\unitlength}{\svgwidth}%
  \fi%
  \global\let\svgwidth\undefined%
  \global\let\svgscale\undefined%
  \makeatother%
  \begin{picture}(1,0.83333333)%
    \lineheight{1}%
    \setlength\tabcolsep{0pt}%
    \put(0,0){\includegraphics[width=\unitlength]{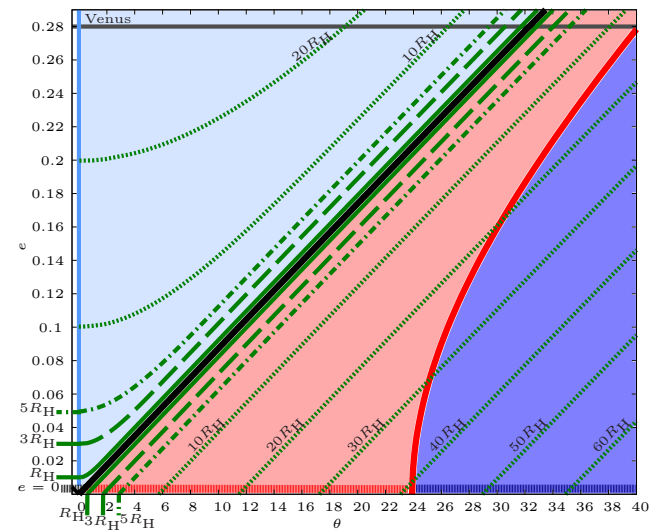}}%
    \put(0.08507834,0.10495333){\color[rgb]{0,0,0}\makebox(0,0)[rt]{\lineheight{1.25}\smash{\begin{tabular}[t]{r}0.02\end{tabular}}}}%
    \put(0.08523206,0.15692092){\color[rgb]{0,0,0}\makebox(0,0)[rt]{\lineheight{1.25}\smash{\begin{tabular}[t]{r}0.04\end{tabular}}}}%
    \put(0.08512355,0.20888851){\color[rgb]{0,0,0}\makebox(0,0)[rt]{\lineheight{1.25}\smash{\begin{tabular}[t]{r}0.06\end{tabular}}}}%
    \put(0.08504217,0.2608561){\color[rgb]{0,0,0}\makebox(0,0)[rt]{\lineheight{1.25}\smash{\begin{tabular}[t]{r}0.08\end{tabular}}}}%
    \put(0.08210807,0.31293944){\color[rgb]{0,0,0}\makebox(0,0)[rt]{\lineheight{1.25}\smash{\begin{tabular}[t]{r}0.1\end{tabular}}}}%
    \put(0.08507834,0.36490703){\color[rgb]{0,0,0}\makebox(0,0)[rt]{\lineheight{1.25}\smash{\begin{tabular}[t]{r}0.12\end{tabular}}}}%
    \put(0.08523206,0.41687462){\color[rgb]{0,0,0}\makebox(0,0)[rt]{\lineheight{1.25}\smash{\begin{tabular}[t]{r}0.14\end{tabular}}}}%
    \put(0.08512355,0.46884221){\color[rgb]{0,0,0}\makebox(0,0)[rt]{\lineheight{1.25}\smash{\begin{tabular}[t]{r}0.16\end{tabular}}}}%
    \put(0.08504217,0.5208098){\color[rgb]{0,0,0}\makebox(0,0)[rt]{\lineheight{1.25}\smash{\begin{tabular}[t]{r}0.18\end{tabular}}}}%
    \put(0.08508297,0.57277739){\color[rgb]{0,0,0}\makebox(0,0)[rt]{\lineheight{1.25}\smash{\begin{tabular}[t]{r}0.2\end{tabular}}}}%
    \put(0.08507834,0.62474499){\color[rgb]{0,0,0}\makebox(0,0)[rt]{\lineheight{1.25}\smash{\begin{tabular}[t]{r}0.22\end{tabular}}}}%
    \put(0.08523206,0.67671258){\color[rgb]{0,0,0}\makebox(0,0)[rt]{\lineheight{1.25}\smash{\begin{tabular}[t]{r}0.24\end{tabular}}}}%
    \put(0.08512355,0.72868018){\color[rgb]{0,0,0}\makebox(0,0)[rt]{\lineheight{1.25}\smash{\begin{tabular}[t]{r}0.26\end{tabular}}}}%
    \put(0.08504217,0.78064777){\color[rgb]{0,0,0}\makebox(0,0)[rt]{\lineheight{1.25}\smash{\begin{tabular}[t]{r}0.28\end{tabular}}}}%
    \put(0.10491587,0.03458296){\color[rgb]{0,0,0}\makebox(0,0)[lt]{\lineheight{1.25}\smash{\begin{tabular}[t]{l}0\end{tabular}}}}%
    \put(0.14797142,0.03458296){\color[rgb]{0,0,0}\makebox(0,0)[lt]{\lineheight{1.25}\smash{\begin{tabular}[t]{l}2\end{tabular}}}}%
    \put(0.19091124,0.03458296){\color[rgb]{0,0,0}\makebox(0,0)[lt]{\lineheight{1.25}\smash{\begin{tabular}[t]{l}4\end{tabular}}}}%
    \put(0.2339668,0.03458296){\color[rgb]{0,0,0}\makebox(0,0)[lt]{\lineheight{1.25}\smash{\begin{tabular}[t]{l}6\end{tabular}}}}%
    \put(0.27702235,0.03458296){\color[rgb]{0,0,0}\makebox(0,0)[lt]{\lineheight{1.25}\smash{\begin{tabular}[t]{l}8\end{tabular}}}}%
    \put(0.31481171,0.03458296){\color[rgb]{0,0,0}\makebox(0,0)[lt]{\lineheight{1.25}\smash{\begin{tabular}[t]{l}10\end{tabular}}}}%
    \put(0.35786726,0.03458296){\color[rgb]{0,0,0}\makebox(0,0)[lt]{\lineheight{1.25}\smash{\begin{tabular}[t]{l}12\end{tabular}}}}%
    \put(0.40080707,0.03458296){\color[rgb]{0,0,0}\makebox(0,0)[lt]{\lineheight{1.25}\smash{\begin{tabular}[t]{l}14\end{tabular}}}}%
    \put(0.44386263,0.03458296){\color[rgb]{0,0,0}\makebox(0,0)[lt]{\lineheight{1.25}\smash{\begin{tabular}[t]{l}16\end{tabular}}}}%
    \put(0.48691818,0.03458296){\color[rgb]{0,0,0}\makebox(0,0)[lt]{\lineheight{1.25}\smash{\begin{tabular}[t]{l}18\end{tabular}}}}%
    \put(0.529858,0.03458296){\color[rgb]{0,0,0}\makebox(0,0)[lt]{\lineheight{1.25}\smash{\begin{tabular}[t]{l}20\end{tabular}}}}%
    \put(0.57291355,0.03458296){\color[rgb]{0,0,0}\makebox(0,0)[lt]{\lineheight{1.25}\smash{\begin{tabular}[t]{l}22\end{tabular}}}}%
    \put(0.6159691,0.03458296){\color[rgb]{0,0,0}\makebox(0,0)[lt]{\lineheight{1.25}\smash{\begin{tabular}[t]{l}24\end{tabular}}}}%
    \put(0.65890892,0.03458296){\color[rgb]{0,0,0}\makebox(0,0)[lt]{\lineheight{1.25}\smash{\begin{tabular}[t]{l}26\end{tabular}}}}%
    \put(0.70196447,0.03458296){\color[rgb]{0,0,0}\makebox(0,0)[lt]{\lineheight{1.25}\smash{\begin{tabular}[t]{l}28\end{tabular}}}}%
    \put(0.74502003,0.03458296){\color[rgb]{0,0,0}\makebox(0,0)[lt]{\lineheight{1.25}\smash{\begin{tabular}[t]{l}30\end{tabular}}}}%
    \put(0.78795984,0.03458296){\color[rgb]{0,0,0}\makebox(0,0)[lt]{\lineheight{1.25}\smash{\begin{tabular}[t]{l}32\end{tabular}}}}%
    \put(0.8310154,0.03458296){\color[rgb]{0,0,0}\makebox(0,0)[lt]{\lineheight{1.25}\smash{\begin{tabular}[t]{l}34\end{tabular}}}}%
    \put(0.87407095,0.03458296){\color[rgb]{0,0,0}\makebox(0,0)[lt]{\lineheight{1.25}\smash{\begin{tabular}[t]{l}36\end{tabular}}}}%
    \put(0.91701076,0.03458296){\color[rgb]{0,0,0}\makebox(0,0)[lt]{\lineheight{1.25}\smash{\begin{tabular}[t]{l}38\end{tabular}}}}%
    \put(0.96006632,0.03458296){\color[rgb]{0,0,0}\makebox(0,0)[lt]{\lineheight{1.25}\smash{\begin{tabular}[t]{l}40\end{tabular}}}}%
    \put(0.02502351,0.438831){\color[rgb]{0,0,0}\rotatebox{90}{\makebox(0,0)[lt]{\lineheight{1.25}\smash{\begin{tabular}[t]{l}$e$\end{tabular}}}}}%
    \put(0.49890045,0.00703667){\color[rgb]{0,0,0}\makebox(0,0)[lt]{\lineheight{1.25}\smash{\begin{tabular}[t]{l}$\theta$\end{tabular}}}}%
    \put(0.9081413,0.12367623){\rotatebox{47.420299}{\makebox(0,0)[lt]{\lineheight{1.25}\smash{\begin{tabular}[t]{l}$60R_\rH$\end{tabular}}}}}%
    \put(0.77967165,0.12367737){\rotatebox{47.420299}{\makebox(0,0)[lt]{\lineheight{1.25}\smash{\begin{tabular}[t]{l}$50R_\rH$\end{tabular}}}}}%
    \put(0.65467416,0.12367848){\rotatebox{47.420299}{\makebox(0,0)[lt]{\lineheight{1.25}\smash{\begin{tabular}[t]{l}$40R_\rH$\end{tabular}}}}}%
    \put(0.52967666,0.12367959){\rotatebox{47.420299}{\makebox(0,0)[lt]{\lineheight{1.25}\smash{\begin{tabular}[t]{l}$30R_\rH$\end{tabular}}}}}%
    \put(0.40467917,0.1236807){\rotatebox{47.420299}{\makebox(0,0)[lt]{\lineheight{1.25}\smash{\begin{tabular}[t]{l}$20R_\rH$\end{tabular}}}}}%
    \put(0.28315382,0.12368178){\rotatebox{47.420299}{\makebox(0,0)[lt]{\lineheight{1.25}\smash{\begin{tabular}[t]{l}$10R_\rH$\end{tabular}}}}}%
    \put(0.66236254,0.78113129){\rotatebox{47.420299}{\makebox(0,0)[rt]{\lineheight{1.25}\smash{\begin{tabular}[t]{r}$10R_\rH$\end{tabular}}}}}%
    \put(0.49770146,0.77723576){\rotatebox{38.757107}{\makebox(0,0)[rt]{\lineheight{1.25}\smash{\begin{tabular}[t]{r}$20R_\rH$\end{tabular}}}}}%
    \put(0.16984722,0.01479334){\makebox(0,0)[lt]{\lineheight{1.25}\smash{\begin{tabular}[t]{l}$5R_\rH$\end{tabular}}}}%
    \put(0.14474256,0.01132112){\makebox(0,0)[t]{\lineheight{1.25}\smash{\begin{tabular}[t]{c}$3R_\rH$\end{tabular}}}}%
    \put(0.11878792,0.02521001){\makebox(0,0)[rt]{\lineheight{1.25}\smash{\begin{tabular}[t]{r}$R_\rH$\end{tabular}}}}%
    \put(0.07364903,0.08771001){\makebox(0,0)[rt]{\lineheight{1.25}\smash{\begin{tabular}[t]{r}$R_\rH$\end{tabular}}}}%
    \put(0.07364903,0.13979334){\makebox(0,0)[rt]{\lineheight{1.25}\smash{\begin{tabular}[t]{r}$3R_\rH$\end{tabular}}}}%
    \put(0.07364903,0.18840445){\makebox(0,0)[rt]{\lineheight{1.25}\smash{\begin{tabular}[t]{r}$5R_\rH$\end{tabular}}}}%
    \put(0.11627865,0.79290558){\makebox(0,0)[lt]{\lineheight{1.25}\smash{\begin{tabular}[t]{l}Venus\end{tabular}}}}%
    \put(0.07831904,0.06529963){\color[rgb]{0,0,0}\makebox(0,0)[rt]{\lineheight{1.25}\smash{\begin{tabular}[t]{r}$e=0$\end{tabular}}}}%
  \end{picture}%
\endgroup%

		\caption{Map of the co-orbital motion for a Sun-Earth like system.}
\label{fig:7}
	\end{center}
\end{figure}

		Our theoretical results allowed us to understand the co-orbital motion
			(1:1 mean-motion resonance),
			that is, the quasi-satellite, the tadpole and the horseshoe orbits,
			that comprise
		 		the family $f$, 
				the short-periodic and long-periodic families that originate from $L_4$ and $L_5$,
				and the Lyapunov family associated with $L_3$. 
		In particular, in the framework of the circular-planar case, 
			we propose a method,
				illustrated by a “map'' of the co-orbital motion,
			that takes advantage of the averaged problem
			in order to compute co-orbital trajectories in the synodic reference frame.
		The results are presented in the case of a Sun-Jupiter like system, but the “map'' of the co-orbital motion, 
			plotted in the Fig.~\ref{fig:5}, 
			is independent of the small parameter $\eps$, 
			that is, the mass ratio of the planet over the total masses of the system.
		Hence, only the elements for which the minimal mutual distance $\Delta$ is equal to ${ NR_\rH}$,
			must be computed in order to apply the method to a different Sun-planet system 
			(or planet-moon system).
		For example, Fig.~\ref{fig:7} displays the map of the co-orbital motion for a Sun-Earth like system 
		($\eps = 1/333333$).
		We recall that,
			since the accuracy of the averaged problem depends on $\eps$,
				the larger $\eps$ is, the less the map of figure~\ref{fig:5} is reliable. 
		
		A practical application of our theoretical results interests the design of space missions.
		Let us consider a spacecraft affected by the gravitational forces of a Sun-Earth like system. 
		The selected orbits are usually remarkable solutions in the synodic reference frame
			that is,
			the Lagrange fixed points
			and periodic or quasi-periodic trajectories, as well as their associated hyperbolic manifolds (when existing).		
		However, except for the dynamics associated with $L_1$ and $L_2$, most of these solutions are located at a remote distance from the Earth, and thus the cost in terms of energy to reach them is usually not affordable.
		For instance, 
				 Fig.~\ref{fig:7} shows that 
				 	$L_3$, $L_4$, $L_5$, the short-periodic and long-periodic families 
					and the Lyapunov family $\sL_3$
				lay at a distance larger than 40 Hill's radius.
		Only the family $f$ provides periodic orbits available at a lower distance, that is why it becomes 
		an important topic for mission design.

		Our idea is the following: 
			since the duration of a mission is limited,
			it is not necessary to target 
			an equilibrium or a periodic solution in the synodic reference frame.
		With the help of our method,
			that characterizes the elliptic and hyperbolic dynamics of the co-orbital motion 
			through the averaged problem
			as well as defined a time of stability of these solutions,
		it is easy to select an initial condition on the map of Fig.~\ref{fig:7}, 
		that is located close enough to the Earth,
			and that satisfies a given co-orbital dynamics for the whole duration of the mission.
		For instance, for a 30-years mission,
			Theorem \ref{theo:stability} 			
			ensures that a co-orbital solution,
				that is located outside 10 Hill's radii of the Earth,
			will be stable at least during the time of the mission.
		As a consequence,  the horseshoe, tadpole and quasi-satellite solutions become possible target trajectories.
		Notice that  co-orbital solutions that experience closest approaches with the Earth
			may be also stable for the considered time of the mission.
		However, Theorem  \ref{theo:stability}
			requires a global numerical investigation of the time of stability of the co-orbital orbits.
		A detailed study will be addressed in a forthcoming work.
 		
		Finally, we point out that the application of the averaged problem as presented here 
			is not restricted to the co-orbital motion in the circular-planar case:
		it can also be applied to inclined co-orbital trajectories,
			or, more generally, to solutions associated with other mean-motion resonances.
		To this aim, a careful study of the averaged phase space must be realized.
			
%
%
%
%
%
%
%
%
%

	\appendix
	\section{Proofs}	
	
\label{sec:appendix}

		\subsection{Remainders}
	
		We recall that the Hamiltonian flow at a time $t$, 
		generated by an auxiliary function $h(\bX)$, 
		satisfies the following property:
			\be 
				\frac{\rd}{\rd t}\left(g\circ \Phi_t^h\right) 	
					= \cL_hg \circ\Phi_t^f
\label{eq:A0_Flow_Prop}
			\ee
			where $g$ is an auxiliary function.
		Thus, we have the following Taylor expansions:
			\begin{align}
\label{eq:A0_TaylorExp0}
				g\circ\Phi^f_t &= g 			+ \int_0^t \cL_f g\circ \Phi_s^f\rd s,\\
\label{eq:A0_TaylorExp1}
				g\circ\Phi^f_t &= g	+ \cL_fg 	+ \int_0^t (1-s)\cL_f(\cL_f g)\circ \Phi_s^f\rd s.
			\end{align}

	\subsection{Proof of Lemma \ref{lem:HKHPS}}
	
			 	For given ${\rho\pleq 1}$, ${\Delta\pleq 1}$, ${\eps \pleq 1}$ and ${\kappa >0}$,
					the domain $\fD_\kappa$ has been designed
					in order to exclude the collision manifold and make the perturbation $H_\rP$ analytic.
				Hence, $H_\rP$  is bounded on $\fD_\kappa$ as well as its partial derivatives with respect to 
				$\theta$, $u$, $\xt$, $x$, $\yt$ or $y$ up to an arbitrarily fixed order ${n\geq 1}$.
				Their estimates are deduced from the following reasonings.
				
				First of all, we recall the thresholds provided by the definition of $\fD_\kappa$:			
				\bes
					\Deltat/\kappa <\norm{\br\circ\Ups}_{\kappa}  \leqp 1, \,\,
					\Delta/\kappa\leq \norm{\br\circ\Ups-\br'}_\kappa \leqp 1,\,\, 
					\norm{\br'}_\kappa \eqp 1,
				\ees
				where
				${\Deltat=\cO(1)}$ since it is an arbitrarily fixed quantity
				that does not depend on $\eps$, $\Delta$ and $\rho$. 
				For $n\geq 1$, 
					the perturbation $\cH_\rP$ in heliocentric cartesian coordinates, Eq.~\eqref{eq:HamPert},
					yields the following estimates on $\fD_\kappa$:
				\be
				\norm{H_\rP}_\kappa = \norm{\cH_\rP\circ\Ups}_\kappa \leqp \frac{\eps}{\Delta},
				\quad
				\norm{\frac{\partial^n\cH_\rP}{\partial \br^n}\circ\Ups}_\kappa \leqp \frac{\eps}{\Delta^{n+1}}.
\label{eq:A1_1}	
				\ee

				Since the transformation $\Upsh$, 
					that introduces the Poincar\'e complex variables, Eq.~\eqref{eq:Poinc_transf},
					is regular when eccentricity and inclination tend to zero,
				it does not have singularities.
				More precisely, it is an analytic transformation on $\fD_\kappa$
					and, for each order, its derivatives 
					can be bounded by a constant that does not depend on $\eps$, $\rho$ and $\Delta$.
				$\Upsc$, 
				 	which introduces the resonant variables, Eq.~\eqref{eq:res_transf}, 
					is an affine transformation that 
				 	fulfills the same properties.
					As a consequence, $\Ups$ is analytic 
					and, for each order
					${n\geq 1}$ and ${(W_i)_{i\leq n} \in\{\theta,u,\xt,x,\yt,y\}}$,
					the following threshold is satisfied:
				\be
					\norm{\frac{\partial^n(\br\circ\Ups)}{\partial{W_1}\ldots\partial{W_n}}}_{\kappa} \leqp 1.
\label{eq:A1_2}	
				\ee

\begin{sloppypar}
				Finally, for a given analytic function $\cF(\br, \lam')$ and ${W\in\{\theta,u,\xt,x,\yt,y\}}$, 
					we recall the chain rule:
				\be
					\dron{\cF\circ\Ups}{W} = \left(\dron{\cF}{\br}\circ\Ups\right) \bigcdot \dron{(\br\circ\Ups)}{W}.
\label{eq:A1_3}
				\ee
				The bounds on the partial derivatives of $H_\rP$
					with respect to $\theta$, $u$, $\xt$, $x$, $\yt$ and $y$,
					and up to an arbitrarily fixed order ${n\geq 1}$,
					are deduced
					from the combination of the chain rule, Eq.~\eqref{eq:A1_3},
					with the thresholds given by  Eq.~\eqref{eq:A1_1} and Eq.~\eqref{eq:A1_2}.
\end{sloppypar}

				Since 
				${\norm{\Hb_\rP}_\kappa \leq \norm{H_\rP}_\kappa}$,
				the results on the averaged perturbation
					is a direct consequence of the previous developement.
\smallskip

\begin{sloppypar}
				On the domain $\fD_2$, $H_\rK$ 
					is analytic,
					is different from zero,
					and thus satisfies $\norm{H_\rK}_2 \eqp 1$
					while, for each order, its derivatives are bounded.	
				More precisely, over the compact ${\modu{u}\leq 2\rho}$, 
				 	there exists a constant ${M>0}$ that does not depend of $\eps$, $\rho$ and $\Delta$ 
					such that
				${\norm{H_\rK'''}_{2} \leq M.}$
				Hence,
					the upper bound of the derivative at second order
					on the smaller domain $\fD_{3/2}$
					can be deduced by the mean value theorem, that is,
				${\norm{H_\rK''}_{3/2} \leq 2 M \rho \leqp \rho.}$
				Finally, 
				\bes
					\modu{H'_\rK} = \frac{3pq^{-1}\sqrt{\tilde{a}}\modu{u}}{\modu{\sqrt{\tilde{a}}+u}^3}\modu{1  + \sqrt{\tilde{a}}u + \frac{u^2}{3\sqrt{\tilde{a}}}}
				\ees
				implies that $\norm{H'_\rK}_{3/2}\leqp \rho$,
				while
				\bes
					\modu{H_\rK - H_\rK(0)} 
					= \frac{3u^2}{2\tilde{a}\modu{\sqrt{\tilde{a}} + u }^2}\modu{1 + \frac{2}{3}\frac{u}{\sqrt{\tilde{a}}}}
				\ees
				provides the estimates of Eq~\eqref{eq:Lem1_HK_born}.
\end{sloppypar}

		\subsection{Proof of Theorem \ref{theo:averaging}}	
	
			First of all, we recall some results about the construction of the transformation of averaging:
			${\Upsb=\Phi_1^S}$ with $S$ that reads:
			\be
			\begin{aligned}
			&S(\theta, u, \xt,x,\yt,y,\lam') \\
			&\phantom{S}=
				\frac{1}{2\pi} \int_0^{2\pi} s (H_\rP - \Hb_\rP)_{(\theta, u, \xt,x,\yt,y,\lam' + s)} \rd s,
\label{eq:A_S}
			\end{aligned}
			\ee
			in order to satisfy the following property:
			\be
				\cL_S\Xi = -\dron{S}{\lam'} =   \Hb_\rP - H_\rP.
\label{eq:A_Scond}
			\ee
			Under these conditions, the remainder  of the averaging process reads:
			\be
				H_* =  	 (\exp\cL_S - \Iden)(H_\rK + H_\rP) +  (\exp\cL_S - \cL_S - \rI\rd)\Xi.
\label{eq:A_H*}
			\ee

		Most of the estimates required for the proof are computed 
				with the Taylor expansion at zero order, Eq.~\eqref{eq:A0_TaylorExp0}.
				More precisely, 
				if we assume that there exists ${\kappa<3/2}$ such that ${\Upsb(\fD_\kappa)\subset \fD_{3/2}}$,
					then for a given function $g$ that depends on ${(\theta, u, \xt, x, \yt, y, \lam')}$, 
					the following thresholds are ensured: 
			\bes
				\norm{g\circ \Upsb - g}_\kappa	\leqp \norm{\dron{S}{W}}_{3/2} \norm{\dron{g}{W}}_{3/2}
				\quad\mbox{for $W\in\{\theta, u, \xt, x, \yt, y\}$.}
			\ees 
		 For ${n\geq 1}$ and ${(W_i)_{i\leq n}\in \{\theta, u, \xt, x, \yt, y\}}$, 
		 we point out that 
		 	Eq.~\eqref{eq:A_S} and Lemma  \ref{lem:HKHPS}  provide the following thresholds on the partial derivatives of $S$:		
		\be
			\norm{\frac{\partial^n S}{\partial W_1\ldots\partial W_n}}_{3/2} 
				\leqp \frac{\eps}{\Delta^{n+1}}. 
\label{eq:A_derS}
		\ee 
		Hence, in $\fD_\kappa$ and for ${W\in \{\theta, u, \xt, x, \yt, y\}}$,
			we have the following:
		\bes
				\begin{aligned}
				\norm{\br\circ\Upsb - \br'}_\kappa
					&\geq \norm{\br- \br'}_\kappa - \norm{\cL_S(\br\circ\Ups)}_{3/2}\\
				\norm{\br\circ\Upsb}_\kappa 
					&\geq \norm{\br}_\kappa - \norm{\cL_S(\br\circ\Ups)}_{3/2}\\
				\norm{W\circ\Upsb}_{\kappa} 
					&\leq \norm{W}_\kappa + \norm{\cL_S W}_{3/2}
				\end{aligned}
		\ees
		with 
		\bes
			\norm{\cL_S(\br\circ\Ups)}_{3/2} 	
			\leqp \frac{\eps}{\Delta^2},
			\quad
			\norm{\cL_S W}_{3/2} \leqp \frac{\eps}{\Delta^2}.
		\ees		
		As a consequence, 
			for ${\eps \pleq \rho\Delta^2}$ and  ${\eps\pleq \Delta^3}$ with $\eps$, $\rho$ and $\Delta$ small enough,
			we can choose $\kappa = 4/3$ such that 
		the symplectic transformation of averaging satisfies ${\Upsb(\fD_{\kappa})\subset\fD_{3/2}}$
		and is close to identity such that
		\be
				\norm{W\circ\Upsb- W}_{{4/3}} \leqp \frac{\eps}{\Delta^2},
				\quad
				\norm{\dron{\Upsb}{W}}_{4/3} \leqp 1
\label{eq:A_Tranf}
		\ee
		for ${W \in\{\theta, u, \xt, x, \yt, y\}}$.		
	
\begin{sloppypar}		
		 It remains to prove the estimates on the remainder of the averaging process.		
		The Taylor expansions at zero and first order, Eqs.~(\ref{eq:A0_TaylorExp0}-\ref{eq:A0_TaylorExp1}),
			 combined with the condition satisfied by $S$, Eq.~\eqref{eq:A_Scond},
		ensure that the remainder $H_*$, Eq.~\eqref{eq:A_H*}, can be written as
			\bes
			\begin{aligned}
			H_* 	&=  \int_0^1 \cL_SH_\rK\circ\Phi_s^S\rd s  
				+  \int_0^1 s\cL_SH_\rP\circ\Phi_s^S\rd s   \\
				&\phantom{=}+  \int_0^1(1-s)\cL_S\Hb_\rP\circ\Phi_s^S\rd s .
			\end{aligned}
			\ees			
		As a consequence, 
			the thresholds given by Eq.~\eqref{eq:A_derS} and Lemma \ref{lem:HKHPS} provide 
			the following upper bound on the remainder:		
		\bes
				\norm{H_*}_{4/3} 
							\leqp 	\left(
									\norm{\cL_S H_\rK}_{3/2} 		
							+ 		\norm{\cL_S H_\rP}_{3/2}
									\right)
		\ees
		with
		\bes		
				\norm{\cL_S H_\rK}_{3/2} 					
				\leqp \frac{\eps\rho}{\Delta^2}, \quad
				\norm{\cL_S H_\rP}_{3/2} 
				\leqp \frac{\eps^2}{\Delta^4}.
		\ees		
		The upper bound on the derivative of $H_*$ with respect to $\theta$, $u$, $\xt$, $x$, $\yt$ and $y$
			is deduced in the same way. 
		For ${(W_i)_{i\leq 2}\in\{\theta, u ,\xt, x, \yt, y\}}$, 
			Lemma \ref{lem:HKHPS} and Eqs.~(\ref{eq:A_derS}-\ref{eq:A_Tranf})
			provide the following upper bounds:
		\bes
				\norm{\dron{H_*}{W_1}}_{4/3}	
							\leqp 	\left(	\norm{\dron{\cL_S H_\rK}{W_1}}_{3/2}
							+ 		\norm{\dron{\cL_S H_\rP}{W_1}}_{3/2}
									\right)
		\ees
		with
		\bes
				\norm{\dron{\cL_SH_\rK}{W_1}}_{3/2} 
				\leqp \frac{\eps}{\Delta^2}\left(\frac{\rho}{\Delta} +  \rho\right),
				\quad
				\norm{\dron{\cL_SH_\rP}{W_1}}_{3/2}
				\leqp \frac{\eps^2}{\Delta^5}. \quad
		\ees
\end{sloppypar}

		\subsection{Proof of Theorem \ref{theo:stability}}

		The proof follows the classical strategy described in \cite{1989Ar}.
		The aim of the first part of the proof is to bound the Hamiltonian vector field linked to 
		$H_*$ over the domain $\fD_{4/3}$. 
		The second part comes from the size of the transformation of averaging $\Upsb$ 
		and the choice of a time ${\cT>0}$ 
		which gives terms of the same order in the upper bound on the error of approximation.

\smallskip

		For a given initial condition $\bX_0\in\fD_1$ and a time of escape ${\cT_1>0}$, 
			we assume that the solution $\und{\bX}(t)$ 
			governed by ${\Xi + \Hb}$, does not escape of $\fD_1$ 
			for all ${\modu{t}\leq\cT_{1}}$. 
		Hence, for a given ${\mu>0}$ that satisfies 
		\be
			\mu\pleq \Delta,
			 \qtext{and}
		  	\mu\pleq \rho,
\label{eq:A0_mucond}
		\ee
		we can ensure that the neighborhood $${\fE_\mu(t) = \{\norm{	\und{\bX}(t) - \bX}_{4/3} \leq \mu \}}$$ 
		belongs to $\fD_{4/3}$ for $\modu{t}\leq \cT_{1}$.
		For a given initial condition in ${\bXt_0\in\fE_\mu(0)}$, 
		$\bXt(t)$ denotes the solution  at a time $t$  generated by 
		the flow of the original Hamiltonian ${\Xi + \Hb + H_*}$.
		We assume that there exists a time  $\cT>0$ 
			such that $\bXt(t)$ belongs to the neighborhood $\fE_\mu(t)$ for ${\modu{t}\leq \min(\cT, \cT_1)}$.
					
		For ${\bW=(\theta,u,\xt,x,\yt,y)}$, we denote ${\bW_*(t) = \bWt(t) - \und{\bW}(t)}$  
				the ‘‘error" at a time $t$ on the approximation given by the averaged problem
				with respect to the solution of the original one.
		The vector field of $\bW_*(t)$, deduced from Eq.~\eqref{eq:A0_Flow_Prop}, 
		can be written as
		${\dot{\bW}_*(t) 	= \bF_1(t) + \bF_2(t)}$
		with
		\bes
		\begin{aligned}
			\bF_1(t) &= (\cL_{H_*}\bW)(\bWt(t), \lam'(t)),\\
			\bF_2(t) &= (\cL_{\Hb}\bW)(\bWt(t)) -  (\cL_{\Hb}\bW)(\und{\bW}(t)).
		\end{aligned}
		\ees	
		$\bF_1$ corresponds to the vector field of the remainder $H_*$
			whose upper bound has been computed in Theorem \ref{theo:averaging}.
		$\bF_2$ is derived from the difference between 
			the two considered Hamiltonian flow.		
		It has to be estimated.
		For that purpose, we apply the Taylor expansion at first order, Eq.~\eqref{eq:A0_TaylorExp1},
		to 
		\bes
			{\bF_2(t) 	=  \bG\circ\cS_1(t) - \bG\circ\cS_0(t)}\ees 
		where
		\bes
		\begin{gathered}
			\bG(\bW) = (\cL_{\Hb}\bW)(\bW),
			\quad
			\cS_s(t) = \bWt(t)  - s\bW_*(t).
		\end{gathered}
		\ees
		Thus, we deduce the following upper bound from the thresholds of Lemma \ref{lem:HKHPS} 
			and the conditions on $\mu$, Eq.~\eqref{eq:A0_mucond},
		\bes
		\begin{aligned}
		\norm{\bF_2(t)}_{4/3} 
			&\leq \left(\norm{\rd \bG}_{4/3} + \mu\norm{\rd^2 \bG}_{4/3}\right)	\norm{\bW_*(t)}_{4/3}\\
								&\leqp \left(\frac{\eps}{\Delta^3} + \rho\right) \norm{\bW_*(t)}_{4/3}.
		\end{aligned}
		\ees
		Since, we choose ${\rho \eqp \sqrt{\frac{\eps}{\Delta}}}$,
			the upper bound on the vector field of $\bW_*(t)$ can be written:
		\bes
		\norm{\dot{\bW}_*(t)}_{4/3} \leq  a\norm{\bW_*(t)}_{4/3} +  b\\
		\ees
		with
		\bes
			a \eqp  \sqrt{\frac{\eps}{\Delta^3}},
			\quad
			b \eqp \frac{\eps}{\Delta^2} \sqrt{\frac{\eps}{\Delta^3}}.
		\ees		
		As a consequence,		
			for a given ${d>0}$ such that 
		\bes
		 \norm{\dot{\bW}_*(0)}_1\leq d<\mu,
			\qtext{and} 
			0<t<\cT,
		\ees
			the errors on the approximation are bounded by $\mu$ as
		\be
			(d + bt)\exp (at) \leq \mu.
\label{eq:A2_cond}
		\ee	
	
\begin{sloppypar}	
		From now on, 
		we consider the initial condition ${\bX_0\in\fD_1}$ in the (non-averaged) resonant variables.
		The properties on the transformation of averaging impose that there exists a ${\bXt_0\in\fD_{4/3}}$
		such that
		$\norm{\bX_0 - \bXt_0} \leqp \frac{\eps}{\Delta^2}$.
		Hence, it imposes ${d\eqp \frac{\eps}{\Delta^2}}$ in order to get $\bXt_0$ in the neighborhood
		of the initial condition $\bX_0$ in the ‘‘averaged" resonant variables.
		This choice on $d$ fulfills the condition of Eq.~\eqref{eq:A0_mucond}.
		As a consequence, we can choose ${\cT = 2\pi \sqrt{\frac{\eps}{\Delta^3}}}$ in order to deal with terms 
		of the same order as in Eq.~\eqref{eq:A2_cond}, which gives ${\mu \eqp \frac{\eps}{\Delta^2}}$.
\end{sloppypar}

\begin{acknowledgements}
The authors are indebted to Philippe Robutel for key discussions 
concerning the rigorous treatment of the averaging process.\\
They also acknowledge the support of the project entitled 
‘‘\textit{co-orbital motion and three-body regimes in the solar system}",
funded by Fondazione Cariplo through the program:
‘‘\textit{Promozione dell'attrattivit\`a e competitivit\`a dei ricercatori su strumenti
dell'European Research Council -- Sottomisura rafforzamento}".	
\end{acknowledgements}

%
\section*{Declarations}
\subsection*{Funding}
This work was supported by the project entitled 
‘‘\textit{co-orbital motion and three-body regimes in the solar system}",
funded by Fondazione Cariplo through the program:
‘‘\textit{Promozione dell'attrattivit\`a e competitivit\`a dei ricercatori su strumenti
dell'European Research Council -- Sottomisura rafforzamento}"

\subsection*{Conflict of interest}

 The authors declare that they have no conflict of interest that are relevant to the content of this article.

\subsection*{Availability of data and materials}

Data sharing not applicable to this article as no datasets were generated or analyzed during the current study.

\bibliographystyle{spmpsci}      

%
%

\end{document}